\documentclass[10pt,twocolumn,showpacs,preprintnumbers,amsmath,amssymb,aps,prx,longbibliography,superscriptaddress]{revtex4-1}
\usepackage{mathrsfs}
\usepackage{graphicx}
\usepackage{dcolumn}
\usepackage{bm}
\usepackage{amsmath}
\usepackage{amsfonts}
\usepackage{color}
\usepackage{hyperref}

\begin{document}

\title{Twisted Bilayer Graphene: A Phonon Driven Superconductor}
\author{Biao Lian}
\affiliation{Princeton Center for Theoretical Science, Princeton University, Princeton, New Jersey 08544, USA}
\author{Zhijun Wang}
\affiliation{Department of Physics, Princeton University, Princeton, New Jersey 08544, USA}
\author{B. Andrei Bernevig}
\affiliation{Department of Physics, Princeton University, Princeton, New Jersey 08544, USA}
\affiliation{Dahlem Center for Complex Quantum Systems and Fachbereich Physik, Freie Universitat Berlin, Arnimallee 14, 14195 Berlin, Germany}
\affiliation{Max Planck Institute of Microstructure Physics, 06120 Halle, Germany}


\begin{abstract}
We study the electron-phonon coupling in twisted bilayer graphene (TBG), which was recently experimentally observed to exhibit superconductivity around the magic twist angle $\theta\approx 1.05^\circ$. We show that phonon-mediated electron electron attraction at the magic angle is strong enough to induce a conventional intervalley pairing between graphene valleys $K$ and $K'$ with a superconducting critical temperature $T_c\sim1K$, in agreement with the experiment. We predict that superconductivity can also be observed in TBG at many other angles $\theta$ and higher electron densities in higher Moir\'e bands, which may also explain the possible granular superconductivity of highly oriented pyrolytic graphite. We support our conclusions by \emph{ab initio} calculations.
\end{abstract}

\date{\today}

\maketitle

Twisted bilayer graphene (TBG) is a highly tunable condensed matter system that exhibits a rich physics. The TBG system is engineered by stacking one graphene layer on top of another at a relative twist angle $\theta$, a procedure which produces a Moir\'e pattern superlattice potential. In recent experiments \cite{cao2018,cao2018b}, it was observed that TBG at low Moir\'e unit cell filling exhibits unconventional insulator and superconductor phases near the magic angle $\theta=1.05^\circ$. At this angle, the low energy electron bands of the superlattice are predicted to be extremely flat \cite{morell2010,bistritzer2011}. The Fermi energy of the system is below $10$meV, while by comparison the superconductor critical temperature $T_c\sim1$K is relatively high. Since then, some theoretical studies have been devoted to understanding the insulator and superconductor phases of the TBG \cite{yuan2018,po2018,xu2018,roy2018,volovik2018,padhi2018,dodaro2018,baskaran2018,wu2018, isobe2018,huang2018,you2018,wux2018,zhangy2018,kang2018,koshino2018,kennes2018,zhangl2018,
pizarro2018,guinea2018,thomson2018,ochi2018,xux2018,peltonen2018,fidrysiak2018,
zou2018,gonz2018,su2018,guo2018}. A closely related system, the highly oriented pyrolytic graphite (HOPG) which contains numerous twisted interfaces, was also earlier reported showing evidences of granular superconductivity \cite{esquinazi2013,ballestar2013,ballestar2014}, and is suggested to share a similar superconductivity mechanism as that in TBG \cite{esquinazi2014,volovik2018}.

Here we study electron-phonon mediated superconductivity of TBG. We show that the TBG Moir\'e pattern exhibits a strong electron-phonon coupling, which can lead to a conventional superconductivity with high $T_c$ at certain twist angles and electron densities. In particular, our calculation estimates a $T_c$ of order $1$K at the magic angle $\theta=1.05^\circ$ around a filling of $2$ electrons per superlattice unit cell, in agreement with the TBG experiment \cite{cao2018}. Most importantly, we make the falsifiable prediction that $T_c$ can be higher at larger electron densities and certain ranges of the twist angle $\theta$, for instance in the \emph{second Moir\'e} band near $\theta=0.6^\circ$, and in the \emph{second or higher Moir\'e bands} for $\theta\gtrsim 1^\circ$. This may explain the possible superconductivity of HOPG where the interface twist angles are mostly not at the magic angle. 
At last, we conjecture the insulating phase observed at $2$ electrons per unit cell is a Bose Mott insulator \cite{fisher1989}.

\begin{figure}[tbp]
\begin{center}
\includegraphics[width=3.4in]{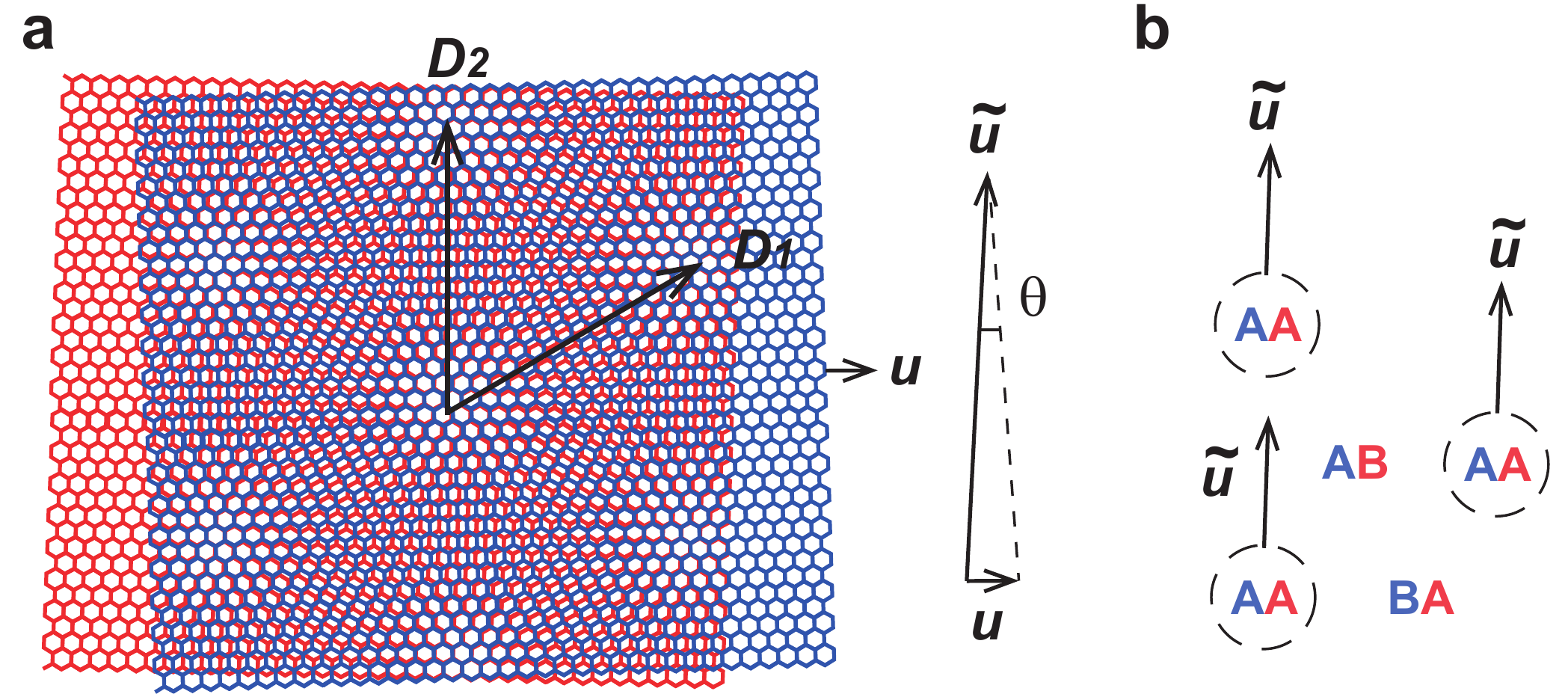}
\end{center}
\caption{{\bf a}. Moir\'e pattern of the TBG, where the AA stacking centers form a triangular superlattice, while AB and BA stacking centers form a dual hexagon superlattice (A and B denote the honeycomb sublattices). {\bf b}. When graphene layer $1$ (blue) undergoes a uniform displacement $\mathbf{u}$ relative to graphene layer $2$ (red), the superlattice sites displaces a distance $|\widetilde{\mathbf{u}}|=\gamma |\mathbf{u}|$ perpendicular to $\mathbf{u}$, with $\gamma=1/2\tan(\theta/2)$.}
\label{fig1}
\end{figure}

The Moir\'e superlattice of TBG is as shown in Fig. \ref{fig1}a, which has two lattice vectors $\bm{D}_1$ and $\bm{D}_2$ of length $|\bm{D}_j|=a_0/[2\sin(\theta/2)]$, where $a_0=0.246$ nm is the graphene lattice constant. The superlattice potential significantly modifies the electron band structure of graphene \cite{bistritzer2011}.

The large electron-phonon coupling of TBG can be intuitively understood from Fig. \ref{fig1}. We denote the phonon field in TBG layer $j$ ($j=1,2$) as $\mathbf{u}^{(j)}(\mathbf{r})$, namely, the atomic displacement at coordinate $\mathbf{r}$ in layer $j$. 
We then define the relative displacement $\mathbf{u}=\mathbf{u}^{(1)}-\mathbf{u}^{(2)}$, and the center of mass displacement $\mathbf{u}_c=(\mathbf{u}^{(1)}+\mathbf{u}^{(2)})/2$. The key observation is that for small $\theta$, a small in-plane relative displacement $\mathbf{u}$ can significantly affect the superlattice. For example, when layer $1$ of the TBG undergoes a uniform translation $\mathbf{u}$ relative to layer $2$ as shown in Fig. \ref{fig1}a, the AA stacking positions will move by $|\widetilde{\mathbf{u}}|=\gamma|\mathbf{u}|$ perpendicular to $\mathbf{u}$, where $\gamma=1/[2\tan(\theta/2)]$ (Fig. \ref{fig1}b). If the in-plane relative displacement $\mathbf{u}$ is nonuniform, it will induce a large superlattice deformation due to the amplification factor $\gamma\approx\theta^{-1}\gg1$ (\cite{video,suppl} Sec. I). Accordingly, the low energy electrons will experience a large variation of superlattice potential, yielding a strong coupling with the in-plane relative displacement phonon field $\mathbf{u}$. In contrast, the center of mass displacement $\mathbf{u}_c$ has no amplified effect on the superlattice, and has much weaker couplings to electrons (\cite{suppl} Sec. I). Therefore, we shall only focus on the electron-phonon coupling of the relative displacement field $\mathbf{u}$.

\begin{figure}[tbp]
\begin{center}
\includegraphics[width=3.4in]{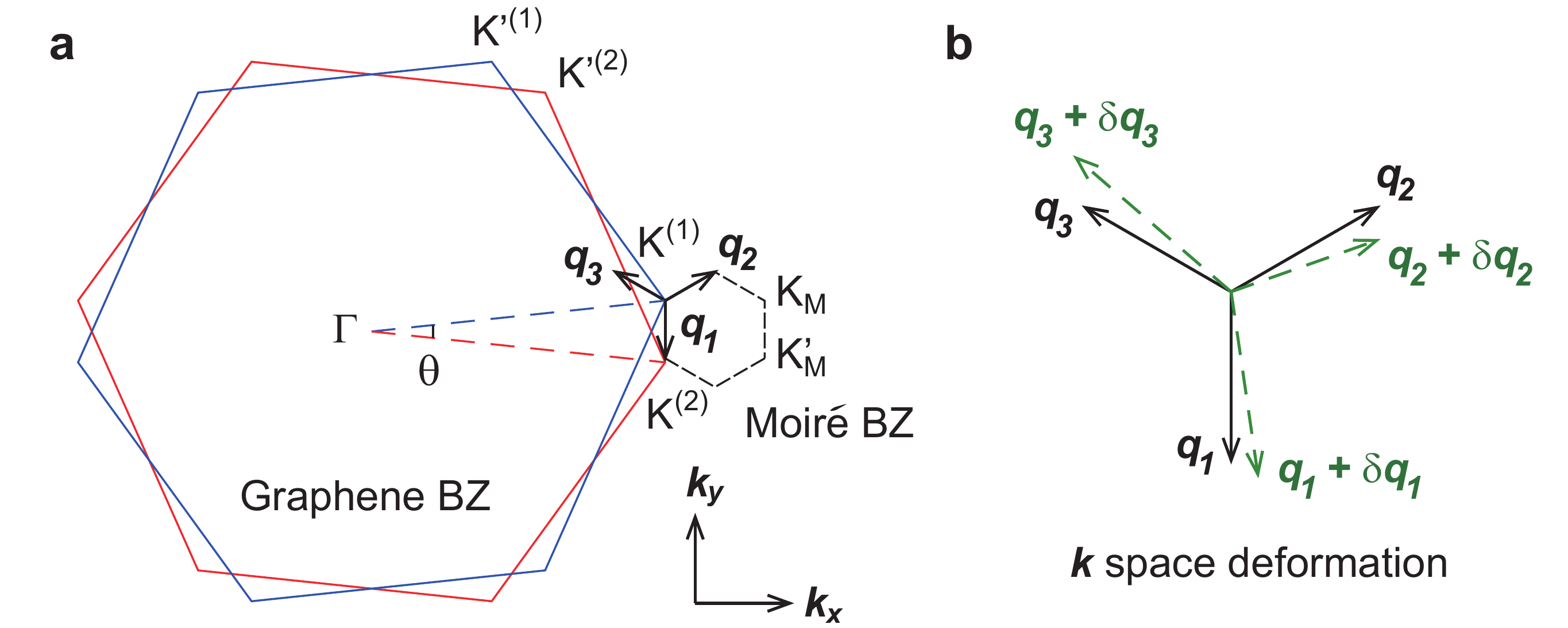}
\end{center}
\caption{{\bf a}. Illustration of the graphene BZs of two layers, and their relation to the Moir\'e BZ. 
{\bf b}. Under phonon induced superlattice deformations, $\mathbf{q}_j$ are deformed, which leads to the change in electron band energies.}
\label{fig2}
\end{figure}

The low energy band structure of the TBG can be calculated using the continuum model in momentum space constructed in Ref. \cite{bistritzer2011}. Fig. \ref{fig2}a shows the hexagonal graphene Brillouin zones (BZs) of the two layers, which are relatively twisted by $\theta$. When the two layers are decoupled, the low energy electrons of each layer are Dirac fermions at $K$ and $K'$ points, which are described by Hamiltonian $h^K(\mathbf{k})=v(\sigma_xk_x-\sigma_yk_y)=\hbar v\bm{\sigma}^*\cdot\mathbf{k}$ and $h^{K'}(\mathbf{k})=-\hbar v\bm{\sigma}\cdot\mathbf{k}$, respectively. Here $\sigma_{x,y,z}$ are the Pauli matrices for sublattice indices, $\hbar v\approx 610$ meV$\cdot$nm is the graphene Fermi velocity, and momentum $\mathbf{k}$ is measured from the Dirac point. In addition, each Dirac band has a 2-fold spin degeneracy, and we assume zero spin orbit coupling. The $K$ ($K'$) points of the two layers differ by momentum vectors $\mathbf{q}_j$ ($-\mathbf{q}_j$) as shown in Fig. \ref{fig2}a ($j=1,2,3$), which constitute the edges of the hexagonal superlattice Moir\'e BZ. Their lengths are given by $|\mathbf{q}_j|=k_\theta=8\pi\sin(\theta/2)/3a_0$.

When the interlayer hopping is introduced, a state of momentum $\mathbf{k}$ in layer $1$ can hop with a state of momentum $\mathbf{p}'$ in layer $2$ if $\mathbf{k}-\mathbf{p}'=\mathbf{q}_j$ or higher superlattice reciprocal vectors \cite{bistritzer2011}. If we only keep the nearest hoppings, to the lowest order, the Hamiltonian at valley $K$ and near Moir\'e BZ $K_M'$ point (Fig. \ref{fig2}a) has the truncated form \cite{bistritzer2011}
\begin{equation}\label{kp-model}
H^K(\mathbf{k})=\left(\begin{array}{cccc}
h^{K}(\mathbf{k})& wT_1& wT_2&wT_3\\
wT_1^\dag&h^{K}(\mathbf{k}_1)&0&0\\
wT_2^\dag&0&h^{K}(\mathbf{k}_2)&0\\
wT_3^\dag&0&0&h^{K}(\mathbf{k}_3)\\
\end{array}\right)\ ,
\end{equation}
where $\mathbf{k}$ is measured from $K_M'$ point, $\mathbf{k}_j=\mathbf{k}-\mathbf{q}_j$ ($j=1,2,3$), the matrices $T_j$ are given by $T_1=1+\sigma_x$, $T_2=1-\frac{1}{2}\sigma_x-\frac{\sqrt{3}}{2}\sigma_y$, $T_3=1-\frac{1}{2}\sigma_x+\frac{\sqrt{3}}{2}\sigma_y$, and $w\approx110$ meV is the nearest momentum hopping amplitude. In the vicinity of $\mathbf{k}=0$ and zero energy, the Hamiltonian (\ref{kp-model}) can be further folded into a $2\times2$ effective Dirac Hamiltonian \cite{bistritzer2011}
\begin{equation}\label{HK2}
\widetilde{H}^K(\mathbf{k})=\left(\frac{1-3\alpha^2}{1+6\alpha^2}\right)\hbar v\bm{\sigma}^*\cdot\mathbf{k}\ ,
\end{equation}
where $\alpha=w/\hbar vk_\theta$. In total, there are $4$ Dirac fermions at $K_M'$ and $4$ Dirac fermions at $K_M$ near zero energy, due to the valley $K,K'$ and spin $\uparrow,\downarrow$ 4-fold degeneracy (further momentum hoppings are needed in Eq. (\ref{kp-model}) to obtain the Dirac fermions at $K_M$). The Dirac fermions at valley $K'$ have an opposite helicity, described by Eq. (\ref{HK2}) with $\bm{\sigma}^*\cdot\mathbf{k}\rightarrow \bm{\sigma}\cdot\mathbf{k}$.
The magic angle $\theta\approx 1.05^\circ$ is given by $\alpha^2=1/3$, where the Fermi velocity of the Dirac band becomes zero. Numerical calculations at the magic angle show the entire band width of the lowest two bands can be as low as $1$ meV \cite{bistritzer2011}.

The coupling between electrons and long wavelength phonons can be obtained by examining the change of electron band energies under uniform lattice deformations. Under the superlattice deformation induced by a relative displacement $\mathbf{u}$,
one can show that the momentum vectors $\mathbf{q}_j$ (Fig. \ref{fig2}b) are deformed by $\delta\mathbf{q}_1=\gamma k_\theta(\partial_xu_x,\partial_yu_x)$, and $\delta\mathbf{q}_{2,3}=\gamma k_\theta(\pm\frac{\sqrt{3}}{2}\partial_xu_y-\frac{1}{2}\partial_xu_x,-\frac{1}{2}\partial_yu_x\pm \frac{\sqrt{3}}{2}\partial_yu_y)$ (\cite{suppl} Sec. I). This induces a change of $\mathbf{k}_j=\mathbf{k}-\mathbf{q}_j$ in the Hamiltonian (\ref{kp-model}), and thus perturbs the electron band energies. The variations of $v$ and $w$ are subleading compared to $\delta\mathbf{q}_j$, and will be ignored here. The variation in the
folded $2\times2$ Hamiltonian (\ref{HK2}), namely, the electron-phonon coupling $H_{\text{ep}}(\overline{\mathbf{k}})=\delta\widetilde{H}(\overline{\mathbf{k}})$, can be derived to be (\cite{suppl} Sec. I)
\begin{equation}\label{ep}
\begin{split}
&H_{\text{ep}}^{\eta,\zeta,s}(\overline{\mathbf{k}}) =H_{C3}^{\eta,\zeta,s}(\overline{\mathbf{k}})+H_{\text{SO}(2)}^{\eta,\zeta,s}(\overline{\mathbf{k}})\ ,\\
&H_{C3}^{\eta,\zeta,s}(\overline{\mathbf{k}})=g_{1\alpha}\eta\gamma \hbar v\psi^\dag_{\mathbf{k}'}[\overline{k}_x(\partial_yu_x+\partial_xu_y)\\
&\qquad\quad\quad +\overline{k}_y(\partial_xu_x-\partial_yu_y)]\psi_{\mathbf{k}}\ ,\\
&H_{\text{SO}(2)}^{\eta,\zeta,s}(\overline{\mathbf{k}})=\gamma \hbar v\psi^\dag_{\mathbf{k}'}[g_{2\alpha}(\eta\sigma_x\overline{k}_x-\sigma_y\overline{k}_y)(\partial_yu_x-\partial_xu_y) \\
&\qquad\quad\quad +g_{3\alpha} (\eta\sigma_x\overline{k}_y+\sigma_y\overline{k}_x)(\partial_xu_x+\partial_xu_y)]\psi_{\mathbf{k}}\ ,
\end{split}
\end{equation}
where index $\eta=\pm1$ is for graphene valley $K,K'$, $\zeta=\pm1$ is for Moir\'e BZ valley $K_M,K_M'$, $s=\pm1$ is for spin $\uparrow,\downarrow$, and we have defined $g_{1\alpha}=\frac{9\alpha^2(1+3\alpha^2)}{(1+6\alpha^2)^2}$, $g_{2\alpha}=\frac{9\alpha^2}{(1+6\alpha^2)^2}$ and $g_{3\alpha}=\frac{3\alpha^2}{1+6\alpha^2}$. $\psi_\mathbf{k}$ and $\psi^\dag_{\mathbf{k}}$ are the Dirac electron annihilation and creation operators, and $\overline{\mathbf{k}}=(\mathbf{k}+\mathbf{k}')/2$ is the average momentum of the initial and final electron state before and after phonon emission (absorption). Note that $H_{\text{ep}}$ is independent of $\zeta$ and $s$, and contains two parts $H_{C3}$ and $H_{\text{SO}(2)}$, which are $C_{3z}$ and SO($2$) rotationally invariant about $z$ axis, respectively. Besides, $H_{\text{ep}}$ respects the TBG 2-fold rotation symmetry $C_{2x}$ about $x$ axis, which transforms $(u_x,u_y)$ to $(-u_x,u_y)$.

We also need to know the phonon spectrum of the TBG. Previous studies show that the coupling between in-plane phonons of the two layers of TBG is extremely small \cite{cocemasov2013}, so we can approximate the interlayer coupling as zero. In this approximation, the TBG in-plane phonon spectrum is simply that of two isolated graphene monolayers folded into the Moir\'e BZ. The lowest bands of phonon field $\mathbf{u}$ is described by Hamiltonian
\begin{equation}\label{Hpha}
H_{\text{ph}}=\sum_\mathbf{p}\left(\hbar \omega_{\mathbf{p},L}a_{\mathbf{p},L}^\dag a_{\mathbf{p},L}+\hbar \omega_{\mathbf{p},T}a_{\mathbf{p},T}^\dag a_{\mathbf{p},T}\right),
\end{equation}
where $a_{\mathbf{p},L}$, $a_{\mathbf{p},L}^\dag$ and $a_{\mathbf{p},T}$, $a_{\mathbf{p},T}^\dag$ are the annihilation and creation operators of longitudinal and transverse polarized phonons, respectively. The frequencies $\omega_{\mathbf{p},L}=c_Lp$ and $\omega_{\mathbf{p},T}=c_Tp$ are acoustic, with $p=|\mathbf{p}|$. $c_L$, $c_T$ are the longitudinal and transverse sound speeds of monolayer graphene. The phonon field $\mathbf{u}$ at long wavelengths is \begin{equation}
\mathbf{u}(\mathbf{r})=\sum_{\mathbf{p}} \frac{e^{i\mathbf{p}\cdot\mathbf{r}}}{\sqrt{N_s\Omega_s}} (i\hat{\mathbf{p}}u_{\mathbf{p},L}+i\hat{\mathbf{z}}\times\hat{\mathbf{p}}u_{\mathbf{p},T})\ ,
\end{equation}
where $u_{\mathbf{p},\chi}= \sqrt{\frac{\hbar\Omega}{2M\omega_{\mathbf{p},\chi}}}(a_{\mathbf{p},\chi}+a_{-\mathbf{p},\chi}^\dag)$ for $\chi=L,T$ polarizations, $\Omega$ and $\Omega_s$ are the unit cell areas of the graphene lattice and Moir\'e superlattice, respectively, and $N_s$ is the number of supercells. There are also many optical phonon bands in the Moir\'e BZ corresponding to short wavelength components of $\mathbf{u}$, but here we will only focus on the lowest acoustic phonon bands in Eq. (\ref{Hpha}), since $H_{\text{ep}}$ in Eq. (\ref{ep}) is derived for long wavelength deformations.

\begin{figure}[tbp]
\begin{center}
\includegraphics[width=3.4in]{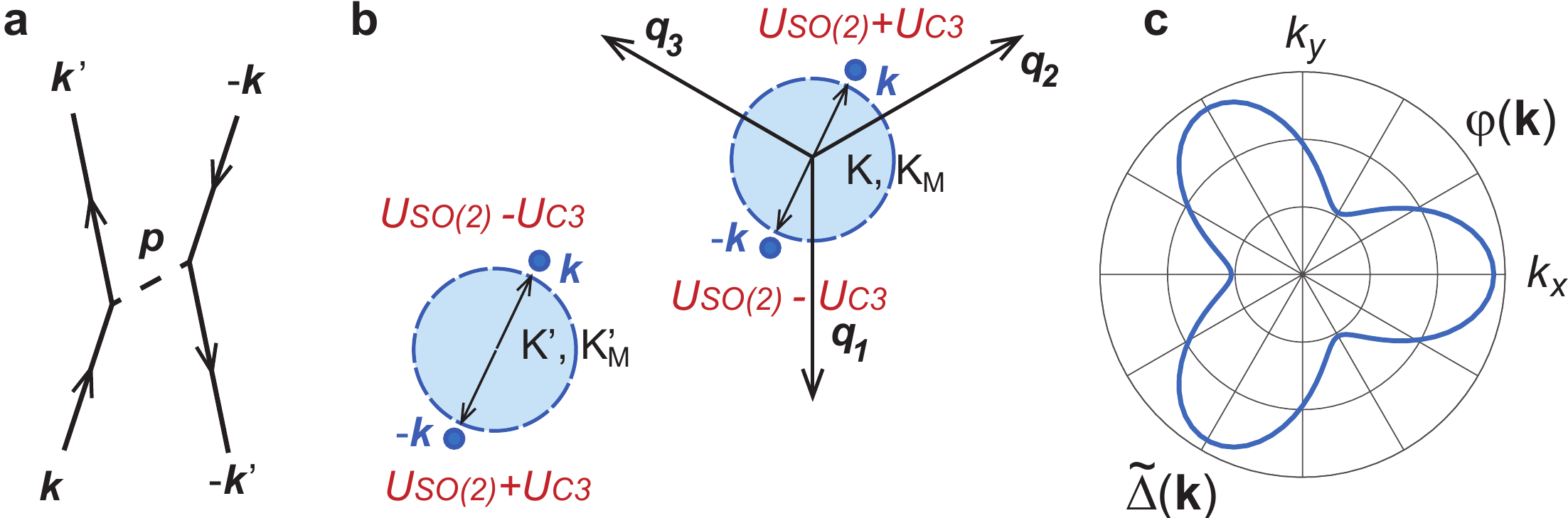}
\end{center}
\caption{{\bf a}. The process two electrons of momentum $\mathbf{k}$ and $-\mathbf{k}$ exchanges a phonon of momentum $\mathbf{p}=\mathbf{k}'-\mathbf{k}$, which mediates the electron-electron interaction in Eq. (\ref{Vkk}). {\bf b}. Illustration of phonon-induced potentials for electrons at valleys $K$ and $K'$ (we have plotted them at Moir\'e valleys $K_M$ and $K_M'$, respectively, but the potentials do not depend on Moir\'e valley), where $U_S$ and $\pm U_{C3}$ are contributed by $H^{\eta,\zeta,s}_{\text{SO}(2)}$ and $H^{\eta,\zeta,s}_{C3}$,
respectively. The dashed circles represent the Fermi surfaces.
{\bf c}. Intervalley pairing $\widetilde{\Delta}(\mathbf{k})$ solved numerically as a function of $\varphi(\mathbf{k})$, which is $s$-wave (\cite{suppl} Sec. III).}
\label{fig3}
\end{figure}

We now assume that the Fermi surfaces are given by $|\mathbf{k}|=k_F$ in the Dirac hole (or electron) bands, and calculate the phonon mediated electron-electron interaction near the Fermi surfaces. The Bardeen-Cooper-Schrieffer (BCS) channel of the interaction takes the generic form:
\begin{equation}
H_{\text{int}}^{(\text{ph})}=\sum_{\mathbf{k},\mathbf{k}'}\frac{V_{\mathbf{k}\mathbf{k}'}^{II'} (\omega)}{N_s\Omega_s} c^{\dag}_{\mathbf{k}',I}c^{\dag}_{-\mathbf{k}',I'} c_{-\mathbf{k},I'}c_{\mathbf{k},I}\ ,
\end{equation}
where $I=(\eta,\zeta,s)$ denotes indices for the eight Dirac cones, the frequency $\omega=(\xi_{\mathbf{k}'}-\xi_\mathbf{k})/\hbar$ with $\xi_\mathbf{k}=-\frac{1-3\alpha^2}{1+6\alpha^2}\hbar v(|\mathbf{k}|-k_F)$ being the band energy at $\mathbf{k}$, while $c_{\mathbf{k},I}$ and $c^{\dag}_{\mathbf{k},I}$ are electron annihilation and creation operators in the Dirac hole band $I$. 
To simplify the result, we take the approximation $c_L=c_T$ (both around $10^4$m/s). For $\theta$ near the magic angle ($\alpha^2\approx 1/3$), we find the interaction in the lowest two Moir\'e bands is
\begin{equation}\label{Vkk}
\frac{V^{II'}_{\mathbf{k}\mathbf{k}'}(\omega)}{\Omega_s}\approx\frac{\hbar^2v^2k_F^2 \varpi_{\mathbf{k}\mathbf{k}'}^{\eta\eta'}}{9Mc_T^2} \frac{\omega_{\mathbf{p},T}^2}{\omega^2-\omega_{\mathbf{p},T}^2} f_{\eta\eta'}(\varphi_\mathbf{k},\varphi_{\mathbf{k}'}),
\end{equation}
where $M$ is the Carbon atomic mass, $\mathbf{p}=\mathbf{k}-\mathbf{k}'$, $\varphi_{\mathbf{k}}=\arg(k_x+ik_y)$ is the polar angle of $\mathbf{k}$, and $\varpi_{\mathbf{k}\mathbf{k}'}^{\eta\eta'}=\phi^{\eta\dag}_{\mathbf{k}'}\phi^{\eta}_{\mathbf{k}} \phi^{\eta'\dag}_{-\mathbf{k}'}\phi^{\eta'}_{-\mathbf{k}}$ with $\phi^{\eta}_{\mathbf{k}}=(1,-\eta e^{-i\eta\varphi_\mathbf{k}})^T/\sqrt{2}$ being the Dirac hole band wave function at valley $\eta$. The function $f_{\eta\eta'}$ is given by
\begin{equation}
f_{\eta\eta'}(\varphi_\mathbf{k},\varphi_{\mathbf{k}'})=
\begin{cases}
&-1-2\cos(\varphi_{\mathbf{k}}-\varphi_{\mathbf{k}'}),\ (\eta=\eta')\\
&\left|1-\eta\frac{e^{2i \varphi_\mathbf{k}} -e^{2i \varphi_{\mathbf{k}'}} }{e^{-i \varphi_\mathbf{k}}-e^{-i \varphi_{\mathbf{k}'}} }\right|^2\ .(\eta=-\eta')\\
\end{cases}
\end{equation}
The interaction $V_{\mathbf{k}\mathbf{k}'}^{II'} (\omega)$ is independent of spin $s$ and Moir\'e valley $\zeta$.

At low energies $|\omega|<\omega_{\mathbf{p},T}$, $f_{\eta,-\eta}>0$ indicates the intervalley interaction between $K$ and $K'$ ($\eta=-\eta'$) is attractive. In contrast, $f_{\eta,\eta}$ is on-average negative, and one can prove that the intravalley interaction ($\eta=\eta'$) is repulsive in all pairing channels (\cite{suppl} Sec. IIC).
This is due to the fact that the hole (or electron) band projections of $H_{C3}^{\eta,\zeta,s}$ and $H_{SO(2)}^{\eta,\zeta,s}$ in Eq. (\ref{ep}) are odd and even under $\mathbf{k},\mathbf{k}'\rightarrow -\mathbf{k},-\mathbf{k}'$, or under $\eta\rightarrow -\eta$, respectively (\cite{suppl} Sec. IIC). Assume an electron state (wave packet) $|\overline{\mathbf{k}}_{K,\zeta,s}\rangle$ around momentum $\overline{\mathbf{k}}$ at valley $K$ experiences a phonon-induced lattice potential $\langle H_{C3}^{K,\zeta,s}(\overline{\mathbf{k}})\rangle+\langle H_{SO(2)}^{K,\zeta,s}(\overline{\mathbf{k}})\rangle=U_{C3}+U_{SO(2)}$. By symmetry, the state $|-\overline{\mathbf{k}}_{K,\zeta',s'}\rangle$ at the same valley $K$ will feel a potential $-U_{C3}+U_{SO(2)}$, while the state $|-\overline{\mathbf{k}}_{K',\zeta',s'}\rangle$ in the opposite valley $K'$ will feel a potential $U_{C3}+U_{SO(2)}$. Therefore, two electrons $|\overline{\mathbf{k}}_{K,\zeta',s'}\rangle$ and $|-\overline{\mathbf{k}}_{K',\zeta',s'}\rangle$ in opposite valleys feel the same phonon-induced lattice potential, which induces an effective attraction between them. In contrast, two electrons $|\overline{\mathbf{k}}_{K,\zeta',s'}\rangle$ and $|-\overline{\mathbf{k}}_{K,\zeta',s'}\rangle$ in the same valley $K$ feel different potentials, so the effective attraction between them is weaker or even absent. Therefore, the intervalley Cooper pairing is preferred.

This does not yet uniquely determine the form of pairing. Since the Dirac bands are degenerate with respect to indices $\zeta$ and $s$, the intervalley pairing could be either spin-singlet Moir\'e valley-triplet, or spin-triplet Moir\'e valley-singlet. Here we shall simply assume the pairing is time reversal invariant, which is generically more robust under non-magnetic disorders \cite{anderson1959}. This forces a pairing between opposite Moir\'e valleys and opposite spins, and yields an intervalley pairing amplitude (\cite{suppl} Sec. III)
\begin{equation}
\Delta_{\mathbf{k}}^{\eta\eta',\zeta\zeta',ss'}= s\delta_{s,-s'} \delta_{\zeta,-\zeta'}\delta_{\eta,-\eta'} \widetilde{\Delta}(\eta\mathbf{k})\ ,
\end{equation}
where $\widetilde{\Delta}(\mathbf{k})$ is a real function of $\varphi_{\mathbf{k}}=\arg(k_x+ik_y)$. The Numerically, $\widetilde{\Delta}(\mathbf{k})$ can be solved and has the shape shown in Fig. (\ref{fig3})c, which is nodeless and dominated by $s$-wave. We note that an earlier phonon study \cite{wu2018} obtained both $s$-wave and $d$-wave, while a recent atomistic study supports $s$-wave \cite{choi2018}.

\begin{figure}[tbp]
\begin{center}
\includegraphics[width=3.4in]{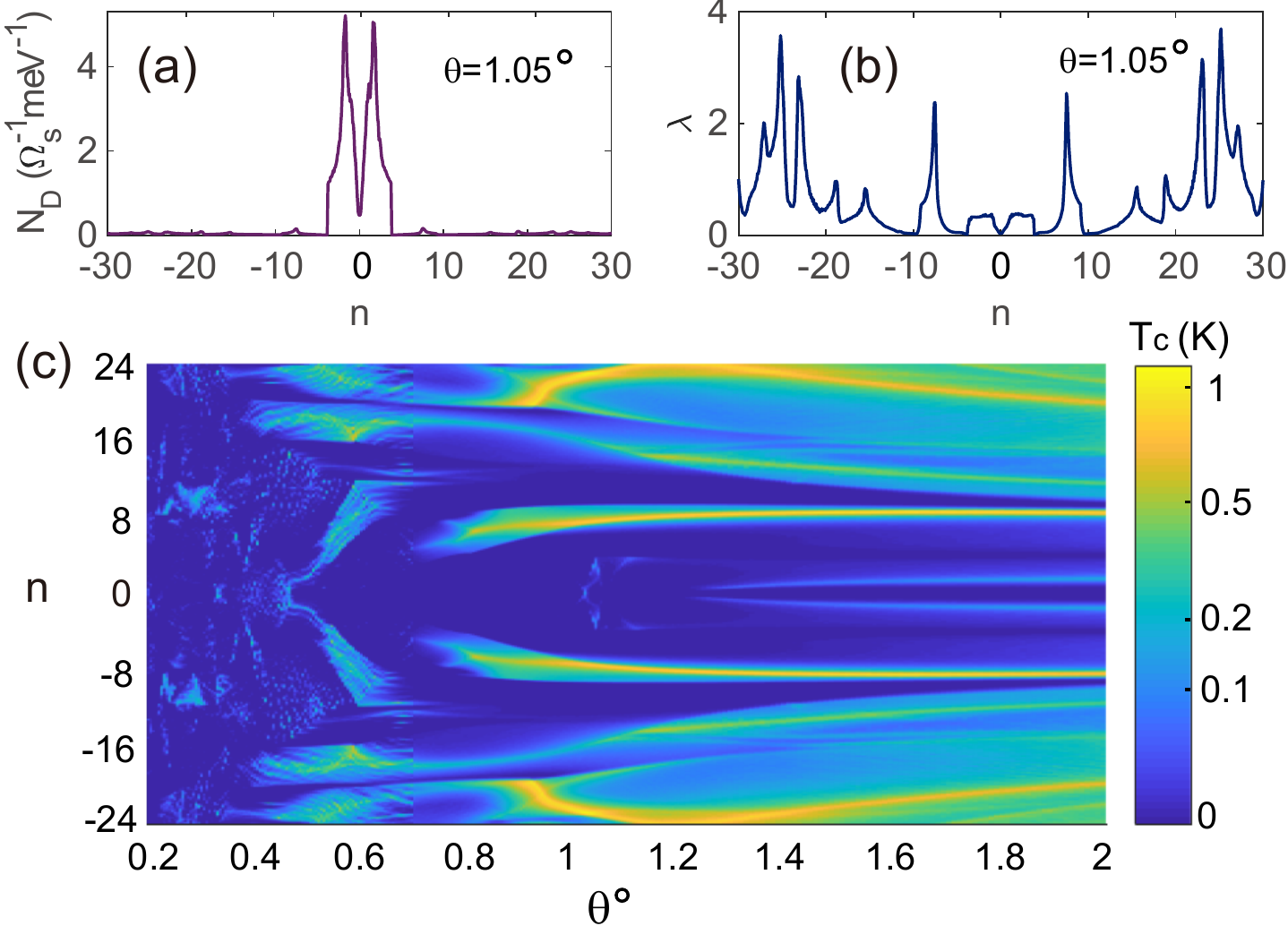}
\end{center}
\caption{{\bf a}. The density of states $N_D$ as a function of number of electrons per superlattice unit cell $n$, showing the lowest conduction and valence bands are pretty flat. {\bf b}. The estimated BCS coupling strength $\lambda\approx N_D|V_{\mathbf{k}\mathbf{k}'}(0)|$ as a function of $n$. {\bf c}. The superconducting $T_c$ with respect to $n$ and $\theta$ estimated from the McMillan formula. 
}
\label{fig4}
\end{figure}

Substituting the realistic parameters into Eq. (\ref{Vkk}) and taking $k_F\sim k_\theta$, we find that the phonon mediated intervalley attraction is of order of magnitude $-V^{II'}_{\mathbf{k}\mathbf{k}'}(0)/\Omega_s\sim 1$ meV around the magic angle, which is comparable to the Fermi energy $\epsilon_F$ and the Debye frequency of acoustic Moir\'e phonon bands $\hbar\omega_D\sim\hbar c_Tk_\theta\approx 2$ meV. When the optical phonon contributions are included, the attraction could be further enhanced. Since the density of states (DOS) is as large as $N_D\gtrsim 1$ meV$^{-1}\cdot\Omega_s^{-1}$ at the magic angle, the BCS coupling strength $\lambda\approx N_D|V^{II'}_{\mathbf{k}\mathbf{k}'}|\gtrsim1$ is strong.
The screened Coulomb interaction takes the form $\mathcal{V}_{e}(q)=2\pi e^2/q\epsilon(q)$, where $\epsilon(q)$ is the screened dielectric function at momentum $q$.
Here we simply adopt the (two-dimensional) Thomas-Fermi approximation (\cite{suppl} Sec. III) $\epsilon(q)\approx\epsilon_I(1+q_{TF}/q)$, where $\epsilon_I\approx 2\sim 10$ is the dielectric constant of undoped graphene, and $q_{TF}\approx 2\pi e^2 (\partial n_e/\partial\mu)/\epsilon_I=2\pi e^2 N_D/\epsilon_I$ is the Thomas-Fermi momentum ($n_e$ and $\mu$ are the electron density and chemical potential, respectively). With $N_D\gtrsim1$ meV$^{-1}\cdot\Omega_s^{-1}$ around the magic angle, $q_{TF}\gtrsim 50 k_\theta\gg q$, so the screened Coulomb potential $\mathcal{V}_{e}(q)\approx 2\pi e^2/\epsilon_Iq_{TF}\sim N_D^{-1}$, yielding a Coulomb coupling strength $\mu_c\approx N_D\mathcal{V}_{e}(q)\sim 1$. \ If we adopt the McMillan formula for superconductor $T_c$ \cite{mcmillan1968,allen1975}, taking $\lambda=1.5$ and $\mu_c=1$, we obtain
\begin{equation}
T_c=\frac{\hbar\omega_D}{1.45 k_B}\exp\left[-\frac{1.04(1+\lambda)}{\lambda-\mu^*_c(1+0.62\lambda)}\right]\approx 0.9 K
\end{equation}
at the magic angle, where $\mu^*_c=\mu_c/[1+\mu_c\ln(\omega_{pe}/\omega_D)]$ is the reduced Coulomb coupling strength, $\omega_{\text{pe}}$ is the plasma frequency, which is roughly $\hbar\omega_{pe}\sim\sqrt{(4\pi n_e)^{1/2}e^2\epsilon_F/\epsilon_I}\sim10\hbar\omega_D$ \cite{hwang2007,das-sarma2009}. This agrees well with the experimentally observed $T_c$. We do emphasize that our $T_c$ estimation is very rough, with inaccuracies from both $\lambda$, $\mu_c^*$ and the McMillan formula itself for large $\lambda$.

The above electron-phonon coupling calculation can be easily generalized to other twist angles and electron densities. We still keep only the nearest momentum hoppings in the continuum model of TBG, but truncate the Hamiltonian at sufficiently high momentum to obtain more accurate band structures. We then numerically calculate the energy change in each electron band under small deformations of $\mathbf{q}_j$, and verify it is comparable to our \emph{ab initio} results (\cite{suppl} Sec. V). Subtracting the contribution from Moir\'e BZ deformations (\cite{suppl} Sec. IV), we can estimate the electron-phonon coupling of each band and the BCS coupling strength $\lambda\approx N_D V_{\mathbf{k}\mathbf{k}}$. Fig. \ref{fig4}a and \ref{fig4}b show the DOS $N_D$ and BCS coupling strength $\lambda$ with respect to the number of electrons filling per superlattice unit cell $n=n_e\Omega_s$ at $\theta=1.05^\circ$. The DOS is predominantly high for the first conduction and valence bands ($|n|<4$). However, the BCS coupling $\lambda$ for $|n|<4$ and for $|n|>4$ are of the same order, despite the fact that the DOS is much lower ($\sim 0.05$ meV$^{-1}\cdot\Omega_s^{-1}$) at $|n|>4$. Numerically, this is because the energy susceptibility to deformations of a Moir\'e band becomes large when the band energy is large (\cite{suppl} Sec. IV).
This implies possible BCS superconductivity at higher $|n|$. Fig. \ref{fig4}c shows $T_c$ from the McMillan formula with respect to angle $\theta$ and number of electrons per unit cell $n$. There is only a narrow superconducting region at $|n|<4$ near the magic angle, which correspond to the superconductivity observed in TBG. In contrast, the superconductivity occurs in a wide range of $\theta\gtrsim 1^\circ$ for $|n|$ near $8$ and higher. There are also parameter spaces for $\theta<1^\circ$ where $T_c$ is of order of $1K$, e.g., $4\lesssim|n|\lesssim8$ near $\theta=0.8^\circ$, $2\lesssim|n|\lesssim12$ near $\theta=0.6^\circ$, and $8\lesssim|n|\lesssim12$ around $\theta=0.3^\circ$. The vast superconductivity region indicates the Moir\'e pattern generically enhances the electron-phonon coupling of all bands. This may explain the possible superconductivity of HOPG, which contains numerous Moir\'e interfaces with different electron densities $|n|$ and twist angles $\theta$.

Lastly, we comment that the strong phonon-mediated attraction may favor a Bose Mott insulator \cite{fisher1989} for the TBG insulating phase observed at $|n|=2$ \cite{cao2018,cao2018b}. This is because the attraction may pair the electrons into charge $2e$ bosons (Cooper pairs), and for $|n|=2$, the system has one boson per unit cell, thus may form a Bose Mott insulator \cite{fisher1989}, with a possible charge density wave order. Such a phase will have a resistivity around $h/(2e)^2\approx 6k\Omega$ at the superconductor-insulator transition, due to charge $2e$ carriers \cite{fisher1990,yazdani1995,steiner2008}. 
However, the experimentally observed quantum oscillations \cite{cao2018} indicates these Cooper pairs have to break down for magnetic fields above $1$T, if this explanation is correct.

In summary, we have shown the electron-phonon coupling is strong in TBG, and can lead to BCS superconductivity with a $T_c$ in agreement with the experiment \cite{cao2018}. We find the intervalley pairing between valleys $K$ and $K'$ is favored in the flat bands near the magic angle, which is topologically trivial.
Besides, we predict that superconductivity can be achieved at many other angles and at higher electron fillings (Fig. \ref{fig4}c), and we expect our prediction to be verified by higher doping TBG experiments in the future \cite{kim2017,wong2015}.

\begin{acknowledgments}
\emph{Acknowledgments.}
BL is supported by Princeton Center for Theoretical Science at Princeton University. ZW and BB are supported by the Department of Energy Grant No. de-sc0016239, the National Science Foundation EAGER Grant No. noaawd1004957, Simons Investigator Grants No. ONRN00014-14-1-0330, No. ARO MURI W911NF-12-1-0461, and No. NSF-MRSEC DMR- 1420541, the Packard Foundation, the Schmidt Fund for Innovative Research.
\end{acknowledgments}

\bibliography{TBG_ref}

\begin{widetext}

\section*{Supplementary Material for "Twisted Bilayer Graphene: A Phonon Driven Superconductor"}

\section{Derivation of electron-phonon coupling at magic angle}\label{sec1}

In this section we first briefly recall the continuum model in momentum space formulated in Ref. \cite{bistritzer2011}, and then use it to derive the coupling between electrons and interlayer phonons is derived.

Fig. \ref{figS1} illustrates the real space configuration of two graphene layers which have a relative twist angle $\theta$. We denote the lattice vectors of layer $j$ ($j=1,2$) as $\mathbf{a}_1^{(j)}$ and $\mathbf{a}_2^{(j)}$, and the reciprocal lattice vectors of layer $j$ as $\mathbf{G}_1^{(j)}$ and $\mathbf{G}_2^{(j)}$, which satisfy $\mathbf{G}_a^{(j)}\cdot \mathbf{a}_b^{(j)}=2\pi\delta_{ab}$. The norms of the lattice vectors $\mathbf{a}_i^{(j)}$ are equal to the lattice constant $a_0=0.246$nm. Each layer $j$ consists of two sublattice positions $A$ and $B$, which are located at $\bm{\tau}_A^{(j)}$ and $\bm{\tau}_B^{(j)}$ in the unit cell of layer $j$. Without loss of generality, we can choose sublattice $A$ in layer $j$ as the origin of the unit cell of layer $j$, so that $\bm{\tau}_A^{(j)}=0$, and $\bm{\tau}_B^{(j)}=\bm{\tau}^{(j)}$ as shown in Fig. \ref{figS1}. The vectors $\mathbf{a}_i^{1}$ and $\mathbf{a}_i^{2}$ are rotated by $\theta$ from one another, and so do $\bm{\tau}^{(1)}$ and $\bm{\tau}^{(2)}$. More explicitly, the above vectors in components are given by
\begin{equation}
\begin{split}
&\mathbf{a}_1^{(j)}=a_0R^{(j)}_{\theta/2}\left(-\frac{1}{2},\frac{\sqrt{3}}{2}\right)^T,\quad
\mathbf{a}_2^{(j)}=a_0R^{(j)}_{\theta/2}\left(-\frac{1}{2},-\frac{\sqrt{3}}{2}\right)^T,\quad
\bm{\tau}_A^{(j)}=(0,0)^T,\quad
\bm{\tau}_B^{(j)}=\frac{a_0}{\sqrt{3}}R^{(j)}_{\theta/2}\left(0,1\right)^T, \\
&\mathbf{G}_1^{(j)}=\frac{4\pi}{\sqrt{3}a_0}R^{(j)}_{\theta/2}\left(-\frac{1}{2},\frac{\sqrt{3}}{2}\right)^T\ , \qquad
\mathbf{G}_2^{(j)}=\frac{4\pi}{\sqrt{3}a_0}R^{(j)}_{\theta/2}\left(-\frac{1}{2},\frac{\sqrt{3}}{2}\right)^T\ ,
\end{split}
\end{equation}
for layer $j=1,2$, where
\begin{equation}\label{Rj}
R^{(1)}_{\theta/2}=
\left(\begin{array}{cc}
\cos(\theta/2)&-\sin(\theta/2)\\
\sin(\theta/2)&\cos(\theta/2)\\
\end{array}\right)\ ,\qquad
R^{(2)}_{\theta/2}=
\left(\begin{array}{cc}
\cos(\theta/2)&\sin(\theta/2)\\
-\sin(\theta/2)&\cos(\theta/2)\\
\end{array}\right)\ ,
\end{equation}
which are the rotation matrices of angle $\pm\theta/2$, respectively.

The interlayer hopping $t(r)$ between two atoms in different layers is generically a function of their in-plane distance $r$. When transformed into momentum space, the electron hopping from momentum $\mathbf{p}'$ and sublattice $\beta$ in layer $2$ to momentum $\mathbf{k}$ and sublattice $\alpha$ in layer $1$ takes the form
\begin{equation}\label{Tab}
\begin{split}
&T^{\alpha\beta}_{\mathbf{k}\mathbf{p}'}=\frac{1}{N}\sum_{\mathbf{R}^{(1)},\mathbf{R}^{(2)}}t\left(\mathbf{R}^{(1)}+\bm{\tau}_\alpha^{(1)}-\mathbf{R}^{(2)}-\bm{\tau}_\beta^{(2)}\right) e^{i\mathbf{p}'\cdot(\mathbf{R}^{(2)}+\bm{\tau}_\beta^{(2)})-i\mathbf{k}\cdot(\mathbf{R}^{(1)}+\bm{\tau}_\alpha^{(1)})}\\
=&\sum_{n_1,m_1}\sum_{n_2,m_2} \frac{t_{\mathbf{k}+\mathbf{G}^{(1)}_{n_1,m_1}}}{\Omega}\delta_{\mathbf{k}+\mathbf{G}^{(1)}_{n_1,m_1},\mathbf{p}'+\mathbf{G}^{(2)}_{n_2,m_2}} e^{i\mathbf{G}^{(2)}_{n_2,m_2}\cdot\bm{\tau}_\beta^{(2)}-i\mathbf{G}^{(1)}_{n_1,m_1}\cdot\bm{\tau}_\alpha^{(1)}}\ ,
\end{split}
\end{equation}
where $\alpha,\beta=A,B$ are sublattice indices, $\mathbf{k}$ and $\mathbf{p}'$ are measured from $\Gamma$ point of the graphene Brillouin zone (BZ), $\Omega=\sqrt{3}a_0^2/2$ is the area of graphene unit cell, $t_{\mathbf{k}}$ is the Fourier transform of $t(\mathbf{r})$, and $\mathbf{G}^{(j)}_{n,m}=n\mathbf{G}^{(j)}_{1}+m\mathbf{G}^{(j)}_{2}$ runs over all $n,m\in\mathbb{Z}$ reciprocal lattices in layer $j$.

\begin{figure}[htbp]
\begin{center}
\includegraphics[width=5in]{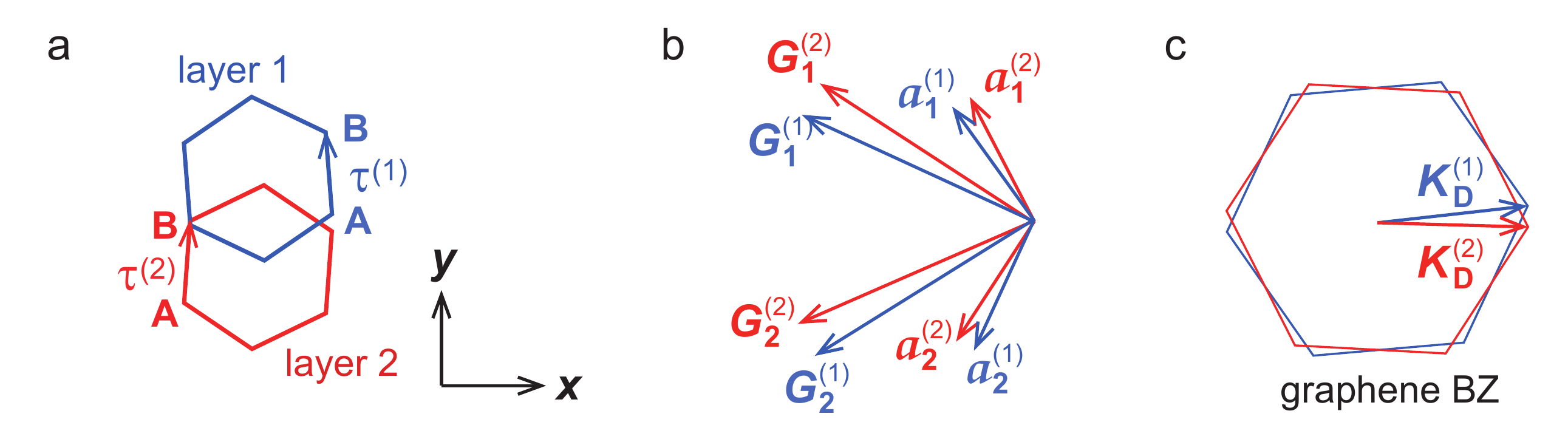}
\end{center}
\caption{{\bf a}. Illustration of the real space TBG lattice, where the twist angle is $\theta=10^\circ$. {\bf b}. The graphene lattice vectors and reciprocal vectors in each layer. {\bf c}. The graphene BZs of the two layers.}
\label{figS1}
\end{figure}

We first consider the low energy physics near the $K$ points of the graphene BZs of the two layers (see main text Fig. 2), namely, $\mathbf{k}$ and $\mathbf{p}'$ near the Dirac point momenta $\mathbf{K}^{(1)}_{D}=-(\mathbf{G}^{(1)}_{1}+\mathbf{G}^{(1)}_{2})/3$ and $\mathbf{K}^{(2)}_{D}=-(\mathbf{G}^{(2)}_{1}+\mathbf{G}^{(2)}_{2})/3$, respectively. It is shown that the hopping $t_{\mathbf{k}}$ decays exponentially with respect to $|\mathbf{k}|$ \cite{bistritzer2011}, so a good approximation is to keep only the $3$ leading nearest hopping terms $t_{\mathbf{k}+\mathbf{G}^{(1)}_{n,m}}$ with $|\mathbf{k}+\mathbf{G}^{(1)}_{n,m}|$ around the magnitude $|\mathbf{K}^{(1)}_{D}|$, and approximate them to $t_{\mathbf{K}^{(1)}_{D}}$.

We then define the three vectors
\[\mathbf{q}_1=\mathbf{K}^{(2)}_{D}-\mathbf{K}^{(1)}_{D}\ ,\quad \mathbf{q}_2=C_{3z}\mathbf{q}_1=\mathbf{K}^{(2)}_{D}+\mathbf{G}^{(2)}_{1}-\mathbf{K}^{(1)}_{D}-\mathbf{G}^{(1)}_{1}\ ,\quad \mathbf{q}_3=C_{3z}^2\mathbf{q}_1=\mathbf{K}^{(2)}_{D}+\mathbf{G}^{(2)}_{2}-\mathbf{K}^{(1)}_{D}-\mathbf{G}^{(1)}_{2}\ ,\]
where $C_{3z}$ is the 3-fold rotation about $z$ axis. Explicitly, $\mathbf{q}_j$ in components are
\[
\mathbf{q}_1=k_\theta(0,-1)^T\ ,\qquad \mathbf{q}_2=k_\theta\left(\frac{\sqrt{3}}{2},\frac{1}{2}\right)^T\ ,\qquad \mathbf{q}_3=k_\theta\left(-\frac{\sqrt{3}}{2},\frac{1}{2}\right)^T\ ,
\]
where $k_\theta=|\mathbf{q}_j|=(8\pi/3a_0)\sin(\theta/2)$.
Under the above nearest hopping approximation, an electron state with momentum $\mathbf{p}'$ in layer $2$ can hop to an electron state with momentum $\mathbf{k}$ in layer $1$ if $\mathbf{k}-\mathbf{p}'=\mathbf{q}_{j}$.

Since we are interested in the low energy band structure near the graphene BZ $K$ (or $K'$) point, hereafter we set the origin of $\mathbf{k}$ and $\mathbf{p}'$ at $K$ (or $K'$) point of each layer. The effective Hamiltonian of each layer is given by $h^K(\mathbf{k})=v(k_x\sigma_x-k_y\sigma_y)=\hbar v\bm{\sigma}^*\cdot\mathbf{k}$ at $K$ point of each layer, and $h^{K'}(\mathbf{k})=-\hbar v\bm{\sigma}\cdot\mathbf{k}$ at $K'$ point of each layer (related to $h^K(\mathbf{k})$ via the time reversal transformation $\mathcal{T}$), where $v$ is the fermi velocity, and $\sigma_{x,y,z}$ are the Pauli matrices for sublattice indices. Note that the Dirac fermions at graphene valleys $K$ and $K'$ have opposite helicities. Therefore, to the lowest order, the TBG Hamiltonian at $K$ point in the vicinity of layer $1$ momentum $\mathbf{k}=0$ (which is the $K_M'$ point of the Moir\'e BZ) truncated at the nearest hoppings is \cite{bistritzer2011}
\begin{equation}\label{SeqH}
H^{K,K_M'}(\mathbf{k})=\left(\begin{array}{cccc}
h^{K}_{\theta/2}(\mathbf{k})& wT_1& wT_2&wT_3\\
wT_1^\dag&h^{K}_{-\theta/2}(\mathbf{k}-\mathbf{q}_1)&0&0\\
wT_2^\dag&0&h^{K}_{-\theta/2}(\mathbf{k}-\mathbf{q}_2)&0\\
wT_3^\dag&0&0&h^{K}_{-\theta/2}(\mathbf{k}-\mathbf{q}_3)\\
\end{array}\right)\ ,
\end{equation}
where the basis is $(\psi_{0,\mathbf{k}}^T,\psi_{1,\mathbf{k}}^T,\psi_{2,\mathbf{k}}^T,\psi_{3,\mathbf{k}}^T)^T$, and $\psi_{0,\mathbf{k}}$ and $\psi_{j,\mathbf{k}}$ ($j=1,2,3$) are the 2-component column spinors in the AB sublattice index basis at momentum $\mathbf{k}$ in layer $1$ and momentum $\mathbf{k}-\mathbf{q}_j$ in layer $2$, respectively. $h^{K}_{\pm\theta/2}(\mathbf{k})$ is $h^{K}(\mathbf{k})$ rotated by a $\pm\theta/2$ angle, and $w=t_{\mathbf{K}^{(1)}_{D}}/\Omega$ is the interlayer hopping which can be chosen as real. The hopping matrix $T_j$ are defined by $(T_j)_{\alpha\beta}=e^{i\mathbf{G}^{(2)}_{n_j,m_j}\cdot\bm{\tau}_\beta^{(2)} -i\mathbf{G}^{(1)}_{n_j,m_j}\cdot\bm{\tau}_\alpha^{(1)}}$ (which is nothing but the phase factor part of Eq. (\ref{Tab})), where $(n_1,m_1)=(0,0)$, $(n_2,m_2)=(1,0)$ and $(n_3,m_3)=(0,1)$, and $\mathbf{G}^{(i)}_{n,m}$ in layer $i$ is defined below Eq. (\ref{Tab}). They satisfy $|\mathbf{K}_D^{(i)}+\mathbf{G}^{(i)}_{n_j,m_j}|=|\mathbf{K}_D^{(i)}|=4\pi/3a_0$, thus correspond to the nearest interlayer hoppings in the momentum space. Explicitly, $T_j$ are given by
\[T_1=1+\sigma_x\ ,\qquad T_2=1-\frac{1}{2}\sigma_x-\frac{\sqrt{3}}{2}\sigma_y\ ,\qquad T_3=1-\frac{1}{2}\sigma_x+\frac{\sqrt{3}}{2}\sigma_y\ .\]

Since the twist angle $\theta$ is small, to the lowest order, we shall ignore the $\pm\theta/2$ rotation of $h_{\pm\theta/2}^K(\mathbf{k})$ in Eq. (\ref{SeqH}). Under this approximation, the system has a particle-hole symmetry, and the low energy eigenstates are given by $\psi_{j,\mathbf{k}}\approx-wh_j^{-1}T_j^\dag\psi_{0,\mathbf{k}}$ ($j=1,2,3$), where $h_j$ is short hand for $h^K(-\mathbf{q}_j)$. The low energy Hamiltonian is a $2\times2$ Hamiltonian in the $\psi_0$ space \cite{bistritzer2011}:
\begin{equation}\label{Hfold}
\widetilde{H}^{K,K_M'}(\mathbf{k})=\frac{\langle \Psi|H^{K,K_M'}(\mathbf{k})|\Psi\rangle }{\langle\Psi|\Psi\rangle}=\frac{\hbar v}{1+6\alpha^2}\left(\bm{\sigma}^*\cdot\mathbf{k}+ w^2\sum_{j=1}^3 T_jh_j^{-1}\bm{\sigma}^*\cdot\mathbf{k}h_j^{-1}T_j^\dag\right)=\frac{1-3\alpha^2}{1+6\alpha^2}\hbar v\bm{\sigma}^*\cdot\mathbf{k}\ ,
\end{equation}
where $\alpha=w/\hbar vk_\theta$. The first magic angle $\theta=1.05^\circ$ is given by $\alpha^2=1/3$, where the Dirac velocity of the above effective Hamiltonian vanishes. Note that $\widetilde{H}^{K,K_M'}(\mathbf{k})$ depends on the momenta $\mathbf{q}_1,\mathbf{q}_2$ and $\mathbf{q}_3$ (which implicitly appear in $h_j$), so we shall also denote it as $\widetilde{H}^{K,K_M'}(\mathbf{k},\mathbf{q}_1,\mathbf{q}_2,\mathbf{q}_3)$ later to emphasize its dependence on $\mathbf{q}_j$.

We note that the Dirac Hamiltonian $\widetilde{H}^{K,K_M'}(\mathbf{k})$ in Eq. (\ref{Hfold}) around layer $1$ momentum $\mathbf{k}=0$ is at the $K_M'$ point of the Moir\'e BZ (MBZ) (see Fig. 2 of the main text), and comes from the $K$ point of both layer $1$ and layer $2$ of TBG. Similarly, there is also a Dirac Hamiltonian around layer $2$ momentum $\mathbf{p}'=0$ coming from the $K$ point of both layer $1$ and layer $2$ of TBG, which is at the $K_M$ point of the MBZ. In total, there are eight Dirac fermions labeled by graphene BZ valley $K,K'$ (which is the same for both layers, no coupling between $K$ and $K'$ exists in the Hamiltonian of Eq. (\ref{Hfold})), Moir\'e valley $K_M,K_M'$ and spin $\uparrow,\downarrow$ indices. The band structure obtained in the above model is thus 4-fold degenerate everywhere in the MBZ with respect to graphene valley $K,K'$ and spin $\uparrow,\downarrow$, and has $4$ Dirac fermions at Moir\'e valley $K_M$ and another $4$ Dirac fermions at Moir\'e valley $K_M'$. In particular, the helicity of the Dirac fermions only depends on whether they come from graphene valley $K$ or $K'$.

When small distortions are added to the graphene lattices, the above effective Hamiltonian will change. Since distortions can be expressed using phonon fields, the change in the effective Hamiltonian simply gives the low energy electron-phonon coupling term. The leading contribution to the change of Hamiltonian is due to the distortion of momentum vectors $\mathbf{q}_j$. In the main text we have argued that interlayer phonon waves change the Moir\'e pattern dramatically. To prove this from microscopics requires some more calculations.
This can be shown explicitly as follows. Denote the in-plane displacement of atoms of layer $j$ at $\mathbf{r}$ as $\mathbf{u}^{(j)}(\mathbf{r})=\left(u^{(j)}_x(\mathbf{r}),u^{(j)}_y(\mathbf{r})\right)^T$. The displacement $\mathbf{u}^{(j)}$ is nothing but the in-plane phonon field in layer $j$. In the continuum limit, the variation of the lattice vectors $\mathbf{a}_{1}^{(j)}$ and $\mathbf{a}_{2}^{(j)}$ under the displacement field $\mathbf{u}^{(j)}$ are simply given by $\delta \mathbf{a}_{1}^{(j)}(\mathbf{r})=(\mathbf{a}_{1}^{(j)}\cdot\nabla)\mathbf{u}^{(j)}(\mathbf{r})$ and $\delta \mathbf{a}_{2}^{(j)}(\mathbf{r})=(\mathbf{a}_{2}^{(j)}\cdot\nabla)\mathbf{u}^{(j)}(\mathbf{r})$. Explicitly, one has
\begin{equation}
\begin{split}
&\delta \mathbf{a}_{b}^{(j)}=\left(\mathbf{a}_{b}^{(j)}\cdot\nabla\right)\mathbf{u}^{(j)} =\left[{\mathbf{a}_{b}^{(j)}}^T\cdot
\left(\begin{array}{c}\partial_x\\ \partial_y\end{array}\right)\right]^T \left(\begin{array}{c}u^{(j)}_x\\u^{(j)}_y\end{array}\right) =a_0 \left(\begin{array}{cc}\partial_xu^{(j)}_x & \partial_yu^{(j)}_x\\ \partial_xu^{(j)}_y& \partial_yu^{(j)}_y\end{array}\right)R^{(j)}_{\theta/2} \left(\begin{array}{c}-\frac{1}{2}\\ \pm\frac{\sqrt{3}}{2}\end{array}\right)\ ,
\end{split}
\end{equation}
where the $\pm$ signs are for $b=1$ (lattice vector $\mathbf{a}_{1}^{(j)}$) and $b=2$ (lattice vector $\mathbf{a}_{2}^{(j)}$), respectively. Accordingly, to linear order the distortion of reciprocal vectors of layer $j$ satisfies $\delta\mathbf{G}_a^{(j)}\cdot \mathbf{a}_b^{(j)}+\mathbf{G}_a^{(j)}\cdot \delta\mathbf{a}_b^{(j)}=0$ ($a,b=1,2$), which has a solution
\[
\delta \mathbf{G}_{a}^{(j)}=-\nabla\left(\mathbf{u}^{(j)}\cdot\mathbf{G}_{a}^{(j)}\right) =-\left(\begin{array}{c}\partial_x\\ \partial_y\end{array}\right)\left(\mathbf{u}^{(j)}\cdot\mathbf{G}_{a}^{(j)}\right) =-\frac{4\pi}{\sqrt{3}a_0} \left(\begin{array}{cc}\partial_xu^{(j)}_x & \partial_xu^{(j)}_y\\ \partial_yu^{(j)}_x& \partial_yu^{(j)}_y\end{array}\right)R^{(j)}_{\theta/2} \left(\begin{array}{c}-\frac{\sqrt{3}}{2}\\ \pm\frac{1}{2}\end{array}\right)\ ,
\]
where the $\pm$ signs are for index $a=1$ (reciprocal vector $\mathbf{G}_{1}^{(j)}$) and $a=2$ (reciprocal vector $\mathbf{G}_{2}^{(j)}$), respectively.
In particular, since $\bm{\tau}^{(j)}=(\mathbf{a}_1^{(j)}-\mathbf{a}_2^{(j)})/3$, one has $\delta \mathbf{G}_{a}^{(j)}\cdot \bm{\tau}^{(j)}+ \mathbf{G}_{a}^{(j)}\cdot \delta\bm{\tau}^{(j)}=0$, which implies $\mathbf{G}_{1,2}^{(j)}\cdot\bm{\tau}^{(j)}=\pm2\pi/3$ remains invariant, so the interlayer hopping matrix $T_j=e^{i\mathbf{G}^{(2)}_{n_j,m_j}\cdot\bm{\tau}_\beta^{(2)} -i\mathbf{G}^{(1)}_{n_j,m_j}\cdot\bm{\tau}_\alpha^{(1)}}$ remains unchanged under the deformation. Only $\mathbf{q}_j$ in the continuum model Hamiltonian change.

With the expressions for $\delta\mathbf{G}_{a}^{(j)}$, it is straightforward to derive the variation of $\mathbf{q}_j$ from their definitions:
\begin{equation}\label{Seqdeltaq}
\begin{split}
&\delta\mathbf{q}_1=\frac{\delta\mathbf{G}_{1}^{(1)}+\delta\mathbf{G}_{2}^{(1)} -\delta\mathbf{G}_{1}^{(2)}-\delta\mathbf{G}_{2}^{(2)}}{3} =\frac{4\pi}{3a_0}\left[
\left(\begin{array}{cc}\partial_xu^{(1)}_x & \partial_xu^{(1)}_y\\ \partial_yu^{(1)}_x& \partial_yu^{(1)}_y\end{array}\right)\left(\begin{array}{c}\cos\frac{\theta}{2}\\
\sin\frac{\theta}{2}\end{array}\right)- \left(\begin{array}{cc}\partial_xu^{(2)}_x & \partial_xu^{(2)}_y\\ \partial_yu^{(2)}_x& \partial_yu^{(2)}_y\end{array}\right)\left(\begin{array}{c}\cos\frac{\theta}{2}\\ -\sin\frac{\theta}{2}\end{array}\right)
\right] \\
&=k_\theta\left(\gamma\partial_xu_x+\partial_xu_y^c\ ,\ \gamma\partial_yu_x+\partial_yu_y^c\right)^T\ , \\
&\delta\mathbf{q}_2=\frac{-2\delta\mathbf{G}_{1}^{(1)}+\delta\mathbf{G}_{2}^{(1)} +2\delta\mathbf{G}_{1}^{(2)}-\delta\mathbf{G}_{2}^{(2)}}{3}\\
&= k_\theta \left[\frac{\sqrt{3}}{2}\left(\gamma\partial_xu_y-\partial_xu_x^c\right) -\frac{1}{2}\left(\gamma\partial_xu_x+\partial_xu_y^c\right),\ -\frac{1}{2}\left(\gamma\partial_yu_x+\partial_yu_y^c\right) +\frac{\sqrt{3}}{2}\left(\gamma\partial_yu_y-\partial_yu_x^c\right)\right]^T,\\
&\delta\mathbf{q}_3=\frac{\delta\mathbf{G}_{1}^{(1)}-2\delta\mathbf{G}_{2}^{(1)} -\delta\mathbf{G}_{1}^{(2)}+2\delta\mathbf{G}_{2}^{(2)}}{3}\\
&= k_\theta \left[-\frac{\sqrt{3}}{2}\left(\gamma\partial_xu_y-\partial_xu_x^c\right) -\frac{1}{2}\left(\gamma\partial_xu_x+\partial_xu_y^c\right),\ -\frac{1}{2}\left(\gamma\partial_yu_x+\partial_yu_y^c\right) -\frac{\sqrt{3}}{2}\left(\gamma\partial_yu_y-\partial_yu_x^c\right)\right]^T,\\
\end{split}
\end{equation}
where we have defined $\gamma=[2\tan(\theta/2)]^{-1}$, the relative displacement $\mathbf{u}=\mathbf{u}^{(1)}-\mathbf{u}^{(2)}$, and the center of mass displacement $\mathbf{u}^c=(\mathbf{u}^{(1)}+\mathbf{u}^{(2)})/2$. For small angles $\theta$, we have $\gamma\approx1/\theta\gg1$, so $\delta \mathbf{q}_j$ is dominated by the relative displacement phonon field $\mathbf{u}$. We therefore will ignore the contribution of center of mass displacement $\mathbf{u}^c$ hereafter.

Fig. \ref{figS2} shows the graphene lattices and Moir\'e patterns for $\theta=5^\circ$ before and after a relative shear deformation $\Sigma_{xx}=-\Sigma_{yy}=0.02$, where $\Sigma_{ab}=(\partial_au_b+\partial_bu_a)/2-(\sum_l\partial_lu_l)\delta_{ab}/2$ ($a,b,l=1,2$) is the relative shear tensor.
One can see that a small relative deformation greatly affects the Moir\'e pattern.

\begin{figure}[tbp]
\begin{center}
\includegraphics[width=4.5in]{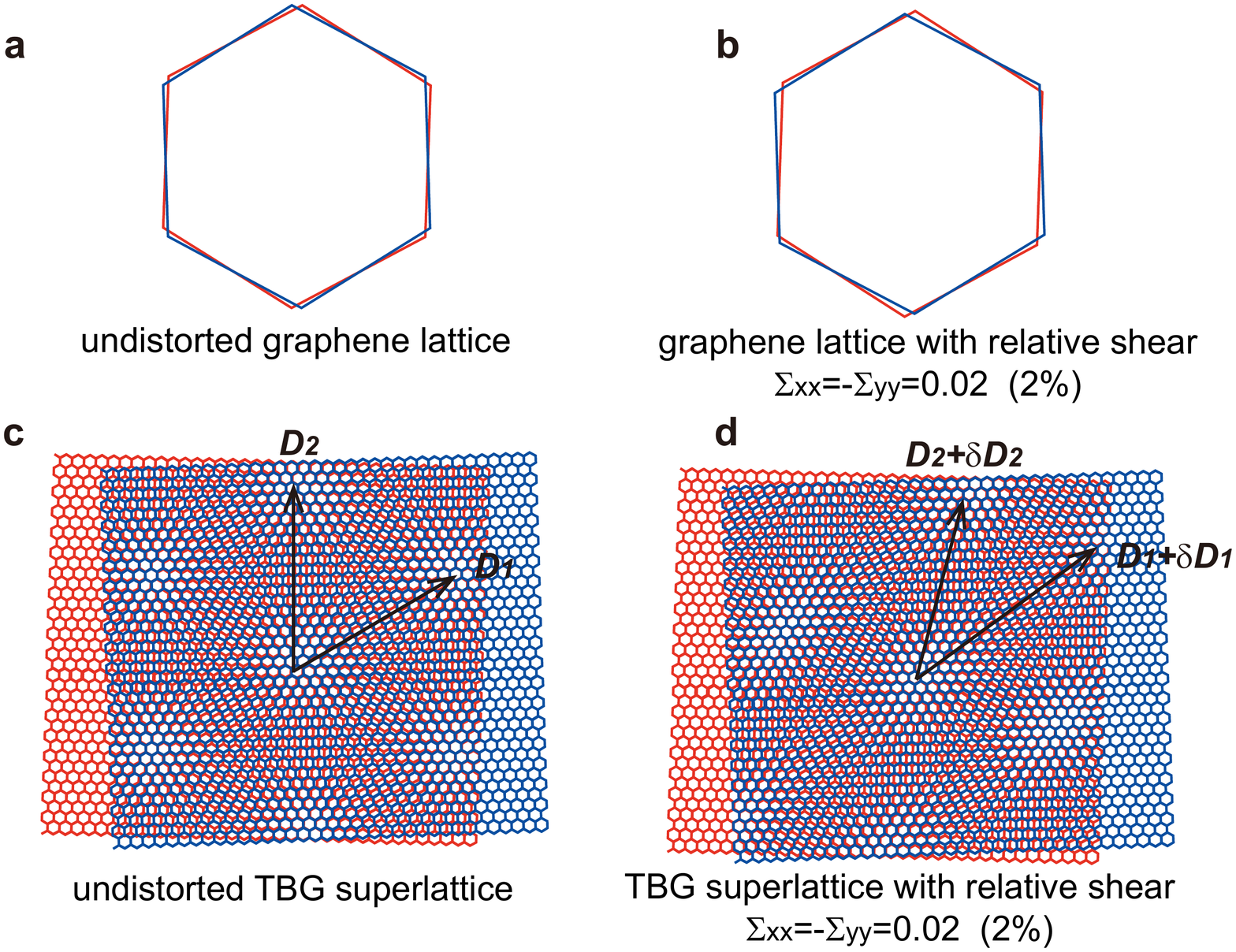}
\end{center}
\caption{{\bf a}. Graphene lattice plaquettes of two layers of TBG (at AA stacking center) before deformation (plotted for $\theta=5^\circ$). {\bf b}. Graphene lattice plaquettes of two layers of TBG (at AA stacking center) with a relative shear deformation $\Sigma_{xx}=-\Sigma_{yy}=\partial_xu_x-\partial_yu_y=0.02$, while expansion and rotation deformations are zero. {\bf c}. The Moir\'e pattern superlattice of TBG before deformation. {\bf d}. The Moir\'e pattern superlattice of TBG after the shear deformation $\Sigma_{xx}=-\Sigma_{yy}=0.02$. One can see the superlattice is greatly affected, although the graphene lattices only have a $2\%$ shear deformation. }
\label{figS2}
\end{figure}

With the deformed vectors $\mathbf{q}_j$, the electron-phonon coupling Hamiltonian is simply given by the variation of the effective $2\times2$ Hamiltonian $\widetilde{H}^{K,K_M}$ in Eq. (\ref{Hfold}), namely, $H_{\text{ep}}^{K,K_M'}(\mathbf{k})=\widetilde{H}^{K,K_M'}(\mathbf{k},\mathbf{q}_1+\delta \mathbf{q}_1, )$
\begin{equation}\label{HepK}
\begin{split}
&H_{\text{ep}}^{K,K_M'}(\mathbf{k})=\widetilde{H}^{K,K_M'}(\mathbf{k},\mathbf{q}_1+\delta \mathbf{q}_1,\mathbf{q}_2+\delta \mathbf{q}_2,\mathbf{q}_3+\delta \mathbf{q}_3)-\widetilde{H}^{K,K_M'}(\mathbf{k},\mathbf{q}_1,\mathbf{q}_2,\mathbf{q}_3)\\
&=\frac{\hbar v w^2}{1+6\alpha^2}\sum_{j=1}^{3}\left(T_jh_j^{-1}\bm{\sigma}^*\cdot\mathbf{k}\delta h_j^{-1}T_j^\dag+h.c.\right)-\frac{1-3\alpha^2}{(1+6\alpha^2)^2}\hbar vw^2 \bm{\sigma}^*\cdot\mathbf{k}\sum_{j=1}^3\psi_0^\dag T_j\delta (h_j^{-2})T_j^\dag\psi_0\ ,
\end{split}
\end{equation}
where we have defined $h_j=h^K(-\mathbf{q}_j)$, and $\delta h_j=h^K(-\mathbf{q}_j-\delta\mathbf{q}_j)-h_j$. The first term and the second term in the above result come from the variations $\delta\langle\Psi|H^{K,K_M'}|\Psi\rangle$ and $\delta\langle\Psi|\Psi\rangle$ of Eq. (\ref{Hfold}), respectively. Explicitly, one can show that
\begin{equation}
\begin{split}
&\delta h_1^{-1}=\gamma (\hbar vk_\theta)^{-1}(\partial_xu_x\sigma_x+\partial_yu_x\sigma_y)\ ,\\
&\delta h_2^{-1}=\gamma (\hbar vk_\theta)^{-1}\left\{ \left[\frac{\sqrt{3}}{4}(\partial_yu_x-\partial_xu_y)+\frac{1}{4}\partial_xu_x -\frac{1}{4}\partial_yu_y\right]\sigma_x+ \left[\frac{1}{4}\partial_yu_x +\frac{3}{4}\partial_xu_y-\frac{\sqrt{3}}{4}(\partial_xu_x+\partial_yu_y)\right]\sigma_y\right\},\\
&\delta h_3^{-1}=\gamma (\hbar vk_\theta)^{-1}\left\{ \left[\frac{\sqrt{3}}{4}(\partial_xu_y-\partial_yu_x)+\frac{1}{4}\partial_xu_x -\frac{1}{4}\partial_yu_y\right]\sigma_x+ \left[\frac{1}{4}\partial_yu_x +\frac{3}{4}\partial_xu_y+\frac{\sqrt{3}}{4}(\partial_xu_x+\partial_yu_y)\right]\sigma_y\right\}.
\end{split}
\end{equation}
If we denote the above expression as $\delta h_j^{-1}=(\hbar vk_\theta)^{-1}(\sigma_x A_{j,x}+\sigma_y A_{j,y})$, we find $w^2T_jh_j^{-1}\bm{\sigma}^*\cdot \mathbf{k}\delta h_j^{-1}T_j^\dag=-2\alpha^2T_j(k_x A_{j,y}+k_y A_{j,x})$. Besides, we note that $h_j^{-2}=(\hbar v)^{-2}|\mathbf{q}_j|^{-2}I_2$ (where $I_2$ is the $2\times2$ identity matrix), so one has $\delta (h_j^{-2})=(\hbar v)^{-2}\delta |\mathbf{q}_j|^{-2}I_2$, and one can show that
\[
\begin{split}
&\delta |\mathbf{q}_1|^{-2}=2\gamma k_\theta^{-2}\partial_y u_x,\qquad \delta |\mathbf{q}_2|^{-2}=\gamma k_\theta^{-2}\left[\frac{1}{2}\partial_yu_x-\frac{3}{2}\partial_xu_y+ \frac{\sqrt{3}}{2}(\partial_xu_x-\partial_yu_y)\right], \\
&\delta |\mathbf{q}_3|^{-2}=\gamma k_\theta^{-2}\left[\frac{1}{2}\partial_yu_x-\frac{3}{2}\partial_xu_y- \frac{\sqrt{3}}{2}(\partial_xu_x-\partial_yu_y)\right].
\end{split}
\]
Further, we set $\psi_0$ in Eq. (\ref{HepK}) to be an eigenstate of $\bm{\sigma}^*\cdot\mathbf{k}$ (with eigenvalue either $+|\mathbf{k}|$ or $-|\mathbf{k}|$). With these results, one then finds the electron-phonon coupling in Eq. (\ref{HepK}) to be
\begin{equation}\label{Hepalpha}
\begin{split}
&H_{\text{ep}}^{K,K_M'}(\mathbf{k})=-\gamma\hbar v\Big\{\frac{9\alpha^2(1+3\alpha^2)}{(1+6\alpha^2)^2} \left[k_x(\partial_yu_x+\partial_xu_y)+k_y(\partial_xu_x-\partial_yu_y)\right] \\
&\qquad +\frac{9\alpha^2}{(1+6\alpha^2)^2}\bm{\sigma}^*\cdot \mathbf{k}(\partial_yu_x-\partial_xu_y) +\frac{3\alpha^2}{1+6\alpha^2} \mathbf{\hat{z}}\cdot(\bm{\sigma}^*\times\mathbf{k})(\partial_xu_x+\partial_yu_y)\Big\}\ ,
\end{split}
\end{equation}
where the momentum $\mathbf{k}$ in the above electron-phonon coupling should be understood as the average momentum of the electron before and after phonon emission (absorption), and explicitly one has $\mathbf{\hat{z}}\cdot(\bm{\sigma}^*\times\mathbf{k})=\sigma_xk_y+\sigma_yk_x$. Note that the three terms
\[
k_x(\partial_yu_x+\partial_xu_y)+k_y(\partial_xu_x-\partial_yu_y)\ ,\qquad \bm{\sigma}^*\cdot \mathbf{k}(\partial_yu_x-\partial_xu_y)\ ,\qquad \mathbf{\hat{z}}\cdot(\bm{\sigma}^*\times\mathbf{k})(\partial_xu_x+\partial_yu_y)
\]
are contributed to by the relative shear tensor $\Sigma_{ab}$, relative rotation $R=\partial_xu_y-\partial_yu_x$, and relative expansion $\Theta=\partial_xu_x+\partial_yu_y$, respectively. Since the shear tensor is a rank $2$ tensor in the 2D space, the first term contributed by shear ($\mathbf{k}$ times the shear tensor) is a rank $3$ tensor. In particular, the first term can be rewritten as
\[
k_x(\partial_yu_x+\partial_xu_y)+k_y(\partial_xu_x-\partial_yu_y)= \mbox{Im}[(k_x+ik_y)(\partial_x+i\partial_y)(u_x+iu_y)]\ .
\]
Under a rotation of angle $\phi$ about $z$ axis, each of the three terms $k_x+ik_y$, $\partial_x+i\partial_y$ and $u_x+iu_y$ gains a phase factor $e^{i\phi}$. Therefore, this term is clearly only $C_{3z}$ rotationally invariant about $z$ axis (i.e., $\phi=2\pi/3$).
In contrast, both $\bm{\sigma}^*\cdot \mathbf{k}$ and $\mathbf{\hat{z}}\cdot(\bm{\sigma}^*\times\mathbf{k})$ are 2D scalars, and both the relative rotation $R$ and the relative expansion $\Theta$ are also 2D scalars, so the second and the third terms contributed by relative rotation and expansion are SO(2) rotationally symmetric about $z$ axis.

One may ask why we only have term $\mbox{Im}[(k_x+ik_y)(\partial_x+i\partial_y)(u_x+iu_y)]$ but not $\mbox{Re}[(k_x+ik_y)(\partial_x+i\partial_y)(u_x+iu_y)]$, both of which respect $C_{3z}$ symmetry. This is because the TBG also has a 2-fold rotation symmetry $C_{2x}$ about the $x$ axis. Under $C_{2x}$, we have $(\partial_x,\partial_y)\rightarrow (\partial_x,-\partial_y)$ and $(k_x,k_y)\rightarrow (k_x,-k_y)$, while the relative displacement field transforms differently as $(u_x,u_y)\rightarrow (-u_x,u_y)$ due to the exchange of two layers. Since $\mbox{Re}[(k_x+ik_y)(\partial_x+i\partial_y)(u_x+iu_y)]$ flips sign under $C_{2x}$, it is forbidden in the TBG electron phonon coupling. Similarly, one can check that terms like $\bm{\sigma}^*\cdot \mathbf{k}(\partial_xu_x+\partial_yu_y)$ and $\mathbf{\hat{z}}\cdot(\bm{\sigma}^*\times\mathbf{k})(\partial_yu_x-\partial_xu_y)$ are also forbidden by $C_{2x}$.

When the TBG is at the magic angle, namely $\alpha^2=1/3$, the second term in Eq. (\ref{HepK}) vanishes, and Eq. (\ref{Hepalpha}) becomes
\begin{equation}\label{HepKKM}
\begin{split}
&H_{\text{ep}}^{K,K_M'}(\mathbf{k})=-\frac{\gamma\hbar v}{3}\left[2k_x(\partial_yu_x+\partial_xu_y)+2k_y(\partial_xu_x-\partial_yu_y)+\bm{\sigma}^*\cdot \mathbf{k}(\partial_yu_x-\partial_xu_y) +(\sigma_xk_y+\sigma_yk_x)(\partial_xu_x+\partial_yu_y)\right].
\end{split}
\end{equation}

In the same way one could obtain the electron-phonon couplings at other valley/Moir\'e valley/spin indices, which we shall not repeat here. Instead, we give a symmetry analysis derivation of electron-phonon coupling at other valley/Moir\'e valley/spin indices in the next section above Eq. (\ref{Hep2}). The results are as shown in Eq. (\ref{Hep2}) as well as in the main text Eq. (3).

\section{The phonon Hamiltonian and phonon mediated electron-electron interaction}
In this section, we first describe the Hamiltonian of the relative in-plane displacement phonon mode, and then consider the phonon mediated electron-electron interaction.

\subsection{Phonon Hamiltonian}

In principle, the relative in-plane displacement phonons are optical phonons. However, if the two graphene lattices twisted by angle $\theta$ are not commensurate, a uniform relative in-plane displacement $\mathbf{u}$ does not cost energy, so these phonons are still acoustic. In fact, since the interlayer Van der Waals interaction between atoms in two graphene layers is much weaker than the intralayer atomic interaction, the in-plane polarized phonons of the two layers are nearly decoupled, and thus the relative in-plane displacement phonons are almost acoustic \cite{cocemasov2013}. As a good approximation, we shall ignore the coupling between phonons in different layers. From elastic dynamics \cite{landau7}, we know the in-plane deformation (i.e., in-plane phonon) energy of the TBG in the continuum limit can be expressed as a quadratic function of the expansion scalar $\Theta^{(j)}=\sum_l\partial_lu_l^{(j)}$ and the traceless shear tensor $\Sigma^{(j)}_{ab}=(\partial_au_b^{(j)}+\partial_bu_a^{(j)})/2- (\sum_l\partial_lu_l^{(j)})\delta_{ab}/2$ of layer $j=1,2$, where $a,b,l=x,y$. In particular, $\Theta^{(j)}$ and $\Sigma^{(j)}_{ab}$ occupy the spin $0$ and spin $2$ representations of the SO($2$) rotation group about $z$ axis, respectively (one can easily verify $\Sigma_{xx}\pm i\Sigma_{xy}$ have spin $\pm2$ under SO(2), respectively, where $\Sigma_{xx}$ and $\Sigma_{xy}$ are the two independent components of $\Sigma_{ab}$). Due to the $C_{6z}$ rotation symmetry of monolayer graphene, the symmetry allowed terms (scalars) in the deformation energy of layer $j$ must occupy the spin $0$ (mod $6$) representations of SO($2$). Therefore, the only allowed quadratic terms (scalars) are ${\Theta^{(j)}}^2$ and $\Sigma_{ab}^{(j)}\Sigma_{ab}^{(j)}$ which have spin $0$ (there are no interlayer terms since we have approximated the two layers as decoupled). Therefore, the Hamiltonian (to quadratic order) for in-plane displacement $\mathbf{u}^{(j)}$ (with two layers decoupled) can be written as
\begin{equation}
H_{ph}=\sum_{j=1}^2 \int d^2\mathbf{r}\left[\frac{M}{\Omega} (\partial_t\mathbf{u}^{(j)})^2+\frac{1}{2}K_{el}{\Theta^{(j)}}^2 +\mu_{el}\Sigma_{ab}^{(j)}\Sigma_{ab}^{(j)}\right]\ ,
\end{equation}
where $M$ is the Carbon atom mass, $\Omega=\sqrt{3}a_0^2/2$ is the graphene unit cell area, while $K_{el}$ and $\mu_{el}$ are the bulk modulus and shear modulus of monolayer graphene, respectively. The first term is the kinetic energy of the Carbon atoms (note that there are $2$ atoms in each graphene unit cell), while the second and the third terms are the elastic potential energy. To separate the relative deformation $\mathbf{u}=\mathbf{u}^{(1)}-\mathbf{u}^{(2)}$ and the center-of-mass deformation $\mathbf{u}^c=(\mathbf{u}^{(1)}+\mathbf{u}^{(2)})/2$, we define the relative expansion scalar and shear tensor as $\Theta=\sum_l\partial_lu_l$ and $\Sigma_{ab}=(\partial_au_b+\partial_bu_a)/2- (\sum_l\partial_lu_l)\delta_{ab}/2$, and the center-of-mass expansion and shear $\Theta^c=\sum_l\partial_lu_l^c$ and $\Sigma_{ab}^c=(\partial_au_b^c+\partial_bu_a^c)/2- (\sum_l\partial_lu_l^c)\delta_{ab}/2$, respectively. The phonon Hamiltonian can then be written as two parts $H_{ph}=H_{ph}^{r}+H_{ph}^{c}$, where the relative phonon wave part is
\begin{equation}\label{Hr}
\begin{split}
&\qquad H_{ph}^r=\int d^2\mathbf{r}\left[\frac{M}{2\Omega}(\partial_t\mathbf{u})^2 +\frac{1}{4}K_{el}\Theta^2+\frac{\mu_{el}}{2}\Sigma_{ab}\Sigma_{ab}\right]\\
&= \int d^2\mathbf{r}\left\{\frac{M}{2\Omega}(\partial_t\mathbf{u})^2 +\frac{K_{el}+\mu_{el}}{4}[(\partial_xu_x)^2+(\partial_yu_y)^2] +\frac{\mu_{el}}{4}[(\partial_xu_y)^2+(\partial_yu_x)^2]+\frac{K_{el}}{2}\partial_xu_x\partial_yu_y\right\}\ ,
\end{split}
\end{equation}
and the center-of-mass phonon wave part is
\begin{equation}
H_{ph}^c=\int d^2\mathbf{r}\left[\frac{2M}{\Omega}(\partial_t\mathbf{u}^c)^2 +K_{el}{(\Theta^c)}^2+2\mu_{el}\Sigma_{ab}^c\Sigma_{ab}^c\right]\ .
\end{equation}
Since the relative phonon wave dominates the electron-phonon coupling, we shall consider only the relative phonon part $H_{ph}^r$ in Eq. (\ref{Hr}). Explicitly, the equation of motion of $H_{ph}^r$ is
\begin{equation}
\frac{M}{\Omega}\partial_t^2\left(\begin{array}{c}u_x\\u_y\end{array}\right)= \frac{1}{2}\left(\begin{array}{cc}K_{el}\partial_x^2+\mu_{el}(\partial_x^2+\partial_y^2) &K_{el}\partial_x\partial_y \\K_{el}\partial_x\partial_y&K_{el}\partial_y^2+\mu_{el}(\partial_x^2+\partial_y^2)\end{array}\right) \left(\begin{array}{c}u_x\\u_y\end{array}\right)=\frac{K_{el}}{2}\nabla(\nabla\cdot \mathbf{u})+ \frac{\mu_{el}}{2}\nabla^2\mathbf{u}\ .
\end{equation}
Given the momentum $-i\nabla=\mathbf{p}=(p_x,p_y)$, there are two eigenvectors: $\mathbf{u}_{\mathbf{p},L}\propto\mathbf{p}$ and $\mathbf{u}_{\mathbf{p},T}\propto\hat{\mathbf{z}}\times\mathbf{p}$, which satisfies $\nabla(\nabla\cdot \mathbf{u}_{\mathbf{p},L})=\nabla^2\mathbf{u}_{\mathbf{p},L}$ and $\nabla(\nabla\cdot \mathbf{u}_{\mathbf{p},T})=0$, respectively. Accordingly, their eigenfrequencies are $\omega_{\mathbf{p},L}=c_Lp$ and $\omega_{\mathbf{p},T}=c_Tp$, respectively, where $c_L=\sqrt{(K_{el}+\mu_{el})\Omega/2M}$ and $c_T=\sqrt{\mu_{el}\Omega/2M}$ being the longitudinal sound speed and transverse sound speed of monolayer graphene, and $p=|\mathbf{p}|$.

The Hamiltonian of relative phonon waves can then be quantized following the standard canonical method. In real space, the canonical momentum $\bm{\pi}(\mathbf{r})=\frac{M}{\Omega}\partial_t\mathbf{u}(\mathbf{r})$ satisfies $[u_a(\mathbf{r}),\pi_b(\mathbf{r}')]=i\hbar\delta_{ab}\delta(\mathbf{r}-\mathbf{r}')$ (where $a,b=x,y$). When transformed into the momentum space, the canonical momentum $\bm{\pi}_{\mathbf{p}}=\int\frac{d^2\mathbf{r}}{\sqrt{A_s}}e^{-i\mathbf{p}\cdot\mathbf{r}}\bm{\pi}(\mathbf{r})$ and displacement $\mathbf{u}_\mathbf{p}=\int\frac{d^2\mathbf{r}}{\sqrt{A_s}}e^{-i\mathbf{p}\cdot\mathbf{r}} \mathbf{u}(\mathbf{r})$ satisfies $[u_{\mathbf{p},a},\pi_{\mathbf{p}',b}]=i\hbar\delta_{ab}\delta_{\mathbf{p},-\mathbf{p}'}$, where $A_s$ is the total area of the sample. We can decompose $\mathbf{u}_\mathbf{p}=\mathbf{u}_{\mathbf{p},L}+\mathbf{u}_{\mathbf{p},T}$ into a longitudinal part $\mathbf{u}_{\mathbf{p},L}$ and a transverse part $\mathbf{u}_{\mathbf{p},T}$ (which are perpendicular to each other), and similarly we can also do so for the canonical momentum $\bm{\pi}_{\mathbf{p}}=\bm{\pi}_{\mathbf{p},L}+\bm{\pi}_{\mathbf{p},T}$ (which are also perpendicular to each other, $\mathbf{u}_{\mathbf{p},T}\cdot \mathbf{u}_{\mathbf{p},L}=0$). One can then show the phonon Hamiltonian in the momentum space becomes
\begin{equation}
H_{\text{ph}}^\text{r}=\sum_\mathbf{p}\left[\frac{\Omega}{2M}\bm{\pi}_\mathbf{p}\cdot\bm{\pi}_{-\mathbf{p}}+ \frac{K_{el}}{2}(\mathbf{p}\cdot \mathbf{u}_\mathbf{p})(\mathbf{p}\cdot \mathbf{u}_{-\mathbf{p}})+ \frac{\mu_{el}}{2}p^2\mathbf{u}_{\mathbf{p}}\cdot \mathbf{u}_{-\mathbf{p}}\right] =\sum_\mathbf{p}\left(\hbar \omega_{\mathbf{p},L}a_{\mathbf{p},L}^\dag a_{\mathbf{p},L}+\hbar \omega_{\mathbf{p},T}a_{\mathbf{p},T}^\dag a_{\mathbf{p},T}\right)\ ,
\end{equation}
where $a_{\mathbf{p},\chi}$, $a_{\mathbf{p},\chi}^\dag$ are the phonon annihilation and creation operators with momentum $\mathbf{p}$ and polarization $\chi$ satisfying $[a_{\mathbf{p},\chi},a_{\mathbf{p},\chi}^\dag]=1$. The relative displacement phonon field $\mathbf{u}$ in the Shr\"odinger picture is given by
\begin{equation}\label{up}
\mathbf{u}(\mathbf{r})=\sum_{\mathbf{p}} \frac{e^{i\mathbf{p}\cdot\mathbf{r}}}{\sqrt{N_s\Omega_s}} (i\hat{\mathbf{p}}u_{\mathbf{p},L}+i\hat{\mathbf{z}}\times\hat{\mathbf{p}}u_{\mathbf{p},T})\ ,\ \   u_{\mathbf{p},L}=\sqrt{\frac{\hbar\Omega}{2M\omega_{\mathbf{p},L}}}(a_{\mathbf{p},L}+a_{-\mathbf{p},L}^\dag)\ ,\ \  u_{\mathbf{p},T}=\sqrt{\frac{\hbar\Omega}{2M\omega_{\mathbf{p},T}}}(a_{\mathbf{p},T}+a_{-\mathbf{p},T}^\dag)\ ,
\end{equation}
where $\hat{\mathbf{p}}=\mathbf{p}/p$ is the unit vector along momentum $\mathbf{p}$, while $\hat{\mathbf{z}}$ is the unit vector along the out of plane direction ($\hat{\mathbf{z}}\times\hat{\mathbf{p}}$ is the transverse direction), and for later convenience we have rewritten the total area of the sample $A_s$ as $N_s\Omega_s$, with $N_s$ being the number of superlattice unit cells, and $\Omega_s=\Omega/[4\sin^2(\theta/2)]$ being the superlattice unit cell area. In the Heisenberg picture, the phonon field $\mathbf{u}$ is time $t$ dependent, namely, $\mathbf{u}(\mathbf{r},t)=e^{iH_{\text{ph}}^\text{r}t}\mathbf{u}(\mathbf{r})e^{-iH_{\text{ph}}^\text{r}t}$, with $\mathbf{u}(\mathbf{r})$ defined in Eq. (\ref{up}) above. Accordingly, one can obtain the canonical momentum as $\bm{\pi}(\mathbf{r},t)=\frac{M}{\Omega}\partial_t\mathbf{u}(\mathbf{r},t)$.

Since the interlayer phonon coupling is ignored, the phonon bands in the MBZ is simply obtained by folding $\mathbf{p}$ into the superlattice MBZ. Among these phonon bands, the lowest two bands (one longitudinal and one transverse) are acoustic (under the approximation that the two layers are decoupled) and cause the long wavelength deformation of the Moir\'e pattern superlattice, while the higher bands are optical and mainly lead to short wavelength deformations within each unit cell of the superlattice. Since our electron phonon coupling is derived in the long wavelength deformation limit, for now we shall restrict ourselves to the lowest acoustic phonon band in the MBZ. We will briefly discuss the contribution of optical phonon bands at the end of this supplementary section.

\subsection{Electron-Phonon Coupling for other valley/Moir\'e valley/spin}

For convenience, hereafter we define the graphene BZ valley $K,K'$ as index $\eta=\pm1$, the Moir\'e valley $K_M,K_M'$ as index $\zeta=\pm1$, and the spin up and down as index $s=\pm1$. Since there is no spin-orbit coupling, the electron-phonon coupling does not flip spin and is independent of spin index $s$. In last section we derived the electron-phonon coupling for $(\eta,\zeta)=(+1,-1)$. Now we use symmetry arguments to obtain the expression of electron-phonon coupling at other indices $(\eta,\zeta)$. Identical results can be obtained by performing brute force calculations.

First, TBG has the time-reversal symmetry. A time-reversal transformation $\mathcal{T}$ changes $(\eta,\zeta,s)\rightarrow (-\eta,-\zeta,-s)$ and $\mathbf{k}\rightarrow -\mathbf{k}$. Since for spinless (no spin-orbit coupling) fermions the time reversal $\mathcal{T}$ is simply complex conjugation, one finds the electron-phonon coupling at $(-\eta,-\zeta)$ is given by
\begin{equation}\label{TR}
H_{\text{ep}}^{-\eta,-\zeta}(\mathbf{k})=\left[H_{\text{ep}}^{\eta,\zeta}(-\mathbf{k})\right]^*\ ,
\end{equation}
where we have found $H_{\text{ep}}^{K,K_M'}(\mathbf{k})$ in Eq. (\ref{HepKKM}), and $z^*$ stands for the complex conjugate of $z$.

Secondly, TBG also has a $C_{2x}$ symmetry, which is the 2-fold rotation about $x$ axis (see Fig. \ref{figS1}a for definitions of $x$ and $y$ axes). From the main text Fig. 2a and supplementary Fig. \ref{figS1}, one can see the $C_{2x}$ transformation interchanges Moir\'e valley $K_M$ and $K_M'$, changes momentum $(k_x,k_y)$ (measured from the $K_M$ point) to $(k_x,-k_y)$ (measured from the $K_M'$ point) in momentum space, interchanges layer $1$ with layer $2$, and interchanges sublattice indices A and B. Meanwhile, the valley $K$ or $K'$ remains invariant. Since the Pauli matrices in Eq. (\ref{SeqH}) (and afterwards) are in the sublattice basis, the interchange of sublattice indices A and B leads to a transformation of Pauli matrices $\bm{\sigma}\rightarrow\sigma_x\bm{\sigma}\sigma_x^{-1}$, namely, $(\sigma_x,\sigma_y)\rightarrow(\sigma_x,-\sigma_y)$. Therefore, the band Hamiltonian of the continuum model at Moir\'e valley $K_M$ and valley $K$ is given by
\begin{equation}
H^{K,K_M}(\mathbf{k})=U\left[H^{K,K_M'}(k_x,-k_y)\right]U^{-1}=\left(\begin{array}{cccc}
h^{K}_{-\theta/2}(\mathbf{k})& wT_1& wT_3&wT_2\\
wT_1^\dag&h^{K}_{\theta/2}(\mathbf{k}+\mathbf{q}_1)&0&0\\
wT_3^\dag&0&h^{K}_{\theta/2}(\mathbf{k}+\mathbf{q}_3)&0\\
wT_2^\dag&0&0&h^{K}_{\theta/2}(\mathbf{k}+\mathbf{q}_2)\\
\end{array}\right)\ ,
\end{equation}
where $H^{K,K_M'}(\mathbf{k})$ is given by Eq. (\ref{SeqH}), and $U=\sigma_x\otimes I_4$ is the transformation matrix for the interchange of sublattices A and B under $C_{2x}$ (i.e., $\psi_{j,\mathbf{k}}\rightarrow\sigma_x \psi_{j,\mathbf{k}}$ for each $\psi_{j,\mathbf{k}}$ ($1\le j\le 4$) in the basis of Hamiltonian $(\psi_{0,\mathbf{k}}^T,\psi_{1,\mathbf{k}}^T,\psi_{2,\mathbf{k}}^T,\psi_{3,\mathbf{k}}^T)^T$, recall that $\psi_{j,\mathbf{k}}$ are defined in the 2D Hilbert space of sublattice A and B). If one approximate $\theta/2=0$ in $h^{K}_{\pm\theta/2}$ (as we have assumed in the first supplementary section), and interchanges electron basis $\psi_{2,\mathbf{k}}\leftrightarrow \psi_{3,\mathbf{k}}$, one finds $H^{K,K_M}(\mathbf{k})$ and $H^{K,K_M'}(\mathbf{k})$ only differ by a sign flip $\mathbf{q}_j\rightarrow-\mathbf{q}_j$ ($j=1,2,3$). From Eq. (\ref{HepK}) one sees the expression of electron-phonon coupling $H_{\text{ep}}^{K,K_M'}(\mathbf{k})$ is quadratic in $\mathbf{q}_j$ (thus invariant under $\mathbf{q}_j\rightarrow-\mathbf{q}_j$), so we conclude $H_{\text{ep}}^{K,K_M'}(\mathbf{k})=H_{\text{ep}}^{K,K_M}(\mathbf{k})$ (where $\mathbf{k}$ of $H_{\text{ep}}^{K,K_M'}(\mathbf{k})$ and $H_{\text{ep}}^{K,K_M}(\mathbf{k})$ are measured from $K_M'$ and $K_M$, respectively), and $H_{\text{ep}}^{\eta,\zeta}(\mathbf{k})$ is invariant under $\zeta\rightarrow-\zeta$. Alternatively, one can achieve this conclusion by directly applying $C_{2x}$ on $H_{\text{ep}}^{K,K_M'}(\mathbf{k})$ we derived in Eq. (\ref{Hepalpha}). In particular, $C_{2x}$ changes the displacement field in layer $j$ as $(u^{(j)}_x,u^{(j)}_y)\rightarrow (u^{(j)}_x,-u^{(j)}_y)$, and interchanges layer $1$ with layer $2$, so the relative displacement field $\mathbf{u}=\mathbf{u}^{(1)}-\mathbf{u}^{(2)}$ changes as $(u_x,u_y)\rightarrow (-u_x,u_y)$. Besides, it changes $(\partial_x,\partial_y)\rightarrow (\partial_x,-\partial_y)$. Therefore, one finds
\begin{equation}
H_{\text{ep}}^{K,K_M}(\mathbf{k}) =\sigma_x\left[H_{\text{ep}}^{K,K_M'}(k_x,-k_y) \Big|_{\partial_y\rightarrow-\partial_y, u_x\rightarrow -u_x}\right]\sigma_x^{-1} =H_{\text{ep}}^{K,K_M'}(\mathbf{k})\ ,
\end{equation}
where $H_{\text{ep}}^{K,K_M'}(\mathbf{k})$ is given in Eq. (\ref{Hepalpha}) (obtained under the approximation $\theta/2=0$ in $h^{K}_{\pm\theta/2}$ in the first supplementary section). This indicates the electron-phonon coupling $H_{\text{ep}}^{\eta,\zeta}(\mathbf{k})$ is also independent of the Moir\'e valley $\zeta$. 
We can then obtain the electron-phonon coupling Hamiltonian for all indices, which in the second quantized language takes the form
\begin{equation}\label{Hep2}
\begin{split}
&H_{\text{ep}}^{\eta,\zeta,s}=H_{C3}^{\eta,\zeta,s}+H_{SO(2)}^{\eta,\zeta,s}\ ,\\
&H_{C3}^{\eta,\zeta,s}=g_{1\alpha}\frac{\gamma\eta \hbar v}{\sqrt{N_s\Omega_s}}\sum_{\mathbf{k},\mathbf{k}'}p\ \psi_{\mathbf{k}}^{\eta,\zeta,s\dag} \left\{\left[2\bar{k}_x\hat{p}_x\hat{p}_y+\bar{k}_y(\hat{p}_x^2-\hat{p}_y^2)\right]u_{\mathbf{p},L}+ \left[2\bar{k}_y\hat{p}_x\hat{p}_y-\bar{k}_x(\hat{p}_x^2-\hat{p}_y^2)\right] u_{\mathbf{p},T}\right\}\psi_{\mathbf{k}'}^{\eta,\zeta,s},\\
&H_{SO(2)}^{\eta,\zeta,s}=-\frac{\gamma\hbar v}{\sqrt{N_s\Omega_s}}\sum_{\mathbf{k},\mathbf{k}'}p\ \psi_{\mathbf{k}}^{\eta,\zeta,s\dag}\left[g_{2\alpha} \left(\eta\sigma_x\bar{k}_x-\sigma_y\bar{k}_y\right)u_{\mathbf{p},T} +g_{3\alpha} \left(\sigma_y\bar{k}_x+\eta\sigma_x\bar{k}_y\right)u_{\mathbf{p},L} \right]\psi_{\mathbf{k}'}^{\eta,\zeta,s}\ ,
\end{split}
\end{equation}
where the numerical factors are from Eq. (\ref{Hepalpha}):
\begin{equation}
g_{1\alpha}=\frac{9\alpha^2(1+3\alpha^2)}{(1+6\alpha^2)^2}\ ,\qquad g_{2\alpha}=\frac{9\alpha^2}{(1+6\alpha^2)^2}\ , \qquad g_{3\alpha}=\frac{3\alpha^2}{1+6\alpha^2}\ ,
\end{equation}
while $\psi_{\mathbf{k}'}^{\eta,\zeta,s}$ is the electron annihilation operator at momentum $\mathbf{k}$ with $\eta,\zeta,s$ indices, $\mathbf{k}-\mathbf{k}'=\mathbf{p}$ (the momentum of the $\partial_au_b$ terms), and we have defined $\bar{\mathbf{k}}=(\mathbf{k}+\mathbf{k}')/2$ is the average electron momentum before and after phonon emission (absorption), which comes from Eq. (\ref{Hepalpha}). Physically, this comes from the Fourier transformation of the real space hopping amplitude $t(\mathbf{r},\mathbf{r}')c^\dag(\mathbf{r})c(\mathbf{r}')$, where $t(\mathbf{r},\mathbf{r}')$ is induced by the displacement field $\mathbf{u}(\mathbf{r})$. In particular, for the magic angle $\alpha^2=1/3$, we have $g_{1\alpha}=2/3$ and $g_{2\alpha}=g_{3\alpha}=1/3$. Besides, as we have shown below Eq. (\ref{Hepalpha}), the term $H_{C3}^{\eta,\zeta,s}$ is only $C_{3z}$ rotationally invariant,
while the term $H_{SO(2)}^{\eta,\zeta,s}$ is SO(2) rotationally invariant. Finally, we note that $H_{C3}^{\eta,\zeta,s}$ is odd in $K,K'$ valley index $\eta$, since the electron band Hamiltonian and thus the electron-phonon coupling undergoes a sign flip and complex conjugate under $\eta\rightarrow-\eta$.

The electron-phonon coupling $H_{\text{ep}}^{\eta,\zeta,s}$ we obtained is independent of the Moir\'e valley $\zeta$ and the spin $s$. However, we note that the independence of Moir\'e valley holds only under the approximation $\theta/2=0$ in $h^{K}_{\pm\theta/2}(\mathbf{k})$. If we do not make this approximation, $H_{\text{ep}}^{\eta,\zeta,s}$ will be weakly $\zeta$ dependent, where the $\zeta$ dependent part will be of a factor $\mathcal{O}(\theta)$ smaller than the $\zeta$ independent part given in Eq. (\ref{Hepalpha}). Since we are interested in small angles $\theta\sim 1^\circ\sim 0.02$ rad, we shall ignore the weak $\zeta$ dependence of $H_{\text{ep}}^{\eta,\zeta,s}$. Besides, since the continuum model of TBG has no coupling between valleys $K$ and $K'$, the electron-phonon coupling we obtained has a definite $K,K'$ valley index $\eta$ (i.e., initial and final states of the electron are in the same valley $\eta$).

\subsection{Phonon-Mediated Electron-Electron Interaction}

We now calculate the phonon mediated electron-electron interaction near the Fermi surface. To be concrete, we shall assume the Fermi surface is at $|\mathbf{k}|=k_F$ in the hole Dirac bands (where superconductivity is observed), and project the electron operators to the vicinity of the Fermi surface. Such a Fermi surface might not be very accurate since the band is quite flat, but we shall simply model the Fermi surface by a circle with a large $k_F$ of a low velocity 2D Dirac fermion. The wave function $\phi^{\eta}_{\mathbf{k}}$ of the Dirac hole band state $|\mathbf{k}_{\eta,\zeta,s}\rangle$ is assumed to be the negative eigenvalue eigenstate of $(\eta\sigma_x k_x-\sigma_yk_y)$ (which is proportional to the Dirac Hamiltonian $\widetilde{H}^{\eta,\zeta}(\mathbf{k})=\frac{1-3\alpha^2}{1+6\alpha^2}\hbar v (\eta\sigma_x k_x-\sigma_yk_y)$), namely, $\phi^{\eta}_{\mathbf{k}}=(1,-\eta e^{-i\eta\varphi_\mathbf{k}})^T/\sqrt{2}$ at valley $\eta$, where $\varphi_\mathbf{k}=\arg(k_x+ik_y)$ is the polar angle of momentum $\mathbf{k}$ (we note that for $\alpha^2<1/3$ this is the valence band of the Dirac Hamiltonian, while for $\alpha^2>1/3$ this in fact becomes the conduction band of the Dirac Hamiltonian, and for $\alpha^2=1/3$ the conduction or valence band becomes ill-defined unless higher order terms in $\mathbf{k}$ are included. Here we shall ignore these complications and take the wave function $\phi^{\eta}_{\mathbf{k}}$). One can then rewrite the Dirac annihilation and creation operators $\psi_{\mathbf{k}}^{\eta,\zeta,s}$, $\psi_{\mathbf{k}}^{\eta,\zeta,s\dag}$ in Eq. (\ref{Hep2}) as $\psi_{\mathbf{k}}^{\eta,\zeta,s}=\phi^{\eta}_{\mathbf{k}}c_{\mathbf{k},\eta,\zeta,s}$ and $\psi_{\mathbf{k}}^{\eta,\zeta,s\dag}=\phi^{\eta\dag}_{\mathbf{k}}c_{\mathbf{k},\eta,\zeta,s}^\dag$, where $c_{\mathbf{k},\eta,\zeta,s}$ and $c_{\mathbf{k},\eta,\zeta,s}^\dag$ are the electron annihilation and creation operators in the hole Dirac band with indices $\eta,\zeta,s$. The Dirac hole band Hamiltonian can then be written as
\[
\widetilde{H}^{\eta,\zeta,s}(\mathbf{k}) =\xi_\mathbf{k}c_{\mathbf{k},\eta,\zeta,s}^\dag c_{\mathbf{k},\eta,\zeta,s}\ ,
\]
where $\xi_\mathbf{k}=-\frac{1-3\alpha^2}{1+6\alpha^2}\hbar v(|\mathbf{k}|-k_F)$ is the the band energy relative to the Fermi level.

We then make the approximation $|\mathbf{k}|\approx|\mathbf{k}'|\approx k_F$, based on which we find the two terms in $H_{SO(2)}^{\eta,\zeta,s}$ are approximately
\[
\phi^{\eta\dag}_{\mathbf{k}}\left(\eta\sigma_x\bar{k}_x-\sigma_y\bar{k}_y\right)\phi^\eta_{\mathbf{k}'} =\phi^{\eta\dag}_{\mathbf{k}}\left[\frac{\left(\eta\sigma_xk_x-\sigma_yk_y\right)}{2} +\frac{\left(\eta\sigma_xk_x'-\sigma_yk_y'\right)}{2}\right]\phi^\eta_{\mathbf{k}'} =\phi^{\eta\dag}_{\mathbf{k}}\left(-\frac{|\mathbf{k}|}{2}-\frac{|\mathbf{k}'|}{2}\right)\phi^\eta_{\mathbf{k}'}
\approx -k_F \phi^{\eta\dag}_{\mathbf{k}}\phi^{\eta}_{\mathbf{k}'}\ ,
\]
\[
\phi^{\eta\dag}_{\mathbf{k}}\left(\eta\sigma_x\bar{k}_y+\sigma_y\bar{k}_x\right)\phi^\eta_{\mathbf{k}'} =\phi^{\eta\dag}_{\mathbf{k}}\left[\frac{\left(\eta\sigma_xk_y+\sigma_yk_x\right)}{2} +\frac{\left(\eta\sigma_xk_y'+\sigma_yk_x'\right)}{2} \right]\phi^\eta_{\mathbf{k}'}=i (|\mathbf{k}'|-|\mathbf{k}|)\phi^{\eta\dag}_{\mathbf{k}}\sigma_z \phi^{\eta}_{\mathbf{k}'} \approx 0\ ,
\]
where we have used the definition $\left(\eta\sigma_xk_x-\sigma_yk_y\right)\phi^\eta_{\mathbf{k}}=-|\mathbf{k}|\phi^\eta_{\mathbf{k}}$, and $\left(\eta\sigma_xk_y+\sigma_yk_x\right)\phi^\eta_{\mathbf{k}} =-i\sigma_z\left(\eta\sigma_xk_x-\sigma_yk_y\right)\phi^\eta_{\mathbf{k}} =i|\mathbf{k}|\sigma_z\phi^{\eta}_{\mathbf{k}}$.
Besides, we note that the two terms in $H_{C3}^{\eta,\zeta,s}$ of Eq. (\ref{Hep2}) satisfy
\begin{equation}
\begin{split}
&-\left[2\bar{k}_y\hat{p}_x\hat{p}_y-\bar{k}_x(\hat{p}_x^2-\hat{p}_y^2)\right] +i\left[2\bar{k}_x\hat{p}_x\hat{p}_y+\bar{k}_y(\hat{p}_x^2-\hat{p}_y^2)\right]=(\hat{p}_x+i\hat{p}_y)^2 (\bar{k}_x+i\bar{k}_y)=\left(\frac{p_x+ip_y}{p_x-ip_y}\right)(\bar{k}_x+i\bar{k}_y) \\
&=\left(\frac{|\mathbf{k}|e^{i\varphi_{\mathbf{k}}} -|\mathbf{k}'|e^{i\varphi_{\mathbf{k}'}}}{ |\mathbf{k}|e^{-i\varphi_{\mathbf{k}}} -|\mathbf{k}'|e^{-i\varphi_{\mathbf{k}'}}}\right)\times \left( \frac{|\mathbf{k}|e^{i\varphi_{\mathbf{k}}} +|\mathbf{k}'|e^{i\varphi_{\mathbf{k}'}}}{2} \right) \approx\frac{k_F}{2} \frac{e^{2i \varphi_\mathbf{k}} -e^{2i \varphi_{\mathbf{k}'}} }{e^{-i \varphi_\mathbf{k}}-e^{-i \varphi_{\mathbf{k}'}} }\ ,
\end{split}
\end{equation}
where we have used $\mathbf{p}=\mathbf{k}-\mathbf{k}'$ and $\bar{\mathbf{k}}=(\mathbf{k}+\mathbf{k}')/2$.
Therefore, they can be written as the real part and imaginary part of the complex quantity $\frac{k_F}{2} \frac{e^{2i \varphi_\mathbf{k}} -e^{2i \varphi_{\mathbf{k}'}} }{e^{-i \varphi_\mathbf{k}}-e^{-i \varphi_{\mathbf{k}'}} }$, respectively.
This enables us to approximate the electron-phonon interaction projected in the Dirac hole band as
\begin{equation}\label{Hepp}
\begin{split}
&H_{\text{ep}}^{\eta,\zeta,s}\approx \frac{\gamma\hbar v k_F}{\sqrt{N_s\Omega_s}}\sum_{\mathbf{k},\mathbf{k}'}p \left\{\left[g_{2\alpha}-\eta\frac{g_{1\alpha}}{2}\text{Re}\left(\frac{e^{2i \varphi_\mathbf{k}} -e^{2i \varphi_{\mathbf{k}'}} }{e^{-i \varphi_\mathbf{k}}-e^{-i \varphi_{\mathbf{k}'}} }\right)\right]u_{\mathbf{p},T}+ \eta \frac{g_{1\alpha}}{2}\text{Im}\left(\frac{e^{2i \varphi_\mathbf{k}} -e^{2i \varphi_{\mathbf{k}'}} }{e^{-i \varphi_\mathbf{k}}-e^{-i \varphi_{\mathbf{k}'}} }\right)u_{\mathbf{p},L}\right\} \\
&\qquad \qquad \times\phi^{\eta\dag}_{\mathbf{k}}\phi^{\eta}_{\mathbf{k}'} c^\dag_{\mathbf{k},\eta,\zeta,s} c_{\mathbf{k}',\eta,\zeta,s}\ .
\end{split}
\end{equation}
In particular, under momentum reversal $\mathbf{k},\mathbf{k}'\rightarrow -\mathbf{k},-\mathbf{k}'$, the polar angles $\varphi_\mathbf{k}\rightarrow \varphi_\mathbf{k}+\pi$ and $\varphi_{\mathbf{k}'}\rightarrow \varphi_{\mathbf{k}'}+\pi$), so we have $\frac{e^{2i \varphi_\mathbf{k}} -e^{2i \varphi_{\mathbf{k}'}} }{e^{-i \varphi_\mathbf{k}}-e^{-i \varphi_{\mathbf{k}'}} }\rightarrow \frac{e^{2i \varphi_\mathbf{k}+2i\pi} -e^{2i \varphi_{\mathbf{k}'}+2i \pi} }{e^{-i \varphi_\mathbf{k}-i\pi}-e^{-i \varphi_{\mathbf{k}'}-i\pi} }= -\frac{e^{2i \varphi_\mathbf{k}} -e^{2i \varphi_{\mathbf{k}'}} }{e^{-i \varphi_\mathbf{k}}-e^{-i \varphi_{\mathbf{k}'}} }$. Therefore, the terms in the projected electron-phonon coupling contributed by $H_{C3}^{\eta,\zeta,s}$ (proportional to $g_{1\alpha}$) and by $H_{SO(2)}^{\eta,\zeta,s}$ (proportional to $g_{2\alpha}$) are odd and even under momentum reversal $\mathbf{k},\mathbf{k}'\rightarrow -\mathbf{k},-\mathbf{k}'$, respectively, which will be useful in later calculations. We note that this odd/evenness results from the projection of electron-phonon coupling onto a single band. In the original expression of Eq. (\ref{Hep2}) before projection, one may thought both $H_{C3}^{\eta,\zeta,s}$ and $H_{SO(2)}^{\eta,\zeta,s}$ are odd in $\mathbf{k}$ since they are both linear in $\mathbf{k}$. However, this naive expectation ignores the effect of the Dirac wave functions $\psi_{\mathbf{k}}^{\eta,\zeta,s}$ in the expression of Eq. (\ref{Hep2}), thus is incorrect.

One can then use the standard second order perturbation theory to calculate the phonon-mediated electron electron interaction. We treat the electron energy $\widetilde{H}^{\eta,\zeta,s}(\mathbf{k})$ and the phonon energy $H_{ph}^r$ as the unperturbed Hamiltonian, and the electron-phonon interaction $H_{\text{ep}}^{\eta,\zeta,s}$ as the perturbation. The electron electron interaction is then induced by the emission and absorption of a phonon between two electrons. Consider the initial state $|\Psi_0\rangle=|\mathbf{k}_{1,I},\mathbf{k}_{2,I'}\rangle =c^\dag_{\mathbf{k}_1,I}c^\dag_{\mathbf{k}_2,I'}|0\rangle$ of two electrons which have momenta $\mathbf{k}_1$ and $\mathbf{k}_2$, respectively, where $I=(\eta,\zeta,s)$ and $I'=(\eta',\zeta',s')$ are short hand for their Dirac cone indices, and $|0\rangle$ is the particle vacuum. The initial state energy is $E_0=\xi_{\mathbf{k}_1}+\xi_{\mathbf{k}_2}$. Assume the final state is $|\Psi_f\rangle=|\mathbf{k}_{3,I},\mathbf{k}_{4,I'}\rangle$, where the total momentum $\mathbf{k}_3+\mathbf{k}_4=\mathbf{k}_1+\mathbf{k}_2$ is conserved, and the final state energy $E_f=\xi_{\mathbf{k}_3}+\xi_{\mathbf{k}_4}=E_0$ remains unchanged. There are two intermediate states with an emitted phonon with polarization $\chi$: one is $|\Psi_{1,\chi}\rangle=|\mathbf{k}_{3,I},\mathbf{k}_{2,I'},\mathbf{p}_{,\chi}\rangle =c^\dag_{\mathbf{k}_3,I}c^\dag_{\mathbf{k}_2,I'}a_{\mathbf{p},\chi}^\dag|0\rangle$ where a phonon with momentum $\mathbf{p}$ and polarization $\chi$ is emitted from the first electron, while the other is $|\Psi_{2,\chi}\rangle=|\mathbf{k}_{1,I},\mathbf{k}_{4,I'},-\mathbf{p}_{,\chi}\rangle$ where a phonon with momentum $-\mathbf{p}$ and polarization $\chi$ is emitted from the second electron, with the momentum $\mathbf{p}=\mathbf{k}_1-\mathbf{k}_3=\mathbf{k}_4-\mathbf{k}_2$. The energy of the two intermediate states are $E_{1,\chi}=\xi_{\mathbf{k}_3}+\xi_{\mathbf{k}_2}+ \hbar\omega_{\mathbf{p},\chi}$ and $E_{2,\chi}=\xi_{\mathbf{k}_1}+\xi_{\mathbf{k}_4}+ \hbar\omega_{-\mathbf{p},\chi}$, respectively, where $\hbar\omega_{\mathbf{p},\chi}$ is the phonon energy. The phonon of the intermediate state is then absorbed by the other electron, resulting in the final state $|\Psi_f\rangle$. According to the second order perturbation theory, the interaction between the two electrons is given by
\begin{equation}\label{SeqVII}
\begin{split}
\frac{V^{II'}_{\mathbf{k}_3\mathbf{k}_4,\mathbf{k}_1\mathbf{k}_2}}{N_s\Omega_s}&=\sum_\chi \left(\frac{\langle \Psi_f|H_{\text{ep}}^{I'}|\Psi_{1,\chi}\rangle\langle \Psi_{1,\chi}|H_{\text{ep}}^{I}|\Psi_0\rangle}{E_0-E_{1,\chi}} +\frac{\langle \Psi_f|H_{\text{ep}}^{I}|\Psi_{2,\chi}\rangle\langle \Psi_{2,\chi}|H_{\text{ep}}^{I'}|\Psi_0\rangle}{E_0-E_{2,\chi}}\right)\\
&=\sum_\chi \left(\frac{\langle \Psi_f|H_{\text{ep}}^{I'}|\Psi_1\rangle\langle \Psi_1|H_{\text{ep}}^{I}|\Psi_0\rangle}{\xi_{\mathbf{k}_3}-\xi_{\mathbf{k}_1}-\hbar\omega_{\mathbf{p},\chi}} +\frac{\langle \Psi_f|H_{\text{ep}}^{I}|\Psi_2\rangle\langle \Psi_2|H_{\text{ep}}^{I'}|\Psi_0\rangle}{\xi_{\mathbf{k}_4}-\xi_{\mathbf{k}_2}-\hbar\omega_{-\mathbf{p},\chi}} \right)\ ,
\end{split}
\end{equation}
where $I=(\eta,\zeta,s)$ and $I'=(\eta',\zeta',s')$ are notations for the Dirac cone indices of the two electrons. Since the electron phonon interaction $H_{\text{ep}}^{I}$ is independent of spin $s$, the above electron-electron interaction is independent of $s$ and $s'$. Besides, in the continuum model $H_{\text{ep}}^{I}$ is also independent of Moir\'e valley $\zeta$ (see Eq. (\ref{Hep2})). Since our goal is to study the BCS superconductivity, we shall focus on the Cooper channel of the interaction, namely, we shall set the momenta of the two electrons as opposite to each other, $\mathbf{k}_1=-\mathbf{k}_2=\mathbf{k}$ and $\mathbf{k}_3=-\mathbf{k}_4=\mathbf{k}'$. The Cooper channel electron-electron interaction is then
\begin{equation}\label{VintS}
\begin{split}
\frac{V^{\eta\eta'}_{\mathbf{k}\mathbf{k}'}(\omega)}{N_s\Omega_s} &=\sum_{\chi}\Big[\frac{\langle\mathbf{k}'_{\eta,\zeta,s},-\mathbf{k}'_{\eta',\zeta',s'} |H_{\text{ep}}^{\eta',\zeta',s'} |\mathbf{k}'_{\eta,\zeta,s},-\mathbf{k}_{\eta',\zeta',s'},\mathbf{p}_{,\chi}\rangle \langle\mathbf{k}'_{\eta,\zeta,s},-\mathbf{k}_{\eta',\zeta',s'},\mathbf{p}_{,\chi} |H_{\text{ep}}^{\eta,\zeta,s}|\mathbf{k}_{\eta,\zeta,s},-\mathbf{k}_{\eta',\zeta',s'}\rangle }{\hbar\omega-\hbar\omega_{\mathbf{p},\chi}}\\
+& \frac{\langle\mathbf{k}'_{\eta,\zeta,s},-\mathbf{k}'_{\eta',\zeta',s'} |H_{\text{ep}}^{\eta,\zeta,s}|\mathbf{k}_{\eta,\zeta,s},-\mathbf{k}'_{\eta',\zeta',s'},-\mathbf{p}_{,\chi}\rangle \langle\mathbf{k}_{\eta,\zeta,s},-\mathbf{k}'_{\eta',\zeta',s'},-\mathbf{p}_{,\chi} |H_{\text{ep}}^{\eta',\zeta',s'}|\mathbf{k}_{\eta,\zeta,s},-\mathbf{k}_{\eta',\zeta',s'}\rangle }{-\hbar\omega-\hbar\omega_{\mathbf{p},\chi}}\Big] \ ,\\
\end{split}
\end{equation}
where $\omega=(\xi_{\mathbf{k}'}-\xi_\mathbf{k})/\hbar$, and we have replaced $\omega_{-\mathbf{p},\chi}$ by $\omega_{\mathbf{p},\chi}$ since they equal to each other. Accordingly, in the second quantized language, the phonon-mediated electron-electron interaction in the Cooper channel is
\begin{equation}
H_{\text{int}}^{(\text{ph})}=\frac{1}{N_s\Omega_s}\sum_{\mathbf{k},\mathbf{k}'}\sum_{\eta,\eta'}\sum_{\zeta,\zeta'}\sum_{s,s'} V^{\eta\eta'}_{\mathbf{k}\mathbf{k}'}(\omega) c^{\dag}_{\mathbf{k}',\eta,\zeta,s}c^{\dag}_{-\mathbf{k}',\eta',\zeta',s'} c_{-\mathbf{k},\eta',\zeta',s'}c_{\mathbf{k},\eta,\zeta,s}\ .
\end{equation}

We now calculate $V^{\eta\eta'}_{\mathbf{k}\mathbf{k}'}$ using the projected electron-phonon Hamiltonian $H_{ep}^{\eta,\zeta,s}$ which we obtained in Eq. (\ref{Hepp}). By Eqs. (\ref{up}) and (\ref{Hepp}), we find the matrix elements for intermediate states (defined as $|\mathbf{k}_{I},\mathbf{k}_{I'}',\mathbf{p}_{,\chi}\rangle =c^\dag_{\mathbf{k},I}c^\dag_{\mathbf{k}',I'}a_{\mathbf{p},\chi}^\dag|0\rangle$) with phonon polarizations $\chi=L$ and $\chi=T$ are:
\begin{equation}\label{MA}
\begin{split}
&\langle\mathbf{k}'_{\eta,\zeta,s},-\mathbf{k}_{\eta',\zeta',s'},\mathbf{p}_{,T} |H_{\text{ep}}^{\eta,\zeta,s}|\mathbf{k}_{\eta,\zeta,s},-\mathbf{k}_{\eta',\zeta',s'}\rangle =\langle\mathbf{k}'_{\eta,\zeta,s},-\mathbf{k}'_{\eta',\zeta',s'} |H_{\text{ep}}^{\eta,\zeta,s} |\mathbf{k}_{\eta,\zeta,s},-\mathbf{k}'_{\eta',\zeta',s'},-\mathbf{p}_{,T}\rangle\\
=&\frac{\gamma\hbar v k_F}{\sqrt{N_s\Omega_s}}\sqrt{\frac{\hbar p\Omega}{2Mc_T}} \phi^{\eta\dag}_{\mathbf{k}'}\phi^{\eta}_{\mathbf{k}} \left[g_{2\alpha}-\eta\frac{g_{1\alpha}}{2}\text{Re}\left(\frac{e^{2i \varphi_\mathbf{k}} -e^{2i \varphi_{\mathbf{k}'}} }{e^{-i \varphi_\mathbf{k}}-e^{-i \varphi_{\mathbf{k}'}} }\right)\right]\ ,\\
&\\
&\langle\mathbf{k}'_{\eta,\zeta,s},-\mathbf{k}_{\eta',\zeta',s'},\mathbf{p}_{,L} |H_{\text{ep}}^{\eta,\zeta,s}|\mathbf{k}_{\eta,\zeta,s},-\mathbf{k}_{\eta',\zeta',s'}\rangle =\langle\mathbf{k}'_{\eta,\zeta,s},-\mathbf{k}'_{\eta',\zeta',s'} |H_{\text{ep}}^{\eta,\zeta,s} |\mathbf{k}_{\eta,\zeta,s},-\mathbf{k}'_{\eta',\zeta',s'},-\mathbf{p}_{,L}\rangle \\
=&\frac{\gamma\hbar v k_F}{\sqrt{N_s\Omega_s}}\sqrt{\frac{\hbar p\Omega}{2Mc_L}} \phi^{\eta\dag}_{\mathbf{k}'}\phi^{\eta}_{\mathbf{k}} \times\eta \frac{g_{1\alpha}}{2}\text{Im}\left(\frac{e^{2i \varphi_\mathbf{k}} -e^{2i \varphi_{\mathbf{k}'}} }{e^{-i \varphi_\mathbf{k}}-e^{-i \varphi_{\mathbf{k}'}} }\right)\ ,\\
&\\
&\langle\mathbf{k}_{\eta,\zeta,s},-\mathbf{k}'_{\eta',\zeta',s'},-\mathbf{p}_{,T} |H_{\text{ep}}^{\eta',\zeta',s'}|\mathbf{k}_{\eta,\zeta,s},-\mathbf{k}_{\eta',\zeta',s'}\rangle
=\langle\mathbf{k}'_{\eta,\zeta,s},-\mathbf{k}'_{\eta',\zeta',s'} |H_{\text{ep}}^{\eta',\zeta',s'} |\mathbf{k}'_{\eta,\zeta,s},-\mathbf{k}_{\eta',\zeta',s'},\mathbf{p}_{,T}\rangle\\
&=\frac{\gamma\hbar v k_F}{\sqrt{N_s\Omega_s}}\sqrt{\frac{\hbar p\Omega}{2Mc_T}} \phi^{\eta'\dag}_{-\mathbf{k}'}\phi^{\eta'}_{-\mathbf{k}} \left[g_{2\alpha}+\eta'\frac{g_{1\alpha}}{2}\text{Re}\left(\frac{e^{2i \varphi_\mathbf{k}} -e^{2i \varphi_{\mathbf{k}'}} }{e^{-i \varphi_\mathbf{k}}-e^{-i \varphi_{\mathbf{k}'}} }\right)\right]\ ,\\
&\\
&\langle\mathbf{k}_{\eta,\zeta,s},-\mathbf{k}'_{\eta',\zeta',s'},-\mathbf{p}_{,L} |H_{\text{ep}}^{\eta',\zeta',s'}|\mathbf{k}_{\eta,\zeta,s},-\mathbf{k}_{\eta',\zeta',s'}\rangle =\langle\mathbf{k}'_{\eta,\zeta,s},-\mathbf{k}'_{\eta',\zeta',s'} |H_{\text{ep}}^{\eta',\zeta',s'} |\mathbf{k}'_{\eta,\zeta,s},-\mathbf{k}_{\eta',\zeta',s'},\mathbf{p}_{,L}\rangle \\
=&-\frac{\gamma\hbar v k_F}{\sqrt{N_s\Omega_s}}\sqrt{\frac{\hbar p\Omega}{2Mc_L}} \phi^{\eta'\dag}_{-\mathbf{k}'}\phi^{\eta'}_{-\mathbf{k}} \times\eta' \frac{g_{1\alpha}}{2}\text{Im}\left(\frac{e^{2i \varphi_\mathbf{k}} -e^{2i \varphi_{\mathbf{k}'}} }{e^{-i \varphi_\mathbf{k}}-e^{-i \varphi_{\mathbf{k}'}} }\right)\ ,
\end{split}
\end{equation}
where we have used the fact that $\varphi_{-\mathbf{k}}=\varphi_{\mathbf{k}}+\pi$, and recall that $\Omega$ and $\Omega_s$ are the graphene unit cell area and superlattice unit cell area, respectively. Since $H_{\text{ep}}^{\eta,\zeta,s}$ is proportional to $u_{\mathbf{p},\chi}\propto(a_{\mathbf{p},\chi}+a^\dag_{-\mathbf{p},\chi})$ (see Eq. (\ref{up})), the amplitude of creating a phonon state $|\mathbf{p}_{,\chi}\rangle$ is always the same as that of annihilating a phonon state $|-\mathbf{p}_{,\chi}\rangle$ (when the other quantum numbers are the same), so we have the above equal relations between every two matrix amplitudes.
To further simplify the result, we approximate $c_L\approx c_T$, both of which are of the order of magnitude $10^4$m/s, so that $\omega_{\mathbf{p},L}\approx \omega_{\mathbf{p},T}=c_Tp$. Under this approximation, the prefactors of all the matrix elements in Eq. (\ref{MA}) become identical, i.e., $\sqrt{\frac{\hbar p\Omega}{2Mc_T}}=\sqrt{\frac{\hbar p\Omega}{2Mc_L}}$. By defining
\begin{equation}\label{wkk}
\varpi_{\mathbf{k}\mathbf{k}'}^{\eta\eta'}=\phi^{\eta\dag}_{\mathbf{k}'}\phi^{\eta}_{\mathbf{k}} \phi^{\eta'\dag}_{-\mathbf{k}'}\phi^{\eta'}_{-\mathbf{k}} =e^{i(\frac{\eta+\eta'}{2})(\varphi_\mathbf{k}-\varphi_{\mathbf{k}'})} \left[\frac{1+\cos(\varphi_\mathbf{k}-\varphi_{\mathbf{k}'})}{2}\right]\ ,
\end{equation}
one can then show the interaction in Eq. (\ref{VintS}) is
\begin{equation}\label{SeqVkk}
\begin{split}
&\ \frac{V^{\eta\eta'}_{\mathbf{k}\mathbf{k}'}(\omega)}{\Omega_s}=\frac{\gamma^2\hbar^2v^2 \Omega k_F^2 p \varpi_{\mathbf{k}\mathbf{k}'}^{\eta\eta'}}{2M c_T\Omega_s}\left( \frac{1}{\omega-\omega_{\mathbf{p},T}}+\frac{1}{-\omega-\omega_{\mathbf{p},T}}\right)\times\\
&\left\{ \left[g_{2\alpha}-\eta\frac{g_{1\alpha}}{2}\text{Re}\left(\frac{e^{2i \varphi_\mathbf{k}} -e^{2i \varphi_{\mathbf{k}'}} }{e^{-i \varphi_\mathbf{k}}-e^{-i \varphi_{\mathbf{k}'}} }\right)\right] \left[g_{2\alpha}+\eta'\frac{g_{1\alpha}}{2}\text{Re}\left(\frac{e^{2i \varphi_\mathbf{k}} -e^{2i \varphi_{\mathbf{k}'}} }{e^{-i \varphi_\mathbf{k}}-e^{-i \varphi_{\mathbf{k}'}} }\right)\right] -\eta\eta' \frac{g_{1\alpha}^2}{4}\left[\text{Im}\left(\frac{e^{2i \varphi_\mathbf{k}} -e^{2i \varphi_{\mathbf{k}'}} }{e^{-i \varphi_\mathbf{k}}-e^{-i \varphi_{\mathbf{k}'}} }\right)\right]^2 \right\}\\
&=\frac{\gamma^2\hbar^2v^2 \Omega k_F^2 \varpi_{\mathbf{k}\mathbf{k}'}^{\eta\eta'}}{2M c_T^2\Omega_s} \frac{\omega_{\mathbf{p},T}^2}{\omega^2-\omega_{\mathbf{p},T}^2} \left\{ g_{2\alpha}^2+\frac{g_{1\alpha}g_{2\alpha}}{2} (\eta'-\eta)\text{Re}\left(\frac{e^{2i \varphi_\mathbf{k}} -e^{2i \varphi_{\mathbf{k}'}} }{e^{-i \varphi_\mathbf{k}}-e^{-i \varphi_{\mathbf{k}'}} }\right) -\eta\eta' \frac{g_{1\alpha}^2}{4}\left|\frac{e^{2i \varphi_\mathbf{k}} -e^{2i \varphi_{\mathbf{k}'}} }{e^{-i \varphi_\mathbf{k}}-e^{-i \varphi_{\mathbf{k}'}} }\right|^2 \right\}\ .\\
\end{split}
\end{equation}
Therefore, the intra valley interaction with $\eta=\eta'$ is
\begin{equation}\label{SeqVintra}
\begin{split}
&\frac{V^{\eta\eta}_{\mathbf{k}\mathbf{k}'}}{\Omega_s}= \frac{\gamma^2\hbar^2v^2 \Omega k_F^2 \varpi_{\mathbf{k}\mathbf{k}'}^{\eta\eta}}{M c_T^2\Omega_s} \frac{\omega_{\mathbf{p},T}^2}{\omega^2-\omega_{\mathbf{p},T}^2} \left\{g_{2\alpha}^2-\frac{g_{1\alpha}^2}{2}[1+\cos(\varphi_\mathbf{k}- \varphi_{\mathbf{k}'})]\right\}\\
=&-\mathcal{N}_0(\omega)e^{i\eta(\varphi_\mathbf{k}-\varphi_{\mathbf{k}'})} \left[1+\cos(\varphi_\mathbf{k}-\varphi_{\mathbf{k}'})\right] \left\{g_{2\alpha}^2-\frac{g_{1\alpha}^2}{2}[1+\cos(\varphi_\mathbf{k}- \varphi_{\mathbf{k}'})]\right\}\ ,
\end{split}
\end{equation}
where we have used the fact that $\left|\frac{e^{2i \varphi_\mathbf{k}} -e^{2i \varphi_{\mathbf{k}'}} }{e^{-i \varphi_\mathbf{k}}-e^{-i \varphi_{\mathbf{k}'}} }\right|^2=\frac{2-2\cos2(\varphi_\mathbf{k}- \varphi_{\mathbf{k}'})}{2-2\cos (\varphi_\mathbf{k}- \varphi_{\mathbf{k}'})}=2[1+\cos(\varphi_\mathbf{k}- \varphi_{\mathbf{k}'})]$, and we have defined $\mathcal{N}_0(\omega)=\frac{\gamma^2\hbar^2v^2 \Omega k_F^2}{M c_T^2\Omega_s} \frac{\omega_{\mathbf{p},T}^2}{\omega_{\mathbf{p},T}^2-\omega^2}$, and used Eq. (\ref{wkk}) for the expression of $\varpi_{\mathbf{k}\mathbf{k}'}^{\eta\eta'}$. We note that for low frequency processes with $|\omega|<\omega_{\mathbf{p},T}$, we have the coefficient $\mathcal{N}_0(\omega)>0$. In a similar way, we find the inter valley interaction with $\eta=-\eta'=1$ is
\begin{equation}\label{SeqVinter}
\begin{split}
&\frac{V^{KK'}_{\mathbf{k}\mathbf{k}'}}{\Omega_s}=\frac{\gamma^2\hbar^2v^2 \Omega k_F^2 \varpi_{\mathbf{k}\mathbf{k}'}^{KK'}}{M c_T^2\Omega_s} \frac{\omega_{\mathbf{p},T}^2}{\omega^2-\omega_{\mathbf{p},T}^2} \left|g_{2\alpha}-\frac{g_{1\alpha}}{2}\frac{e^{2i \varphi_\mathbf{k}} -e^{2i \varphi_{\mathbf{k}'}} }{e^{-i \varphi_\mathbf{k}}-e^{-i \varphi_{\mathbf{k}'}} }\right|^2\\
=&-\mathcal{N}_0(\omega) \left[1+\cos(\varphi_\mathbf{k}-\varphi_{\mathbf{k}'})\right]
\left|g_{2\alpha}-\frac{g_{1\alpha}}{2}\frac{e^{2i \varphi_\mathbf{k}} -e^{2i \varphi_{\mathbf{k}'}} }{e^{-i \varphi_\mathbf{k}}-e^{-i \varphi_{\mathbf{k}'}} }\right|^2\ .
\end{split}
\end{equation}
For $\theta$ near magic angle $\alpha^2\approx1/3$, one has $g_{1\alpha}\approx 2/3$ and $g_{2\alpha}\approx1/3$, and the interaction takes the form shown in main text Eq. (8).

In particular, for low frequencies $|\omega|<\omega_{\mathbf{p},T}$, one finds that the intervalley interaction is attractive, while the intravalley interaction is repulsive. To see this, consider two electrons (near the Fermi surface) at valley $\eta$ and $\eta'$, whose 2-body wave function (with total momentum zero assumed) can generically be written as
\[
|\Psi_{\eta\eta'}\rangle=\sum_{l\in\mathbb{Z}}\sum_{|\mathbf{k}|=k_F}\beta_l e^{il\varphi_{\mathbf{k}}}c^\dag_{\mathbf{k},\eta,\zeta,s}c^\dag_{-\mathbf{k},\eta',\zeta',s'}|0\rangle\ ,
\]
where the physical meaning of $l\in\mathbb{Z}$ is the relative angular momentum between the two electrons. For two electrons in the same valley $\eta=\eta'=K$ (with wave function $|\Psi_{KK}\rangle$), the intravalley interaction energy is given by
\begin{equation}\label{Ekk}
\begin{split}
&E_{KK}=\langle\Psi_{KK}|H_{\text{int}}^{(\text{ph})}|\Psi_{KK}\rangle =\int_{0}^{2\pi}\frac{d\varphi_\mathbf{k}}{2\pi}\int_{0}^{2\pi}\frac{d\varphi_{\mathbf{k}'}}{2\pi} \sum_{l\in\mathbb{Z}}\sum_{l'\in\mathbb{Z}}\frac{V^{KK}_{\mathbf{k}\mathbf{k}'}}{\Omega_s} \beta_l^*\beta_le^{il\varphi_{\mathbf{k}}-il'\varphi_{\mathbf{k}'}}\\
&=-\mathcal{N}_0(\omega)e^{i(\varphi_\mathbf{k}-\varphi_{\mathbf{k}'})} \int_{0}^{2\pi}\frac{d\varphi_\mathbf{k}}{2\pi}\int_{0}^{2\pi}\frac{d\varphi_{\mathbf{k}'}}{2\pi} \left[1+\cos(\varphi_\mathbf{k}-\varphi_{\mathbf{k}'})\right] \left\{g_{2\alpha}^2-\frac{g_{1\alpha}^2}{2}[1+\cos(\varphi_\mathbf{k}- \varphi_{\mathbf{k}'})]\right\} \beta_l^*\beta_le^{il\varphi_{\mathbf{k}}-il'\varphi_{\mathbf{k}'}}\\
&=\mathcal{N}_0(\omega)\left[\left(\frac{3}{4}g_{1\alpha}^2-g_{2\alpha}^2\right)|\beta_{-1}|^2 + \frac{1}{2}\left(g_{1\alpha}^2-g_{2\alpha}^2\right)\left(|\beta_{-2}|^2+|\beta_{0}|^2\right) + \frac{g_{1\alpha}^2}{4}\left(|\beta_{-3}|^2+|\beta_{1}|^2\right)\right]\\
&\approx\mathcal{N}_0(\omega)\left[\frac{2}{9}|\beta_{-1}|^2 + \frac{1}{6}\left(|\beta_{-2}|^2+|\beta_{0}|^2\right) + \frac{1}{9}\left(|\beta_{-3}|^2+|\beta_{1}|^2\right)\right]\ ,
\end{split}
\end{equation}
where in the last line we have assumed $\theta$ is near the first magic angle so that $g_{1\alpha}\approx 2/3$ and $g_{2\alpha}\approx1/3$. For $|\omega|<\omega_{\mathbf{p},T}$, one has $\mathcal{N}_0(\omega)>0$, and one finds the intravalley interaction energy is in fact always positive, i.e., $E_{KK}>0$. Therefore, two electrons in the same valley always repulse each other (similarly one can show this for $\eta=\eta'=K'$).

In contrast, one notes that the intervalley interaction $V^{KK}_{\mathbf{k}\mathbf{k}'}$ in Eq. (\ref{SeqVinter}) is always real and negative for $|\omega|<\omega_{\mathbf{p},T}$, i.e., $\mathcal{N}_0(\omega)>0$. Therefore, if one considers two electrons in valley $\eta=K$ and $\eta'=K'$, respectively, one has the intervalley interaction energy
\begin{equation}
E_{KK'}=\langle\Psi_{KK'}|H_{\text{int}}^{(\text{ph})}|\Psi_{KK'}\rangle<0\ ,
\end{equation}
which is attractive (for $|\omega|<\omega_{\mathbf{p},T}$). Therefore, we conclude the intervalley Cooper pairing will be favored.


Physically, the fact that the intervalley interaction is more attractive than intravalley interaction can be understood as follows (see also the paragraph below main text Eq. (8)). Recall that the electron-phonon interaction projected to the vicinity of the Fermi surface contains two parts, $H_{\text{ep}}^{\eta,\zeta,s}(\overline{\mathbf{k}})=H_{C3}^{\eta,\zeta,s} (\overline{\mathbf{k}})+H_{SO(2)}^{\eta,\zeta,s}(\overline{\mathbf{k}})$, where $H_{C3}^{\eta,\zeta,s}$ and $H_{SO(2)}^{\eta,\zeta,s}$ are odd and even with respect to the electron momentum $\overline{\mathbf{k}}$ (average momentum before and after phonon emission/absorption), respectively. Now we consider an electron wave packet state with average momentum $\overline{\mathbf{k}}$ as $|\overline{\mathbf{k}}_{\eta,\zeta,s}\rangle=\sum_{\mathbf{k}}\beta(\mathbf{k}- \overline{\mathbf{k}})c^\dag_{\mathbf{k},\eta,\zeta,s}|0\rangle$, where $\beta(\mathbf{k}- \overline{\mathbf{k}})$ is a wave packet function peaked at $\overline{\mathbf{k}}$. Under a certain lattice deformation (i.e., given a nonzero configuration of the displacement field $\mathbf{u}(\mathbf{r})$), the electron-phonon interaction will generate a background lattice potential, which we assume is $U_{\text{ep}}=U_{C3}+U_{SO(2)}$ for an electron $|\overline{\mathbf{k}}_{K,\zeta,s}\rangle$ at valley $K$, where $U_{C3}= \langle \overline{\mathbf{k}}_{K,\zeta,s}|H_{C3}^{K,\zeta,s}|\overline{\mathbf{k}}_{K,\zeta,s}\rangle$, and $U_{SO(2)}= \langle \overline{\mathbf{k}}_{K,\zeta,s}|H_{SO(2)}^{K,\zeta,s}|\overline{\mathbf{k}}_{K,\zeta,s}\rangle$. Since $H_{C3}^{\eta,\zeta,s}$ and $H_{SO(2)}^{\eta,\zeta,s}$ are odd and even with respect to $\overline{\mathbf{k}}$, an electron state $|-\overline{\mathbf{k}}_{K,\zeta,s}\rangle$ in the same valley $K$ with opposite momentum would feel a background lattice potential $U_{\text{ep}}'=-U_{C3}+U_{SO(2)}$. On the other hand, an electron state $|-\overline{\mathbf{k}}_{K',\zeta,s}\rangle$ in the other valley $K'$ would feel the same potential $U_{\text{ep}}=U_{C3}+U_{SO(2)}$ as that of the electron $|\overline{\mathbf{k}}_{K,\zeta,s}\rangle$.

In general, if two electrons feel the same background potential due to deformed lattice, they will tend to get closer in the space, thus leading to a phonon-mediated electron-electron attraction. In the above, we have shown that two electrons $|\overline{\mathbf{k}}_{K,\zeta,s}\rangle$ and $|-\overline{\mathbf{k}}_{K',\zeta,s}\rangle$ in opposite valleys $K$ and $K'$ feel the same background potential $U_{C3}+U_{SO(2)}$, thus an effective attraction will be produced between them. Instead, two electrons $|\overline{\mathbf{k}}_{K,\zeta,s}\rangle$ and $|-\overline{\mathbf{k}}_{K,\zeta,s}\rangle$ in the same valley feel different background potentials $U_{C3}+U_{SO(2)}$ and $-U_{C3}+U_{SO(2)}$, thus the effective attraction between them will be weaker or even absent. Therefore, the intervalley interaction is more attractive than the intravalley interaction.

By substituting the realistic system parameters $v\approx 10^6$ m/s, $c_T\approx 10^4$ m/s, $w=110$ meV, $a_0=0.246$ nm, and $\theta\approx 1.05^\circ$ into the phonon-induced inter valley interaction (Eq. (\ref{SeqVinter})), and take $k_F\approx k_\theta$ as an order of magnitude estimation, we find the interaction contributed by the acoustic MBZ phonon bands near magic angle is
\begin{equation}
\frac{V^{KK'}_{\mathbf{k}\mathbf{k}'}(0)}{\Omega_s}\sim -\frac{4\gamma^2\hbar^2v^2 \Omega k_\theta^2}{9M c_T^2\Omega_s}\approx -\frac{4\hbar^2v^2 k_\theta^2}{9M c_T^2} \sim-1\text{meV}\ ,
\end{equation}
which is comparable to both the electron band width around the first magic angle (of order $1\sim10$meV) and the band width of the lowest acoustic MBZ phonon bands $\hbar\omega_D\sim \hbar c_Tk_\theta\approx2$meV. Therefore, the electron phonon coupling is relatively strong.

Furthermore, when the optical phonon bands are taken into account, the total phonon induced attractive interaction should be further enhanced. Since our electron phonon coupling $H_{\text{ep}}$ is derived in the long wavelength phonon limit, it does not apply for optical phonon bands. Nevertheless, here we give a very rough discussion of the optical phonon band contributions to the electron-electron interaction. A very crude approximation is to assume the electron phonon coupling matrix elements $\mathcal{M}_{\text{ep},\chi}$ of all optical phonon bands $\chi$ are roughly the same as that of the lowest acoustic bands. As we have discussed, the optical phonon bands are obtained by folding the phonon bands in the graphene BZ of the two layers into the MBZ. Equivalently, before folding, we can say that the lowest acoustic phonon bands are in the first MBZ, while the post-folding optical phonon bands are in the second, third and higher MBZ. Since the $n$-th MBZ is roughly $nk_\theta$ away from $\Gamma$ point of the first MBZ, the energy of the optical phonon band in the $n$-th MBZ is roughly $n\hbar c_{T} k_\theta$. Besides, the number of $n$-th MBZs is roughly $2\pi n$ (roughly speaking, all the $n$-th MBZs together form a circle at radius $nk_\theta$), so there are roughly $2\pi n$ transverse and longitudinal polarized optical phonon bands with energy $n\hbar c_{T} k_\theta$. The upper limit of $n$ is around $1/\theta$, where $nk_\theta$ reaches the boundary of the original graphene BZ. The modified total electron-electron interaction will then be enhanced to
\[
V^{\text{tot}}_{\mathbf{k}\mathbf{k}'}\sim\sum_\chi \frac{|\mathcal{M}_{\text{ep},\chi}|^2}{-\hbar\omega_{\mathbf{p},\chi}} \sim \sum_{n=1}^{1/\theta}\frac{2\pi n}{n}\frac{|\mathcal{M}_{\text{ep},T}|^2+|\mathcal{M}_{\text{ep},L}|^2}{-\hbar c_Tk_\theta}\approx\frac{2\pi}{\theta} V^{\text{ac}}_{\mathbf{k}\mathbf{k}'} \ ,
\]
where $\chi$ here denotes all the MBZ phonon bands (acoustic and optical), $\omega_{\mathbf{p},\chi}$ is the eigenfrequency of phonon band $\chi$, and $V^{\text{ac}}_{\mathbf{k}\mathbf{k}'}$ is the interaction contributed solely by the acoustic phonon bands we derived earlier in this section. Namely, the optical phonon bands contribute an additional factor $2\pi/\theta$ to the electron-electron interaction if all the optical phonon bands contribute, which is a large number for small angles $\theta$. This is clearly an overestimation, since high energy optical phonon bands are expected not to participate in the low energy superlattice physics. In practice, we expect this factor of interaction enhancement by optical phonons to be much smaller than $2\pi/\theta$ (but obviously greater than $1$).

\section{Screened Coulomb Potential and BCS superconductivity at magic angle}

The phonon-mediated electron-electron interaction can induce conventional BCS superconductivity. The density of states of the lowest bands at magic angle is around $N_D\gtrsim1$meV$^{-1}\cdot\Omega_s^{-1}$ for a $1$ meV band width, which gives a BCS coupling strength $\lambda\approx N_D |V_{\mathbf{k}\mathbf{k}'}(0)|\gtrsim 1$, which is relatively strong.

To determine the superconductor critical temperature, we also need to estimate the screened Coulomb potential between electrons. Here we shall simply adopt the Thomas-Fermi approximation in two dimensions (2D) for an order of magnitude estimation, and do not discuss the accuracy of the approximation. The Coulomb potential without screening is given by $V_e(r)=e^2/\epsilon_Ir$, whose Fourier transform is $\mathcal{V}_e(q)=2\pi e^2/\epsilon_Iq$, where $q=|\mathbf{q}|$ is the Fourier wave vector, and $\epsilon_I\approx 2\sim10$ is the dielectric constant of a charge neutral graphene. The 2D Poisson's equation for the bare Coulomb potential can then be written as
\begin{equation}
q\epsilon_I\mathcal{V}_e(q)=2\pi \rho(q),
\end{equation}
where $\rho(q)$ is the bare charge density. In the presence of free electrons, a Coulomb potential $V_e$ will induce a local charge density $\rho_f(r)\approx -e^2(\partial n_e/\partial\mu)V_e(r)$, or in Fourier space $\rho_f(q)\approx -e^2(\partial n_e/\partial\mu)\mathcal{V}_e(q)$, where $n_e$ is the electron number density, $\mu$ is the chemical potential, and thus $\partial n_e/\partial\mu$ is the density of states $N_D$. One can then write the charge density as $\rho=\rho_s+\rho_f$, where $\rho_s$ is the screened charge. From the Poisson's equation we have
\begin{equation}
2\pi \rho_s(q)=q\epsilon_I\mathcal{V}_e(q)-2\pi\rho_f(q)=[q\epsilon_I+2\pi e^2(\partial n_e/\partial\mu)]\mathcal{V}_e(q)=q\epsilon(q)\mathcal{V}_e(q)\ ,
\end{equation}
so we find the screened dielectric function of the form
\begin{equation}
\epsilon(q)\approx\epsilon_I\left(1+\frac{q_{\text{TF}}}{q}\right)\ ,
\end{equation}
where $q_{\text{TF}}=2\pi e^2 (\partial n_e/\partial\mu)/\epsilon_I=2\pi e^2 N_D/\epsilon_I$ is the Thomas-Fermi screening momentum. Near the magic angle, $N_D\gtrsim 1$ meV$^{-1}\cdot\Omega_s^{-1}$, which yields a $q_{\text{TF}}\gtrsim50 k_\theta\gg q$, so the screened Coulomb potential is approximately $\mathcal{V}_{e}(q)\approx 2\pi e^2/\epsilon_Iq_{TF}\sim N_D^{-1}$, and the Coulomb coupling strength $\mu_c\approx N_D\mathcal{V}_{e}(q)\sim 1$. We note that due to large density of states $N_D$, the screened Coulomb potential is much smaller than the bare Coulomb potential. In particular, the screened Coulomb potential is comparable to $V_{\mathbf{k}\mathbf{k}'}$. However, the phonon induced attraction $V_{\mathbf{k}\mathbf{k}'}(\omega)$ is frequency $\omega$ dependent and large as $\omega$ approaches $\omega_{T,\mathbf{p}}$, while the screened Coulomb potential can be approximated as frequency independent. This is because the 2D plasma frequency $\hbar\omega_{pe}=\sqrt{(4\pi n_e)^{1/2}e^2\epsilon_F/\epsilon_I}\gtrsim20$meV at the magic angle is much larger than the Debye frequency $\hbar\omega_D\approx \hbar c_Tk_\theta\approx2$meV, where $\epsilon_F$ is the Fermi energy (of order $1\sim10$meV around magic angle) \cite{das-sarma2009}. We then adopt the McMillan formula \cite{mcmillan1968,allen1975} to give a proper estimation of the BCS superconductivity critical temperature taking into account both the phonon induced attraction and the Coulomb repulsion:
\begin{equation}
k_BT_c=\frac{\hbar\omega_D}{1.45}\exp\left[-\frac{1.04(1+\lambda)}{\lambda-\mu^*_c(1+0.62\lambda)}\right]\ ,
\end{equation}
where $\mu_c^*=\mu_c/[1+\mu_c\ln(\omega_{pe}/\omega_D)]$ is the reduced Coulomb coupling strength, and $\omega_{pe}/\omega_D\gtrsim10$ around the magic angle. If we take the BCS coupling strength $\lambda\approx N_D|V_{\mathbf{k}\mathbf{k}'}|\approx1.5$, and $\mu_c\approx 1$, we find the superconductivity critical temperature from the McMillan formula is $T_c\approx0.9K$, close to the experimentally measured value.

Next, we discuss the pairing amplitude of the inter valley pairing. The pairing amplitude is defined as
\begin{equation}
\Delta_{\mathbf{k}}^{\eta\eta',\zeta\zeta',ss'}=\sum_{\mathbf{k}'} V_{\mathbf{k}\mathbf{k}'}^{\eta\eta'}\langle c_{-\mathbf{k}',\eta',\zeta',s'}c_{\mathbf{k}',\eta,\zeta,s}\rangle\ ,
\end{equation}
where $V_{\mathbf{k}\mathbf{k}'}^{\eta\eta'}$ is as given in Eq. (\ref{SeqVkk}). At zero temperature, the BCS self-consistency gap equation is given by
\begin{equation}
\Delta_{\mathbf{k}}^{\eta\eta',\zeta\zeta',ss'} =-\frac{1}{2}\sum_{|\xi_{\mathbf{k}'}|<\hbar \omega_D} \frac{V_{\mathbf{k}\mathbf{k}'}^{\eta\eta'}\Delta_{\mathbf{k}'}^{\eta\eta',\zeta\zeta',ss'}} {E_{\mathbf{k}'}^{BdG}}\ ,
\end{equation}
where $E_{\mathbf{k}'}^{BdG}$
is the Bogoliubov-de Gennes (BdG) band energy, while $\xi_{\mathbf{k}}=\hbar v_F(|\mathbf{k}|-k_F)$ is the band energy with the Fermi velocity $v_F=-\frac{1-3\alpha^2}{1+6\alpha^2}v$. Since we have shown the intervalley pairing is favored, we shall ignore the intravalley interaction (which is in fact repulsive as we have shown in Eq. (\ref{Ekk})), and only keep the intervalley interaction $V_{\mathbf{k}\mathbf{k}'}^{KK'}$. Furthermore, for simplicity we shall adopt the BCS approximation which assumes the electron-electron interaction is a constant function of $\omega$ for frequency $|\omega|<\omega_D$ and zero otherwise ($\omega_D$ is the Debye frequency), namely, we approximate the frequency dependence factor $\frac{\omega_{\mathbf{p},T}^2}{\omega^2-\omega_{\mathbf{p},T}^2}$ in Eq. (\ref{SeqVinter}) to $-1$ for $|\omega|<\omega_D$, and $0$ for $|\omega|\ge\omega_D$. Qualitatively, this does not affect the shape of the pairing function. We shall also assume $\theta$ is near the magic angle, so that $\alpha^2\approx 1/3$, and accordingly $g_{1\alpha}\approx 2/3$ and $g_{2\alpha}\approx1/3$. The intervalley interaction in Eq. (\ref{SeqVinter}) is then approximated as
\begin{equation}
\begin{split}
&V_{\mathbf{k}\mathbf{k}'}^{KK',\zeta\zeta',ss'}\approx -V_0 \varpi_{\mathbf{k}\mathbf{k}'}^{KK'} \left|1-\frac{e^{2i \varphi_\mathbf{k}} -e^{2i \varphi_{\mathbf{k}'}} }{e^{-i \varphi_\mathbf{k}}-e^{-i \varphi_{\mathbf{k}'}} }\right|^2 \\
=&-V_0  \left[\frac{1+\cos(\varphi_\mathbf{k}-\varphi_{\mathbf{k}'})}{2}\right] \left|1-\frac{e^{2i \varphi_\mathbf{k}} -e^{2i \varphi_{\mathbf{k}'}} }{e^{-i \varphi_\mathbf{k}}-e^{-i \varphi_{\mathbf{k}'}} }\right|^2 \ ,
\end{split}
\end{equation}
where $V_0\approx \frac{\gamma^2\hbar^2v^2 \Omega k_F^2 }{9M c_T^2\Omega_s}$, and we have used Eq. (\ref{wkk}).

With the 4-fold degeneracy from Moir\'e valley $\zeta$ and spin $s$, the pairing can be either Moir\'e valley triplet spin singlet, or Moir\'e valley singlet spin triplet. Here we shall simply assume the pairing is time-reversal invariant (which is more robust than time-reversal violating pairings in the presence of disorder \cite{anderson1959}). Since the time-reversal symmetry $\mathcal{T}$ brings valley $K\rightarrow K'$, Moir\'e valley $K_M\rightarrow K_M'$, and spin $s\rightarrow -s$ (see Eq. (\ref{TR})), a time reversal invariant pairing will be between opposite spins and opposite Moir\'e valleys. Such a time reversal invariant intervalley pairing then takes the form
\begin{equation}
\Delta_{\mathbf{k}}^{\eta\eta',\zeta\zeta',ss'}\sim s\delta_{s,-s'} \delta_{\zeta,-\zeta'}\delta_{\eta,-\eta'} \widetilde{\Delta}(\eta\mathbf{k})\ ,
\end{equation}
where $\widetilde{\Delta}(\mathbf{k})$ satisfies
\begin{equation}\label{BCSself}
\begin{split}
&\widetilde{\Delta}(\mathbf{k})=N_DV_0\int_{-\hbar\omega_D}^{\hbar\omega_D}d\xi \int_0^{2\pi}\frac{d\varphi_{\mathbf{k}'}}{2\pi} \left[\frac{1+\cos(\varphi_\mathbf{k}-\varphi_{\mathbf{k}'})}{2}\right] \left|1-\frac{e^{2i \varphi_\mathbf{k}} -e^{2i \varphi_{\mathbf{k}'}} }{e^{-i \varphi_\mathbf{k}}-e^{-i \varphi_{\mathbf{k}'}} }\right|^2 \frac{\widetilde{\Delta}({\mathbf{k}'})}{2\sqrt{|\widetilde{\Delta}({\mathbf{k}'})|^2+\xi^2}}\\
=& N_DV_0\int_0^{2\pi}\frac{d\varphi_{\mathbf{k}'}}{2\pi} \left[\frac{1+\cos(\varphi_\mathbf{k}-\varphi_{\mathbf{k}'})}{2}\right] \left|1-\frac{e^{2i \varphi_\mathbf{k}} -e^{2i \varphi_{\mathbf{k}'}} }{e^{-i \varphi_\mathbf{k}}-e^{-i \varphi_{\mathbf{k}'}} }\right|^2 \widetilde{\Delta}({\mathbf{k}'})\sinh^{-1}\frac{\hbar\omega_D}{|\widetilde{\Delta}({\mathbf{k}'})|}\ ,
\end{split}
\end{equation}
where in this case the BdG band energy is $E_{\mathbf{k}'}^{BdG}=\sqrt{|\widetilde{\Delta}({\mathbf{k}'})|^2+\xi_{\mathbf{k}'}^2}$.

Eq. (\ref{BCSself}) can then be numerically solved by iteration. As an example, for $N_DV_0=0.5$, we find the pairing amplitude $\widetilde{\Delta}({\mathbf{k}})$ is real and has the shape as shown in the main text Fig. [3c]. For generic values of $N_DV_0>0$, we find the pairing amplitude is always real and nodeless, thus is dominated by $s$-wave pairing and is topologically trivial. Finally, we note that in the absence of disorders, the pairing is not necessarily time-reversal invariant, and the pairing amplitude could be either Moir\'e valley singlet spin triplet or Moir\'e valley singlet spin triplet, which are degenerate.

\section{Numerical calculation for other angles and electron densities: prediction of other superconducting angles}

The above calculations of electron-phonon coupling can be generically applied to any twist angle $\theta$ and electron density.
We still adopt the continuum model with nearest hoppings \cite{bistritzer2011}, namely, an electron with momentum $\mathbf{k}$ in layer $1$ can hop with an electron with momentum $\mathbf{p}'$ in layer $2$ iff $\mathbf{k}-\mathbf{p}'=\mathbf{q}_j$ ($j=1,2,3$). However, instead of truncating at the smallest four momenta ($\mathbf{k}$ and $\mathbf{k}-\mathbf{q}_j$) in Eq. (\ref{SeqH}), we truncate the momentum of the Hamiltonian to sufficiently high momenta (momentum shells, i.e., second, third and higher MBZs), so that the band structure (which can be solved numerically) is more accurate. Besides, in the calculation we do not approximate $\theta/2$ in $h^K_{\pm\theta/2}(\mathbf{k})$ to zero, and this leads to a small particle-hole asymmetry to the band structure \cite{song2018}. Fig. \ref{figS3}a and \ref{figS3}b show two examples of Moir\'e BZ band structure calculated from the continuum model for $\theta=1.05^\circ$ and $\theta=1.20^\circ$, respectively. We can then calculate the density of states $N_D$ from the band structure. In the main text Fig. [4a] we have shown $N_D$ at $\theta=1.05^\circ$ as a function of number of electrons per superlattice unit cell $n=n_e\Omega_s$. Here in the upper panel of Fig. \ref{figS3}c, we show $N_D$ at $\theta=1.20^\circ$ as another example. Fig. \ref{figS3}d shows the $\text{Log}_{10}(N_D\cdot $meV$\cdot\Omega_s$) in a wide range of $\theta$ and $n$.

We then add small deformations to $\mathbf{q}_j$ induced by $\partial_au_b$ ($a,b=x,y$) (according to supplemental Eq. (\ref{Seqdeltaq})), and numerically calculate the band structure under different deformations. To first order, the change of band energy of a band $E_m$ generically takes the form
\begin{equation}\label{bj}
\delta E_m(\mathbf{k})=b_1(\mathbf{k})\left(\frac{\partial_xu_y+\partial_yu_x}{2}\right)+ b_2(\mathbf{k})\left(\frac{\partial_xu_x-\partial_yu_y}{2}\right) +b_3(\mathbf{k})\left(\frac{\partial_xu_y-\partial_yu_x}{2}\right) +b_4(\mathbf{k})\left(\frac{\partial_xu_x+\partial_yu_y}{2}\right)\ ,
\end{equation}
which is contributed by the band energy response to relative shear (first two terms), rotation and expansion between the two layers, respectively. This is nothing but the electron-phonon coupling of band $E_m$ in the long wavelength limit. The four coefficients $b_j$ ($1\le j\le4$) are real and can be numerically extracted out from band structure variation under small deformations.

\begin{figure}[htbp]
\begin{center}
\includegraphics[width=2.5in]{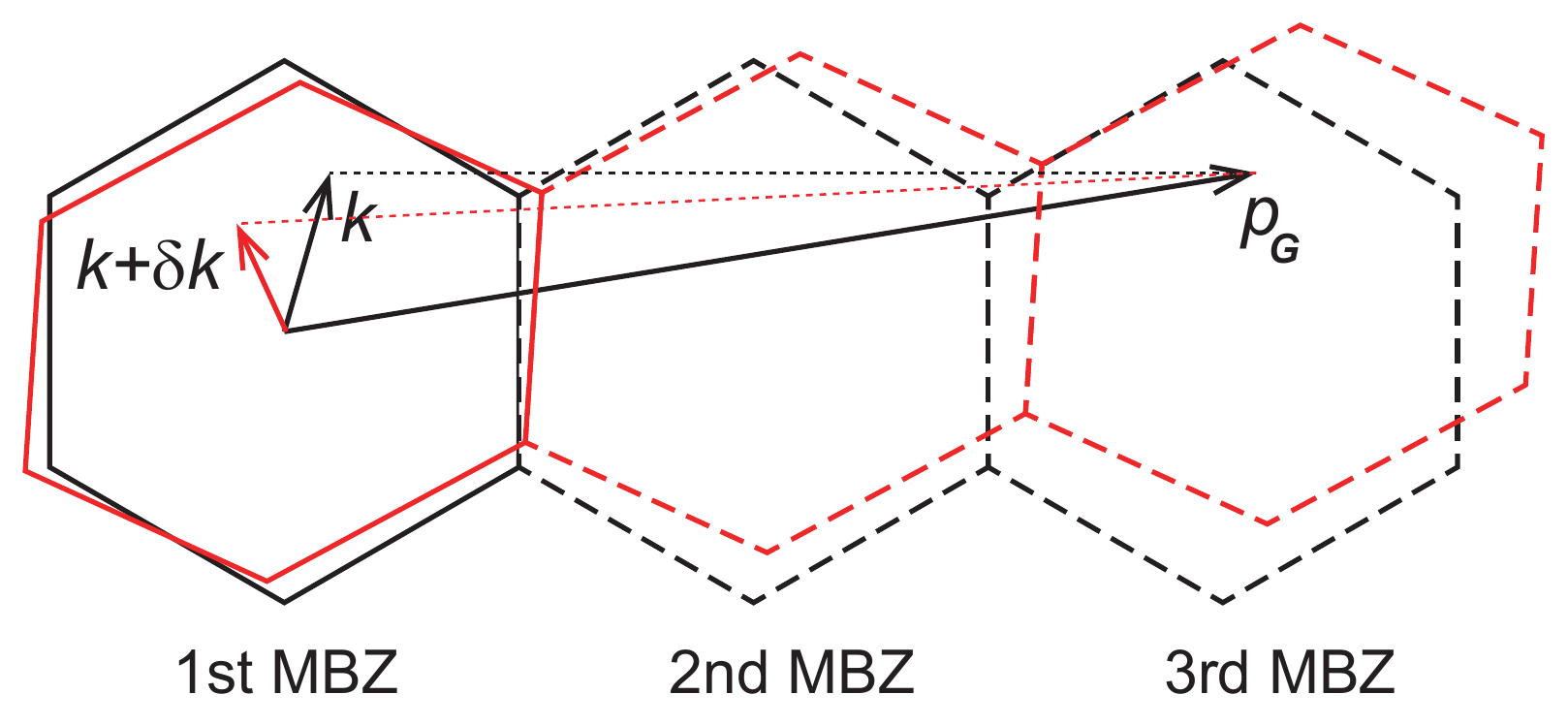}
\end{center}
\caption{A higher Moir\'e band state with Moir\'e momentum $\mathbf{k}$ in the 1st Moir\'e BZ (MBZ, black solid hexagon) originates from the folding of the state with graphene momentum $\mathbf{p}_G$ in the graphene BZ. Under the deformation $\mathbf{q}_j\rightarrow\mathbf{q}_j+\delta\mathbf{q}_j$, the MBZ is deformed (into the red hexagon); accordingly, the graphene momentum $\mathbf{p}_G$ will be folded to a different Moir\'e momentum $\mathbf{k}+\delta\mathbf{k}$. This implies a redefinition of $\mathbf{k}$ in higher Moir\'e bands.}
\label{figSfold}
\end{figure}

One effect we need to exclude in calculating $\delta E_m(\mathbf{k})$ is the following: when one makes a change $\mathbf{q}_j\rightarrow\mathbf{q}_j+\delta\mathbf{q}_j$, the Moir\'e BZ is deformed, so the folding of the original graphene BZ momentum into the Moir\'e BZ is also changed (Fig. \ref{figSfold}). Therefore, for higher Moir\'e bands (which originates from folding of the graphene BZ), the definition of momentum $\mathbf{k}$ in the Moir\'e BZ is changed under deformations. This additional change of electron momentum is simply a redefinition and is unphysical, thus should be eliminated. To be precise, consider the electron state in the $m$-th Moir\'e band (at graphene valley $K$) which takes a generic form
\begin{equation}
|\mathbf{k},m\rangle_M=\sum_{l_1,l_2\in\mathbb{Z}}\sum_{j=1,2} W_{\mathbf{k},l_1,l_2,j}^m|\mathbf{k}+l_1\bm{g}_1+ l_2\bm{g}_2 -(j-1)\mathbf{q}_1\rangle_j\ ,
\end{equation}
where $\mathbf{k}$ is the momentum in the Moir\'e BZ, $j$ is the layer index, $\bm{g}_1=\mathbf{q}_2-\mathbf{q}_3$ and $\bm{g}_2=\mathbf{q}_3-\mathbf{q}_1$ are the reciprocal vectors of the Moir\'e superlattice, and $|\mathbf{p}\rangle_j$ is the basis of Dirac electron in layer $j$ with an original graphene BZ momentum $\mathbf{p}$ measured from graphene valley $K$. Besides, $W_{\mathbf{k},l_1,l_2,j}^m$ denotes the coefficient of basis $|\mathbf{k}+l_1\bm{g}_1+ l_2\bm{g}_2 -(j-1)\mathbf{q}_1\rangle_j$. Therefore, the state $|\mathbf{k},m\rangle_M$ carries an average graphene momentum
\begin{equation}
\langle \mathbf{p}_G\rangle
=\langle\mathbf{k},m|\hat{\mathbf{p}}_G|\mathbf{k},m\rangle_M
=\sum_{l_1,l_2\in\mathbb{Z}}\sum_{j=1,2} |W_{\mathbf{k},l_1,l_2,j}^m|^2\Big(\mathbf{k}+l_1\bm{g}_1+ l_2\bm{g}_2 -(j-1)\mathbf{q}_1\Big)\ .
\end{equation}
Upon the uniform deformation $\mathbf{q}_j\rightarrow\mathbf{q}_j+\delta\mathbf{q}_j$, one should keep the electron's graphene momentum $\langle \mathbf{p}_G\rangle$ invariant and then examine the energy variation, because the absorption or emission of a long wave length phonon should not change the electron momentum. Namely, one need to shift the Moir\'e BZ momentum $\mathbf{k}$ to $\mathbf{k}+\delta\mathbf{k}$ so that
\begin{equation}\label{zerogk}
0=\delta \langle \mathbf{p}_G\rangle\approx\delta\mathbf{k}+\sum_{l_1,l_2\in\mathbb{Z}}\sum_{j=1,2} |W_{\mathbf{k},l_1,l_2,j}^m|^2\Big(l_1\delta\bm{g}_1+ l_2\delta\bm{g}_2 -(j-1)\delta\mathbf{q}_1\Big)\ ,
\end{equation}
where $\delta\bm{g}_1=\delta\mathbf{q}_2-\delta\mathbf{q}_3$ and $\delta\bm{g}_2=\delta\mathbf{q}_3-\delta\mathbf{q}_1$ (see Fig. \ref{figSfold}). Since the wave function $W_{\mathbf{k},l_1,l_2,j}^m$ can be calculated numerically, we are able to compute $\delta\mathbf{k}$. Accordingly, one should evaluate the energy variation in Eq. (\ref{bj}) as (Fig. \ref{figSfold})
\begin{equation}
\delta E_m(\mathbf{k})=E_m'(\mathbf{k}+\delta\mathbf{k})-E_m(\mathbf{k})\ ,
\end{equation}
where $E_m(\mathbf{k})$ and $E_m'(\mathbf{k})$ are the energy dispersion of the $m$-th band before and after deformation, respectively. We note that in our analytical calculations of electron-phonon coupling near the Dirac points of the lowest two Moir\'e flat bands (Eq. (\ref{Hep2})), one has $\delta\mathbf{k}\approx -\frac{2\alpha^2}{1+6\alpha^2}(\delta\mathbf{q}_1+\delta\mathbf{q}_2+\delta\mathbf{q}_3)=0$, so we need not consider this momentum shift $\delta\mathbf{k}$ problem there. For numerical calculations of higher bands, however, $\delta\mathbf{k}$ is in general nonzero.

\begin{figure}[htbp]
\begin{center}
\includegraphics[width=7in]{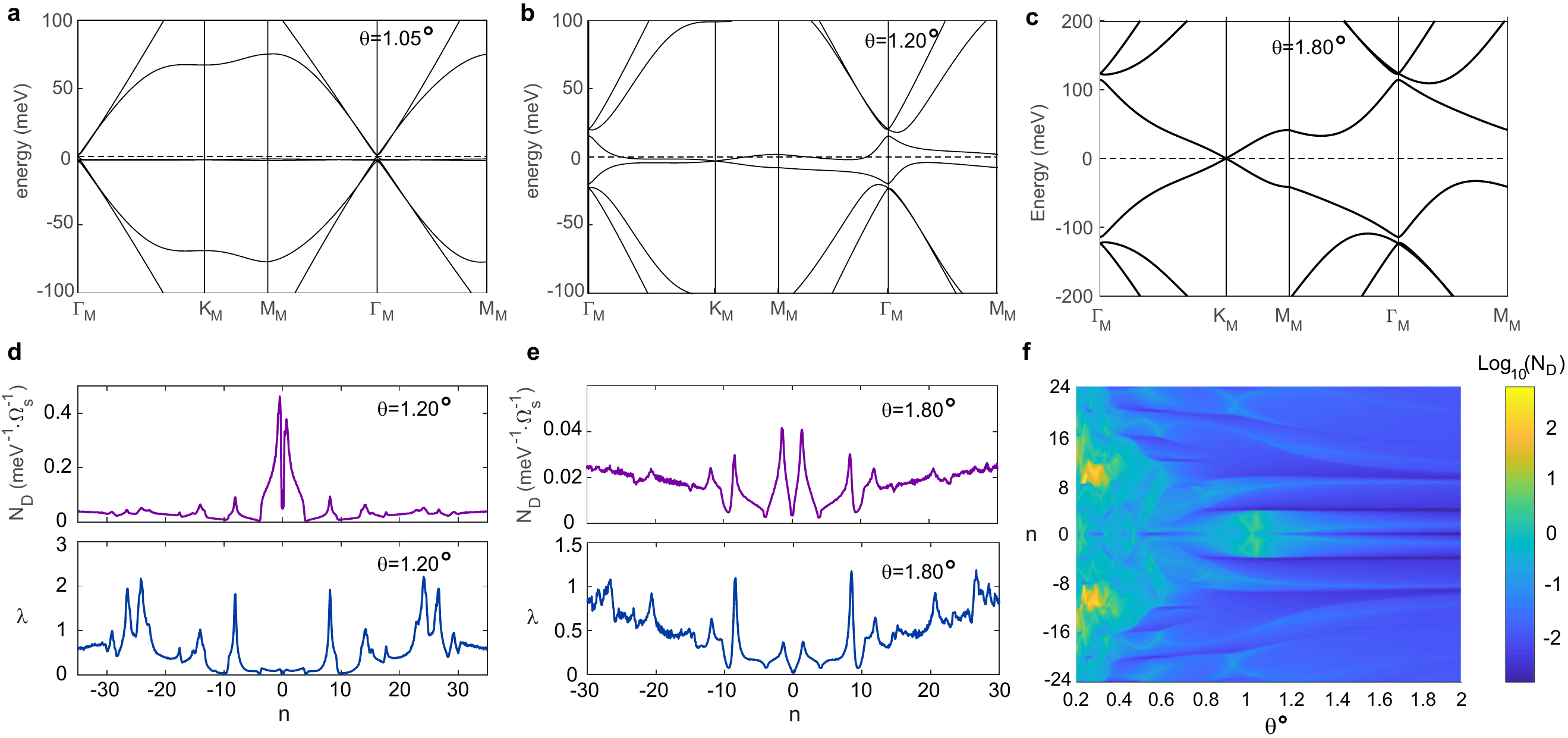}
\end{center}
\caption{{\bf a}. TBG band structure at $\theta=1.05^\circ$ from the continuum model. 
{\bf b}. TBG band structure at $\theta=1.20^\circ$ from the continuum model. {\bf c}. TBG band structure at $\theta=1.80^\circ$ from the continuum model. {\bf d}. Density of states $N_D$ and BCS coupling strength $\lambda$ with respect to $n$ calculated for $\theta=1.20^\circ$. In particular, the peaks in $N_D$ around $|n|=8$ indicates a Van Hove singularity in the second or third conduction (valence) bands (the second band and third band have a significant overlap in energy, as shown in panels {\bf a} and {\bf b} here, so $|n|=8$ is not an in-gap filling factor). {\bf e}. Density of states $N_D$ and BCS coupling strength $\lambda$ with respect to $n$ calculated for $\theta=1.80^\circ$. The shape of the $N_D$ function agrees well with the STM measurement of large angle TBG \cite{wong2015}. {\bf f}. $\text{Log}_{10}(N_D\cdot $meV$\cdot\Omega_s$) plotted as a function of $\theta$ and $n$, where one can clearly see $N_D$ is large at the magic angle, and is generically larger for smaller angles.}
\label{figS3}
\end{figure}


From Eq. (\ref{up}) we have the displacement field after quantization $\mathbf{u}(\mathbf{r})=\sum_{\mathbf{p},\chi} \frac{e^{i\mathbf{p}\cdot\mathbf{r}}\bm{\epsilon}_\chi u_{\mathbf{p},\chi}}{\sqrt{N_s\Omega_s}}$, where $\mathbf{p}$ is the phonon momentum, $u_{\mathbf{p},\chi}=\sqrt{\frac{\hbar\Omega}{2M\omega_{\mathbf{p},\chi}}} (a_{\mathbf{p},\chi}+a_{-\mathbf{p},\chi}^\dag)=\sqrt{\frac{\hbar\Omega}{2Mc_\chi p}} (a_{\mathbf{p},\chi}+a_{-\mathbf{p},\chi}^\dag)$ for polarization $\chi$, and $\bm{\epsilon}_\chi$ is the polarization vector. The electron-phonon coupling in band $E_m$ is then simply the expression of $\delta E_m(\overline{\mathbf{k}})$ in Eq. (\ref{bj}) with the displacement field $\mathbf{u}(\mathbf{r})$ expressed in terms of phonon operators (as above), which in the momentum space takes the form
\begin{equation}\label{Hepb}
H_\text{ep}(\overline{\mathbf{k}})\approx\sum_{j=1}^{4}\sum_{\chi=L,T}\mathcal{N}_{j,\chi}(\hat{\mathbf{p}}) b_j(\overline{\mathbf{k}}) \sqrt{\frac{\hbar\Omega p}{2N_s\Omega_sMc_\chi}}(a_{\mathbf{p},\chi}+a_{-\mathbf{p},\chi}^\dag)\ ,
\end{equation}
where $\overline{\mathbf{k}}$ should be understood as the average of initial momentum $\mathbf{k}$ and final momentum $\mathbf{k}'$ of the electron, $\Omega$ and $\Omega_s$ are the graphene unit cell area and the superlattice unit cell area, and $\mathcal{N}_{j,\chi}(\hat{\mathbf{p}})$ are dimensionless coefficients depending on the unit vector $\hat{\mathbf{p}}=\mathbf{p}/p$ (direction of phonon momentum) and are of order $1$. These coefficients $\mathcal{N}_{j,\chi}(\hat{\mathbf{p}})$ can be obtained by substituting the expression of $\mathbf{u}(\mathbf{r})$ in Eq. (\ref{up}) into Eq. (\ref{bj}), and then rewrite the results into the form of Eq. (\ref{Hepb}).

The electron-electron interaction $V_{\mathbf{k}\mathbf{k}'}(\omega)$ can then be estimated from Eq. (\ref{SeqVII}). For zero frequency $\omega=0$, the interaction  $V_{\mathbf{k}\mathbf{k}'}$ can be crudely estimated as
\begin{equation}\label{Vkkhigher}
\frac{V_{\mathbf{k}\mathbf{k}'}}{\Omega_s}= \sum_{\chi=L,T}\frac{2N_s}{-\hbar\omega_{\mathbf{p},\chi}} \frac{\hbar\Omega p}{2N_s\Omega_sMc_\chi}\sum_{j,j'} \mathcal{N}_{j,\chi}(\hat{\mathbf{p}})\mathcal{N}_{j',\chi}(\hat{\mathbf{p}}) b_j(\overline{\mathbf{k}}) b_{j'}(\overline{\mathbf{k}}) \approx-\frac{\Omega}{M c_T^2 \Omega_s}\left(\sum_{j=1}^4 b_j(\bar{\mathbf{k}})^2\right)\ ,
\end{equation}
where for order of magnitude estimation we have simply approximated $\mathcal{N}_{j',\chi}\sim 1$ and ignored the cross terms between $b_j(\overline{\mathbf{k}})$ and $b_{j'}(\overline{\mathbf{k}})$. We also approximated $c_L=c_T$.
Given a Fermi energy $\epsilon_F$, the BCS coupling strength $\lambda\approx N_D|V_{\mathbf{k}\mathbf{k}'}|$ can be numerically estimated as
\begin{equation}
\lambda\sim \frac{1}{2\hbar\omega_D}\int_{|\xi_\mathbf{k}- \epsilon_F|\le\hbar\omega_D}\frac{d^2\mathbf{k}}{(2\pi)^2} [-V(\mathbf{k})]\ ,
\end{equation}
where $\omega_D$ is the Debye frequency, and we have written $V_{\mathbf{k}\mathbf{k}'}=V(\overline{\mathbf{k}})$ for short. Note that $\lambda$ is well-defined in the limit $\omega_D\rightarrow 0$.

The main text Fig. 4b shows the BCS coupling strength $\lambda$ we estimated for $\theta=1.05^\circ$, while the lower panels of Fig. \ref{figS3}d and Fig. \ref{figS3}e here shows $\lambda$ for $\theta=1.20^\circ$ and $\theta=1.80^\circ$, respectively. We find the BCS coupling strength $\lambda$ in higher bands $|n|>4$ can also be generically as large as order $1$, although the density of states therein is much lower ($\sim 0.05$meV$^{-1}\cdot\Omega_s^{-1}$).
Here we give a heuristic understanding and estimation. First, for angles $\theta$ near $1^\circ$, numerical calculation shows the second and higher MBZ bands have band widths $\sim100$ meV, and this is determined by the two characteristic TBG energy scales which are of the same order: the interlayer hopping $w=110$ meV and the graphene kinetic energy $\hbar vk_\theta \sim 200$ meV. Roughly speaking, the Moir\'e bands are obtained by folding the original graphene band structure into the MBZ, while the interlayer hopping $w$ couples different Moir\'e bands. If we treat $w$ as a perturbation, the energy of the $m$-th band $E_m(\mathbf{k})$ in the 2nd perturbation theory is roughly
\begin{equation}\label{roughEm}
E_m(\mathbf{k})\sim \hbar vk_G^{(m)}+z\frac{w^2}{E_W},
\end{equation}
where $k_G^{(m)}$ is the original graphene momentum, $z$ is a numerical factor, and $E_W\sim 100$ meV is the energy scale of the band width as well as the energy separation between neighbouring bands. Since each Moir\'e band is coupled to three other bands via $\mathbf{q}_j$ ($j=1,2,3$), we can estimate the numerical factor $z$ as $|z|\sim3$.

Under a lattice deformation, the three vectors $\mathbf{q}_j$ changes to $\mathbf{q}_j+\delta \mathbf{q}_j$ ($j=1,2,3$), where $\delta \mathbf{q}_j\sim \gamma k_\theta\partial_a u_b$ ($a,b=x,y$) as given in Eq. (\ref{Seqdeltaq}). This changes the energy separation between neighbouring bands $E_W$ by an amount $\sim \hbar v\delta\mathbf{q}_j$. Keeping the graphene momentum $k_G^{(m)}$ invariant in Eq. (\ref{roughEm}) (which is the requirement of Eq. (\ref{zerogk})), we have the energy variation of band $E_m(\mathbf{k})$ to be roughly
\begin{equation}
\delta E_m(\mathbf{k})\sim z\frac{w^2}{E_W^2}\hbar v\delta\mathbf{q}_j\sim z\frac{w^2}{E_W^2}\hbar v \gamma k_\theta\partial_a u_b\ .
\end{equation}
This gives an estimation of the coefficients in Eq. (\ref{bj}) as $b_j(\mathbf{k})\sim z\frac{w^2}{E_W^2}\hbar v \gamma k_\theta$. By Eq. (\ref{Vkkhigher}), we then find the phonon mediated interaction is of order
\begin{equation}
\frac{V_{\mathbf{k}\mathbf{k}'}}{\Omega_s}\sim -\frac{\Omega z^2\gamma^2 (\hbar v k_\theta)^2}{M c_T^2 \Omega_s} \frac{w^4}{E_W^4}\sim -\frac{z^2}{M c_T^2} \frac{w^4}{E_W^2}\ ,
\end{equation}
where we have used $\hbar vk_\theta\sim E_W$, and $\gamma^2\approx \Omega_s/\Omega$ (recall that $\gamma\sim 1/\theta$, while $\Omega$ and $\Omega_s$ are the graphene unit cell area and the Moir\'e supercell area, respectively).
On the other hand, the density of states $N_D$ also has a generical grow trend as the Fermi energy increases (for electron densities $|n|>4$), as can be seen in the upper panels of Fig. \ref{figS3}d and \ref{figS3}e. This is because as the energy increases, there are more and more Moir\'e bands falling into the same energy interval. This can also be understood in the $w\rightarrow0$ limit, in which case the TBG becomes two decoupled monolayer graphene, and the density of state $N_D$ will be that of the graphene Dirac electrons, which grows as the energy increases. From the numerical results of the upper panel of Fig. \ref{figS3}d, we simply estimate $N_D$ as $N_D\sim 5 E_W^{-1}\cdot\Omega_s^{-1}\sim 0.05$ meV$^{-1}\cdot\Omega_s^{-1}$ for second or higher bands near the magic angle. Therefore, if we take $|z|\approx3$ (for the three nearest neighbour couplings of momenta $\mathbf{q}_j$), we find the BCS coupling strength in higher bands to be around
\begin{equation}
\lambda\sim -N_D V_{\mathbf{k}\mathbf{k}'}\sim \frac{5z^2}{M c_T^2} \frac{w^4}{E_W^3}\sim 0.5\ .
\end{equation}
The density of states $N_D$ could be higher near von Hove singularities, where the BCS coupling strength could be higher. This estimation indicates that the electron-phonon coupling is generically enhanced by the Moir\'e pattern (by the amplification factor $\gamma$), and does not necessarily require the presence of flat bands.

Based on $\lambda$ estimated in the above, we use the McMillan formula to estimate $T_c$ in a wide range of $\theta$ and $n$, as is shown in the main text Fig. 4c. Despite the fact that $T_c$ is estimated in a very rough way, the parameter space where we predict superconductivity may occur at large $|n|$ is stable.

\section{{\em ab initio } calculations}

To verify the continuum model gives the correct order of magnitude of electron-phonon coupling, we also run \emph{ab initio} calculations of the variation of band structure of TBG under deformations where the lattices of the two layers remain commensurate (which we shall explain below), and compare it with the spectrum calculated from the continuum model. We performed \emph{ab initio} calculations based on the density functional theory with the projector augmented wave (PAW) method~\cite{paw1,paw2} as implemented in VASP package~\cite{vasp1,vasp2}. The local density approximation (LDA) was adopted for the exchange-correlation functional~\cite{lda}. The kinetic energy cutoff of the plane wave basis was set to 300 eV. Only the convergence of band energy at $\Gamma$ point in the MBZ is used as the criteria for the convergence of self-consistent calculations.

\begin{figure}[htbp]
\begin{center}
\includegraphics[width=2.5in]{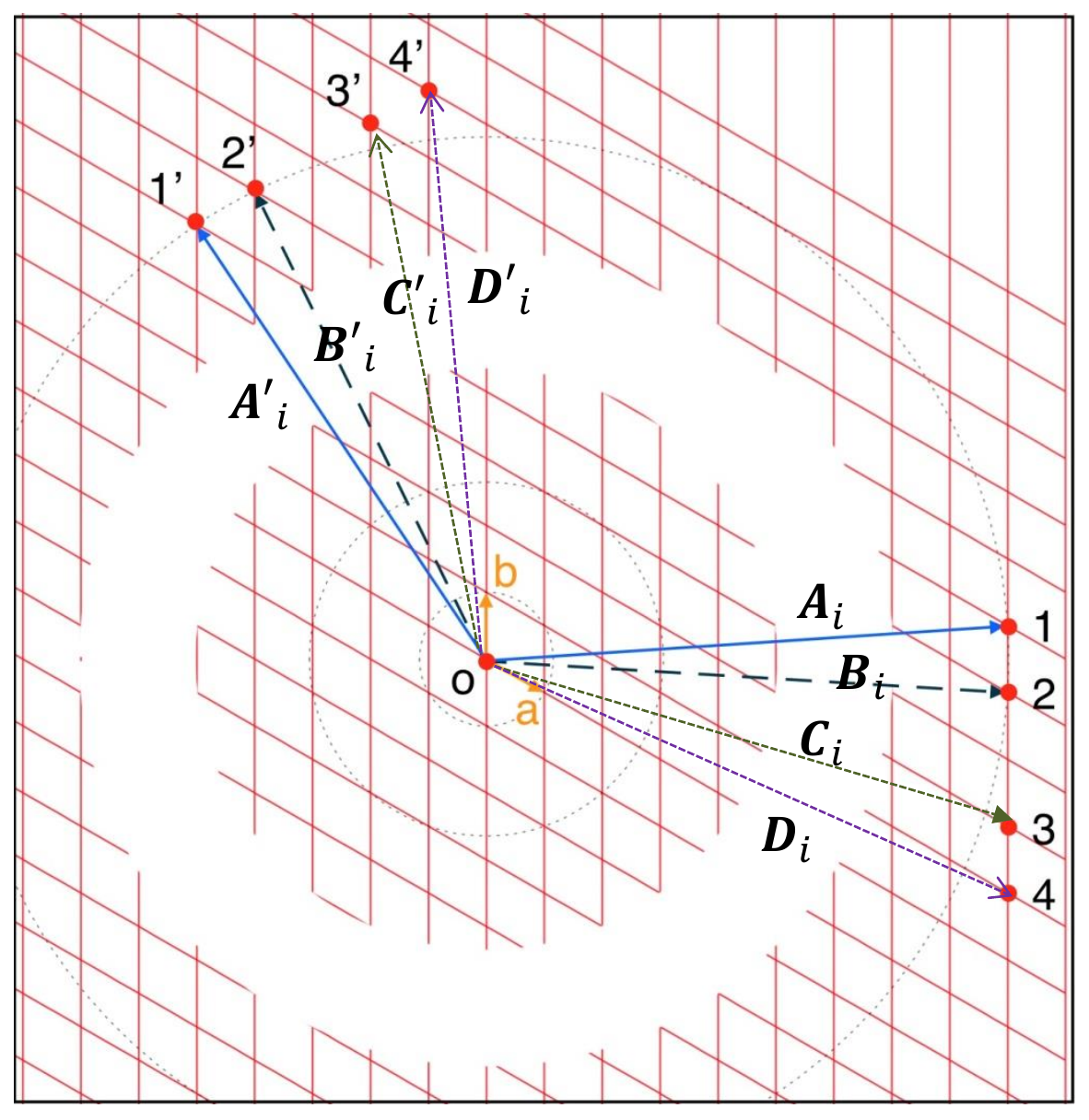}
\end{center}
\caption{Illustration of the commensurate Moir\'e pattern constructed from supercells, and the construction of slightly deformed Moir\'e pattern. The numbers $1,2,3,4$ and $1',2',3',4'$ label several particular sites (red solid points) useful in defining the commensurate configuration.}
\label{figS4}
\end{figure}

Fig.~\ref{figS4} illustrates a graphene lattice, where each site here denotes the center of a hexagon plaquette of graphene. The primitive lattice vectors are denoted as $\bf a$ and $\bf b$, and the lattice constant is set to be $a_0=2.456$ \AA. We define the lattice vector $\mathbf{R}_{m,n}=m{\bf a}+n{\bf b}$. Now assume this graphene lattice is the lattice of layer $1$ of TBG. By stacking the graphene lattice of layer $2$ on top of it (which is not shown in Fig.~\ref{figS4}), we arrive at a TBG. To construct a commensurate TBG (undistorted), we first assume the lattices of layer $1$ and layer $2$ differ by a rotation of angle $\theta$ about point $o$ in Fig.~\ref{figS4} (which we define as the origin), namely, the hexagon plaquette centers of layer $1$ and layer $2$ coincide at point $o$. Now consider two lattice vectors $\mathbf{A}_i=\mathbf{R}_{2i+1,i+1}$ and $\mathbf{B}_i=\mathbf{R}_{2i+1,i}$ away from the origin $o$ with $i\ge0$ being an integer, which correspond to points $1$ and $2$ as illustrated in Fig.~\ref{figS4}. The two vectors have equal lengths
\[
|\mathbf{A}_i|=|\mathbf{B}_i|=L_i^A=\frac{\sqrt{3(2i+1)^2+1}}{2}a_0\ .
\]
Similarly, the vectors $\mathbf{A}_i'=\mathbf{R}_{-i-1,i}$ and $\mathbf{B}_i'=\mathbf{R}_{-i,i+1}$ from the origin $o$ correspond to points $1'$ and $2'$ in Fig.~\ref{figS4}, which also have equal lengths and are related to $\mathbf{A}_i$ and $\mathbf{B}_i$ by a $2\pi/3$ rotation. We then assume the layer $2$ lattice is given by rotating the layer $1$ lattice so that $\mathbf{B}_i$ ($\mathbf{B}_i'$) is rotated to $\mathbf{A}_i$ ($\mathbf{A}_i'$). Such a TBG then forms a commensurate Moir\'e pattern superlattice which is periodic in real space. As shown in Ref.~\cite{TBGstr}, the twist angle $\theta=\theta_i$ of such a commensurate configuration satisfies $\cos\theta_i=\frac{3i^2+3i+0.5}{3i^2+3i+1}$, or equivalently,
\[
\theta_i=2\arctan\frac{1}{\sqrt{3}(2i+1)}\ .
\]
The spacial periods, namely, the superlattice vectors, are then given by $\mathbf{A}_i$ and $\mathbf{A}_i'$, and the superlattice unit cell is the parallelogram with edges $\mathbf{A}_i$ and $\mathbf{A}_i'$. We call such a Moir\'e pattern the $(11'22')$ configuration, of which the definition involves the four points $1$, $1'$ and $2$, $2'$.

As an example, we consider the commensurate undistorted TBG configuration $(11'22')$ with $i=10$, and calculate the band structure with \emph{ab initio}. The distance between the two layers is $d_0$ ($\sim 3.35$ \AA). The electronic band structure from \emph{ab initio} for $i=10$ is shown as red solid lines in both Fig.~\ref{figS5}c and \ref{figS5}d (identical between these two figures), where the twist angle $\theta=\theta_{10}=3.15^\circ$, and the high symmetry points in the figure should be understood as those of the MBZ.

To generate an example of deformed Moir\'e pattern, we consider two different lattice vectors $\mathbf{C}_i=\mathbf{R}_{2i+1,i-2}$ and $\mathbf{D}_i=\mathbf{R}_{2i+1,i-3}$ away from the origin $o$, which correspond to points $3$ and $4$ in Fig. \ref{figS4}, respectively (The reason we choose points $3$ and $4$ will be explained later). We also define another two vectors $\mathbf{C}_i'=\mathbf{R}_{-i+2,i+3}$ and $\mathbf{D}_i'=\mathbf{R}_{-i+3,i+4}$, which are $\mathbf{C}_i$ and $\mathbf{D}_i$ rotated by $2\pi/3$, and correspond to points $3'$ and $4'$, respectively. We shall still assume the lattice plotted in Fig.~\ref{figS4} is the lattice of layer $1$. Next, we assume the lattice of layer $2$ is \emph{ rotated and deformed} relative to the lattice of layer $1$, so that vector $\mathbf{D}_i$ and $\mathbf{D}_i'$ of layer $2$ coincide with $\mathbf{C}_i$ and $\mathbf{C}_i'$ of layer $1$, respectively. This is again a commensurate configuration, and the superlattice unit cell are the parallelogram with edges $\mathbf{C}_i$ and $\mathbf{C}_i'$ in layer $1$.
(which coincide with the parallelogram with edges $\mathbf{D}_i$ and $\mathbf{D}_i'$ in layer $2$, after layer $2$ is rotated and deformed).
We call this configuration $(33'44')$. In particular, the lengths of vectors are
\[
|\mathbf{C}_i|=|\mathbf{C}_i'|=L_i^C=\frac{\sqrt{3(2i+1)^2+5^2}}{2}a_0\ ,\qquad |\mathbf{D}_i|=|\mathbf{D}_i'|=L_i^D=\frac{\sqrt{3(2i+1)^2+7^2}}{2}a_0\ ,
\]
which are not equal, and the twist angle is now the angle between $\mathbf{C}_i$ and $\mathbf{D}_i$, which is
\[
\theta_i'=\arctan\frac{7}{\sqrt{3}(2i+1)}-\arctan\frac{5}{\sqrt{3}(2i+1)}\ .
\]
Therefore, this configuration ($33'44'$) involves both a relative rotation $\delta\theta=\theta_i'-\theta_i$ and a relative expansion $\Theta\approx2(L_i^D-L_i^C)/L_i^C$ compared to the undistorted configuration ($11'22'$). The relative shear $\Sigma_{ab}$ is, however, zero, since the deformation in this case is isotropic. We can then write down the relative deformation field $\mathbf{u}$ for ($33'44'$) (compared to the undistorted TBG configuration) as
\begin{equation}\label{u3344}
\left(\begin{array}{cc}\partial_xu_x&\partial_xu_y\\ \partial_yu_x&\partial_yu_y\end{array}\right)\approx \left(\begin{array}{cc}\frac{L_i^D-L_i^C}{L_i^C}&\theta_i-\theta_i'\\ \theta_i'-\theta_i&\frac{L_i^D-L_i^C}{L_i^C}\end{array}\right)\ .
\end{equation}

\begin{figure}[tbp]
\begin{center}
\includegraphics[width=6.5in]{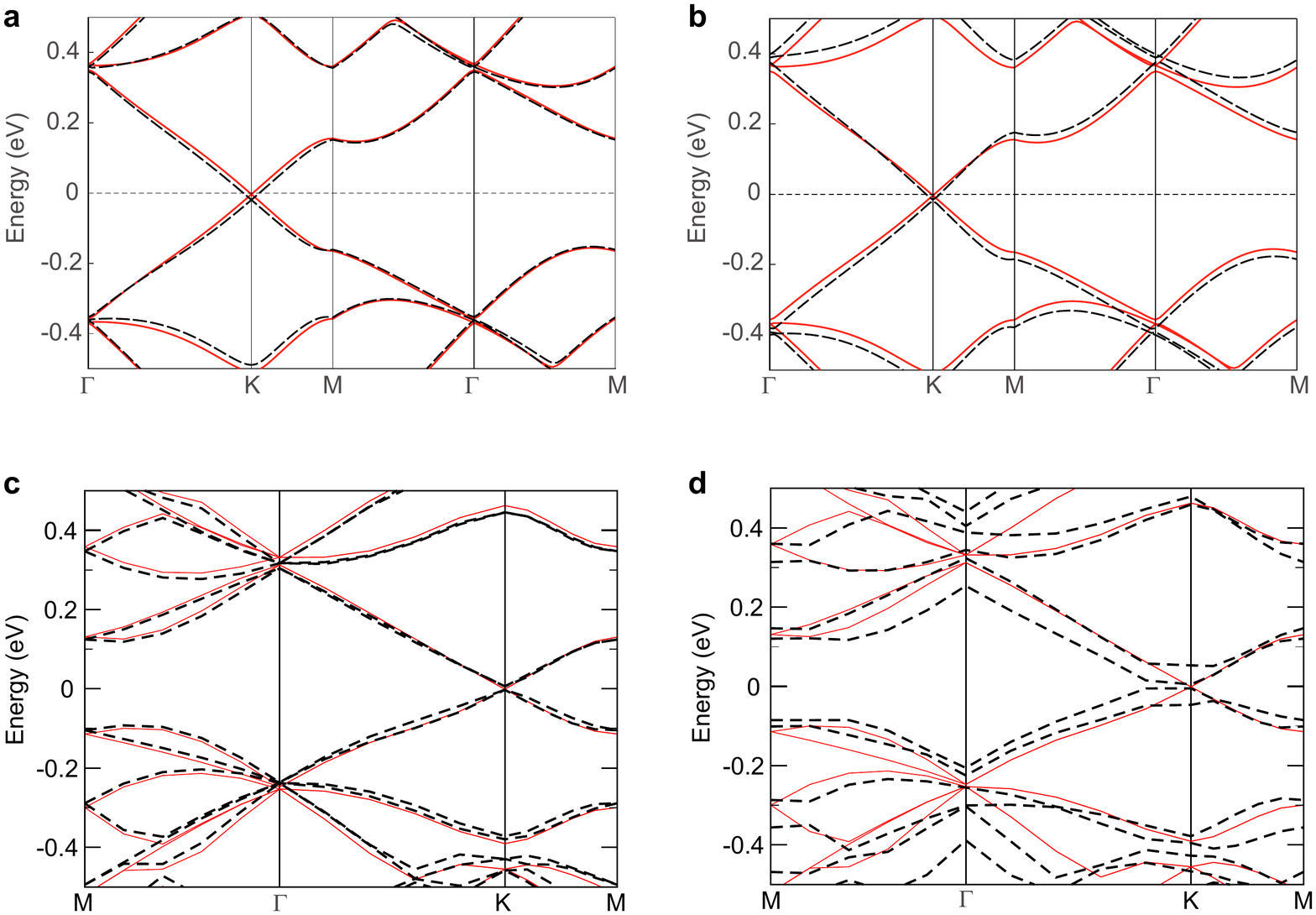}
\end{center}
\caption{Variation of band energies under two deformation configurations ``3344" and ``3142" at $\theta=3.15^\circ$ from the continuum model ({\bf a} and {\bf b}), and from commensurate TBG \emph{ab initio} calculations ({\bf c} and {\bf d}). The energy variations are comparable in order of magnitude in the continuum model and in \emph{ab initio} (slightly larger in \emph{ab initio}).}
\label{figS5}
\end{figure}

We can also construct a configuration with nonzero relative shear. To do this, we can assume the lattice of layer $2$ is \emph{ rotated and deformed} relative to the lattice of layer $1$ so that vector $\mathbf{D}_i$ and $\mathbf{B}_i'$ of layer $2$ coincide with $\mathbf{C}_i$ and $\mathbf{A}_i'$ of layer $1$, respectively, which we shall name as configuration $(31'42')$. The superlattice unit cell is then given by the parallelogram with edges $\mathbf{C}_i$ and $\mathbf{A}_i'$ in layer $1$ (which coincide with the parallelogram with edges $\mathbf{D}_i$ and $\mathbf{B}_i'$ in layer $2$ after relative deformation). This configuration then has relative rotation, shear and expansion, which can be seen in the following calculation of relative displacement field $\mathbf{u}$. The relative displacement field $\mathbf{u}$ of ($31'42'$) relative to the undistorted TBG ($11'22'$) can be solved as follows: Define $\mathbf{U}(\mathbf{r})=\mathbf{u}(\mathbf{r})+\theta_i\mathbf{r}\times\hat{\mathbf{z}}$, which the relative displacement compared to the untwisted (AA stacking) bilayer graphene (The undistorted TBG ($11'22'$) has displacement $\theta_i\mathbf{r}\times\hat{\mathbf{z}}$ relative to the untwisted bilayer graphene). The deformation field $\mathbf{U}$ then satisfies $(\mathbf{C}_i\cdot\nabla)\mathbf{U}\approx \mathbf{D}_i-\mathbf{C}_i$, and $(\mathbf{A}_i\cdot\nabla)\mathbf{U}\approx \mathbf{B}_i-\mathbf{A}_i$. We then find the relative displacement field compared to undistorted TBG ($11'22'$) for large $i$ is
\begin{equation}\label{u3142}
\left(\begin{array}{cc}\partial_xu_x&\partial_xu_y\\ \partial_yu_x&\partial_yu_y\end{array}\right)\approx \left(\begin{array}{cc}\frac{L_i^D-L_i^C}{L_i^C}&\theta_i-\theta_i'\\ \frac{L_i^D-L_i^C}{\sqrt{3}L_i^C}&\frac{\theta_i-\theta_i'}{\sqrt{3}}\end{array}\right)\ .
\end{equation}

We now explain why we choose lattice points $3$ and $4$ ($3'$ and $4'$) to define the deformed TBG. In our \emph{ab initio} calculations, this is to ensure the $K$ point of graphene BZ of layers $1$ and $2$ coincide with the $K_M'$ and $K_M$ points of MBZ when folded into the deformed MBZ, respectively. This allows the band structure to be compared with that of the continuum model (in which $K$ point of two layers always coincide with $K_M'$ and $K_M$). To see this, consider a configuration that has superlattice vectors (which spans the superlattice unit cell parallelogram) $\mathbf{R}_1=m_1\mathbf{a}+n_1\mathbf{b}$ and $\mathbf{R}_2=m_2\mathbf{a}+n_2\mathbf{b}$, where $\mathbf{a}$ and $\mathbf{b}$ are the lattice vectors of the lattice of layer $1$. The reciprocal vectors of the superlattice $\bm{g}_i$ ($i=1,2$) satisfies $\bm{g}_i\cdot\mathbf{R}_j=2\pi\delta_{ij}$. On the other hand, the momentum of the $K$ point of layer $1$ is $\mathbf{K}_{D}$ as we defined in the first section, which satisfies $\mathbf{K}_{D}\cdot\mathbf{a}=\mathbf{K}_{D}\cdot\mathbf{b}=2\pi/3$. Therefore, we find
\[
\mathbf{K}_{D}\cdot\mathbf{R}_1=2\pi(m_1+n_1)/3\ ,\qquad  \mathbf{K}_{D}\cdot\mathbf{R}_2=2\pi(m_2+n_2)/3\ .
\]
This shows $\mathbf{K}_{D}=[(m_1+n_1)\bm{g}_1+(m_2+n_2)\bm{g}_2]/3$. Note that $K_M'$ point of the MBZ is located at momentum $K_M'=-(\bm{g}_1+\bm{g}_2)/3$. Therefore, in order for $K_{D}$ to coincide with $K_M'$ point, one has to have $m_1+n_1\equiv m_2+n_2\equiv-1\ (\text{mod}\ 3)$ (so that $K_{D}$ and $K_M'$ differ by integer multiples of superlattice reciprocal vectors). One can easily see this is satisfied for configuration $(11'22')$ where $\mathbf{R}_1=\mathbf{A}_i$ and $\mathbf{R}_2=\mathbf{A}_i'$. In order to find another configuration satisfying this condition, one has to change $m_1+n_1$ and $m_2+n_2$ by multiples of $3$. The configurations $(33'44')$ and $(31'42')$ are two such configurations which have small deformations.

We then use \emph{ab initio} to calculate the band structure for configurations $(33'44')$ and $(31'42')$ with $i=10$ (see the black dashed lines in Fig.~\ref{figS5}c and \ref{figS5}d), respectively, and compare with that of the undistorted configuration $(11'22')$ (the red solid lines in Fig.~\ref{figS3}c and \ref{figS3}d). Accordingly, we calculate the deformation of TBG band structure from the continuum model (at graphene valley $K$) for configurations (33'44') and (31'42') (using deformations in Eqs. (\ref{u3344}) and (\ref{u3142})) at $\theta=\theta_{10}=3.15^\circ$, respectively, and the results are shown in Fig.~\ref{figS5}a and \ref{figS5}b, where the red solid lines are the original band structure, and the black dashed lines are the deformed band structure. In the figure, the high symmetry points should be understood as those of the MBZ.

First, we note the band structures before deformations match well between the continuum model and the \emph{ab initio}. The continuum model bands appear to be less in number than that of \emph{ab initio}, which is because we have only plotted the bands of continuum model at valley $K$. Besides, for both configurations, one can see that the deformed band energies in the continuum model also have the same order of magnitude as that in \emph{ab initio}.
Therefore, our estimation of electron-phonon coupling strength from the continuum model in the previous sections has the correct order of magnitude.

\end{widetext}

\end{document}